\documentclass[11pt]{article}
\usepackage{cite}
\usepackage{amsfonts,amsmath,amssymb}
\usepackage{mathtools}
\usepackage{stackengine}\stackMath
\usepackage{bbm}
\usepackage{xspace}

\usepackage[subrefformat=parens,labelformat=parens]{subfig}
\usepackage{graphicx}
\usepackage{xcolor}
\usepackage{tikz}
\usetikzlibrary{calc,shapes}
\usepackage{scalerel}

\usepackage[scale=0.8]{geometry}
\numberwithin{equation}{section}

\usepackage{hyperref}

\newcommand\myoverset[2]{\overset{\textstyle #1\mathstrut}{#2}}
\newcommand\myunderset[2]{\underset{\textstyle #1\mathstrut}{#2}}
\newcommand{\fu}{\mathfrak{u}}
\newcommand{\fsu}{\mathfrak{su}}
\newcommand{\fso}{\mathfrak{so}}
\newcommand{\fsp}{\mathfrak{sp}}
\newcommand{\fg}{\mathfrak{g}}
\newcommand{\ff}{\mathfrak{f}}
\newcommand{\fe}{\mathfrak{e}}
\newcommand{\Dim}{\text{dim}}
\newcommand{\DimCB}{\text{dim(CB)}}
\newcommand{\rk}{\text{rk}}
\newcommand{\LST}{\text{LST}\xspace}
\newcommand{\Pic}{\text{Pic}}

\begin{document}
~\vspace{2cm}
\begin{center}
{\huge\bfseries T-Duality and Flavor Symmetries\\[6pt] in Little String Theories}\\[10mm]

Hamza Ahmed$^{a,}$\footnote{\href{mailto:ahmed.ha@northeastern.edu}{ahmed.ha@northeastern.edu}}, Paul-Konstantin Oehlmann$^{a,}$\footnote{\href{mailto:p.oehlmann@northeastern.edu}{p.oehlmann@northeastern.edu}}, Fabian Ruehle$^{a,b,c,}$\footnote{\href{mailto:f.ruehle@northeastern.edu}{f.ruehle@northeastern.edu}} \\[10mm]
\bigskip
{
	{\it ${}^{\text{a}}$ Department of Physics, Northeastern University, Boston, MA 02115}\\[.5em]
	{\it ${}^{\text{b}}$ Department of Mathematics, Northeastern University, Boston, MA 02115}\\[.5em]
	{\it ${}^{\text{c}}$ NSF Institute for Artificial Intelligence and Fundamental Interactions}\\[.5em]
}
\end{center}
\setcounter{footnote}{0} 
\bigskip\bigskip

\begin{abstract}
We explore the T-duality web of 6D Heterotic Little String Theories, focusing on flavor algebra reducing deformations. A careful analysis of the full flavor algebra, including Abelian factors, shows that the flavor rank is preserved under T-duality. This suggests a new T-duality invariant in addition to the Coulomb branch dimension and the two-group structure constants. We also engineer Little String Theories with non-simply laced flavor algebras, whose appearance we attribute to certain discrete 3-form fluxes in M-theory. Geometrically, these theories are engineered in F-theory with non-K\"ahler favorable K3 fibers. This geometric origin leads us to propose that freezing fluxes are preserved across T-duality. Along the way, we discuss various exotic models, including two inequivalent $\text{Spin(32)}/\mathbb{Z}_2$ models that are dual to the same $\text{E}_8 \times \text{E}_8$ theory, and a family of self-T-dual models. 
\end{abstract}

\clearpage
\tableofcontents

%%%%%%%%%%%%%%%%%%%%%%%%%%%%%%%%%%%%%%%%%%%%%%%%%%%%%%%%%%%%%%%%%%%%%%%%%%%%
\section{Introduction}   
Recent advances in understanding generalized symmetries have substantially improved our understanding of supersymmetric six-dimensional theories and their ultraviolet (UV) completions. These theories are special since they can have non-critical BPS strings in their spectrum that become tensionless at certain loci of moduli space~\cite{Witten:1995ex,Seiberg:1996vs}. Moreover, their gauge sectors are strongly constrained by anomalies \cite{Green:1984bx,Sagnotti:1992qw}. 

The advent of F-theory and the geometrization of strongly coupled IIB backgrounds lead to thorough explorations of 6D supergravities (SUGRAs) \cite{Kumar:2010ru} (see \cite{Taylor:2011wt} for a review) and supersymmetric quantum field theories that flow to superconformal field theories (SCFTs) in the UV (see \cite{Heckman:2018jxk} and references therein). In six dimensions however, there is a third class of UV consistent theories, so called little string theories (LSTs) that stayed rather unexplored until recently~\cite{Sagnotti:1987tw,Bianchi:1990yu,Blum:1997mm,Intriligator:1997dh,Aspinwall:1997ye,Font:2017cya,Font:2017cmx,Bhardwaj:2015oru}. LSTs are an interesting intermediate case, since they combine characteristic features of both SUGRAs and SCFTs. For example, they exhibit T-duality, but can also have global symmetries. Furthermore, recent explorations of generalized symmetries \cite{Gaiotto:2014kfa} showed that LSTs have a unique continuous 2-group symmetry \cite{Cordova:2018cvg,Cordova:2020tij,DelZotto:2020sop,Apruzzi:2021mlh} that is absent in SUGRAs and SCFTs. 
Based on the seminal work by Aspinwall and Morrison \cite{Aspinwall:1996vc}, generalized symmetries have renewed interest in geometric realizations of LSTs~\cite{Font:2016odl,Font:2017cya,DelZotto:2022xrh,DelZotto:2023ahf}, explorations of their Higgs branches \cite{DelZotto:2023myd,DelZotto:2023nrb}, and an effort to map out their T-duality structures \cite{DelZotto:2022ohj,DelZotto:2022xrh}. 

Heterotic LSTs capture the degrees of freedom of the world-volume theories of stacks of NS5-branes with flavor groups originating from the 10D $\text{E}_8\times \text{E}_8$ or $\text{Spin}(32)/\mathbbm{Z}_2$ gauge sector probing a transverse $\mathbbm{C}^2/\Gamma_{\fg}$ singularity, with $\Gamma_{\fg} \in SU(2)$. This landscape of theories is enriched by the possibility to include flat connection(s) at infinity that break the 10D gauge groups to residual flavor groups $G_F$. One expects 6D LSTs to inherit a duality map $T_D$ from their 10D origin, which reduces the two 6D theories on a circle and moves to a common locus in the 5D moduli space. 

To map out the space of 6D LSTs and their web of dualities, we should study $T_D$ and its action on the theories. For cases with trivial holonomies, $T_D$ is well understood~\cite{Aspinwall:1997ye}. In this paper, we explore the properties of $T_D$ in the presence of non-trivial flavor holonomies. To do so, we identify invariants of the 6D theories under $T_D$. Two types of invariants that have been discussed previously are the 5D Coulomb branch dimension, and the recently proposed universal 2-group structure constants~\cite{DelZotto:2020sop}
\begin{align}
\label{eq:invars}
    \DimCB  \quad \text{ and } \quad (\kappa_R,\kappa_P) \, .
\end{align} 
Compactifications to 5D come with an additional modulus that specifies the flavor holonomy along the circle and can break the flavor algebra. Since this is part of the common 5D moduli space of T-dual circle-reduced LSTs, the flavor algebras can differ but their rank
\begin{align}
    \rk(G_F)
\end{align}
is expected to be invariant under $T_D$. However, matching flavor ranks across T-duals is complicated by the presence of Abelian symmetries, which can be originate as linear combinations of flavor, baryonic and E-string symmetries. Moreover, they can be broken by ABJ anomalies~\cite{Lee:2018ihr,Apruzzi:2020eqi}. 

The geometric engineering approach using F-theory and its duality to M-theory provides a powerful tool to study such theories. Beyond delivering a consistent construction of strongly coupled and non-Lagrangian theories, T-duality simply corresponds to the existence of inequivalent elliptic fibration structures within a single Calabi-Yau three-fold $X_3$. While there exist general theorems to determine inequivalent elliptic fibrations for compact threefolds~\cite{Kollar:2012pv, Anderson:2016cdu,Grimm:2019bey,Huang:2019pne}, they do not say much about properties of $T_D$. However, once inequivalent fibrations have been identified, one can use birational geometry arguments to show that the invariants~\eqref{eq:invars} are indeed preserved under $T_D$. The flavor group $G_F$, on the other hand, need not be fully encoded in the geometry~\cite{Bertolini:2015bwa}. Furthermore, flavor symmetries may also include non-simply laced algebra factors, as in the case of so-called frozen conformal matter~\cite{Mekareeya:2017jgc}, which further obscures their geometric interpretation. 

The paper is organized as follows:
In Section~\ref{sec:review}, we review basic properties of 6D LSTs and their geometric engineering via F/M-theory. In Section~\ref{sec:flavmatch}, we discuss the	 flavor algebra of Heterotic little strings with emphasis on the Abelian parts required to show flavor rank matching across T-duality. In Section~\ref{sec:nsl}, we discuss models with non-simply laced flavor groups, including their construction via toric geometry, their dualities, and their gravity decoupling limits. We highlight some particular exotic models in Section~\ref{sec:exotic}, before we conclude in Section~\ref{sec:conclusion} . Appendix~\ref{appendix:tables} summarizes novel T-dual families of Heterotic LSTs.

\section{Review of Little String Theories and F-theory}
\label{sec:review}
This section provides a short review of 6D theories with eight supercharges and serves to introduce the basic concepts and notations. Experts may skip this part, while readers who are interested in more details can consult recent reviews~\cite{Weigand:2018rez,Heckman:2018jxk,DelZotto:2022xrh,DelZotto:2023ahf}.

\subsection{General properties of LSTs and their 2-group symmetries}
Six-dimensional (minimally) supersymmetric Quantum Field Theories (6D SQFTs) admit tensor multiplets in their spectrum. Parts of the bosonic field content of such multiplets are self-dual 2-form fields $b^{(2)}$ that couples to 6D non-critical BPS strings. At high energies, these strings can become tensionless and support new degrees freedom required for a UV completion of the theory, i.e., in 6D one does not need gravity for a consistent UV completion~\cite{Seiberg:1996vs,Witten:1995zh}. Two kinds of such UV completions are known: 6D SCFTs and 6D LSTs \cite{Seiberg:1996qx,Seiberg:1997zk}. The difference between the two is that SCFTs have no intrinsic scale, while LSTs have a scale proportional to the tension of a string that does not trivialize under RG flow. This means that LSTs are not like usual QFTs, since they cannot have RG flows to a point where the scale dependence drops out. This intrinsic scale is known as the little string scale, $M_\text{LS}$, and plays a similar role to the Planck scale $M_{P}$ in gravity theories. It is believed that every 6D supersymmetric theory has a tensor branch, where some or all of the scalars in the tensor multiplets have been given a non-zero vacuum expectation value (vev), rendering all BPS strings tensionful. 

The data required to specify a minimal supersymmetric field theory with rank $n_{T}$ (number of tensor multiplets) is given by:
\begin{itemize}
    \item The gauge algebra $\prod_{I} \mathfrak{g}_{I}$.
    \item The flavor algebra $\prod_{k=1}^{n_{f}} \mathfrak{f}_{k}$.
    \item A symmetric matrix $\eta^{IJ}$ of rank $n_{T}$ which encodes the Dirac self-pairings of the $n_{T}$ BPS strings of the theory. The entries of this matrix are integers due to Dirac quantization. The BPS strings source the corresponding 2-form gauge fields $b_{I}^{(2)}$. 
    \item A matrix $\eta_{Ik}$ which encodes information about the flavor algebra $\mathfrak{f}_{k}$ associated to the matter charged under a gauge algebra $\mathfrak{g}_{I}$. Matter representations have to be assigned such that the pure and mixed quartic gauge anomalies cancel.
\end{itemize}
This data can be summarized neatly in a so-called quiver,
\begin{align}
\label{eq:quiverF}  
{\overset{\mathfrak{g}_{1}}{n_{1}}}  \, \,     
{\overset{\mathfrak{g}_2}{n_{2}}} \, \,   
\cdots \, \,
\underset{\left[\mathfrak{f}_{A}\right]_{(\eta_{kA})}}
{\overset{\mathfrak{g}_k}{n_{k}}} \, \,
\cdots \, \,
{\overset{\mathfrak{g}_{n_{T}}}{n_{n_{T}}}} \,,
\end{align}
where we defined $n_{I}=\eta^{II}$, and associate a flavor factor $\ff_{A}$ (in square brackets) to the node where $\eta_{kA} \neq 0$. This is the notation we will follow throughout the paper. The shape of the quiver is such that two nodes $n_{I}$ and $n_{J}$ are adjacent if $\eta_{IJ}=-1$ and non-adjacent otherwise. Other values of $\eta_{IJ}$ are also possible~\cite{Bhardwaj:2015oru,Bhardwaj:2018jgp}, but not relevant to our study here. The matter content of the theory described by the quiver is not written explicitly but can be inferred from demanding absence anomaly freedom, see \cite{Weigand:2018rez,Taylor:2011wt,Johnson:2016qar} for reviews. Typically, one simply has bifundamental matter between adjacent gauge algebras,but sometimes additional matter is required to cancel the 6D anomalies, which leads to the flavor factors $\ff_{A}$. Throughout this paper, we will also suppress $\eta_{kA}$ on the flavor nodes. The geometric interpretation of these quivers will become apparent when we consider the F-theory construction.

The difference between SCFTs and LSTs is encoded in properties of the matrix $\eta^{IJ}$; it is positive definite for the former and positive semi-definite with exactly one zero eigenvalue for the latter. Hence, when the matrix $\eta^{IJ}$ does not have full rank and a null space of dimension one, we get an LST. This means that when we write down the kinetic terms for the 2-form fields, one linear combination vanishes and we have a non-dynamical background field left \cite{Bhardwaj:2015xxa}. We will call this field $B_{\LST}^{(2)}$ field and its associated string the \textit{little string}. Hence 
\begin{align}
\label{tensorrelation}
B_{\LST}^{(2)}=\sum_{I=1}^{n_{T}}l_{\LST,I}\; b_{I}^{(2)} \, ,
\end{align}
where $n_T$ is the number of tensor multiplets, and $l_{\LST,I}$ are the components of the unique null vector of $\eta^{IJ}$, also called the LST \textit{charge vector}. This background field (and the associated little string) generates an Abelian 1-form symmetry $U(1)^{(1)}_{\LST}$ with $l_{\LST,I}$
being the respective charges of the strings that couple to the fields $b_{I}^{(2)}$. The presence of such a symmetry is an important point of differentiation between LSTs and SCFTs. In more detail, the two theories are distinguished by the structure of their 2-form defect group $\mathbbm{D}^{(2)}$ \cite{DelZotto:2015isa,Cordova:2018cvg}, which is determined by the intersection matrix $\eta^{IJ}$: For LSTs, $\mathbbm{D}^{(2)}$ has an extra $U(1)$ factor contributing to the higher form symmetry \cite{Bhardwaj:2020phs,Cordova:2020tij}. This leads to the presence of \textit{continuous} 2-group symmetries in LSTs. 

The characteristic feature of 2-group symmetries is a \textit{mixing} of a 0-form and a 1-form global symmetry. Physically, this mixing is characterized by the mixed anomaly coefficients between the gauge fields corresponding to the gauge algebra $\mathfrak{g}_{I}$ and 0-form global symmetries in our theory. In our case, we have the Poincare symmetry, $R$-symmetry, and the flavor symmetry as global 0-form symmetries of the theory, so we have three anomaly coefficients which we will call $\kappa_{P},\kappa_{R}, \kappa_{\mathfrak{f}_{k}}$. In 6D, the anomalies are encoded in the eight-form anomaly polynomial $\mathcal{I}^{(8)}$, which involves products of topological quantities associated to the curvature forms for gauge and background fields associated to the local and global symmetries, respectively. We will denote these curvature forms by $f_{g_{I}}$ and $f_{n}$, where $n \in \{P,R,\mathfrak{f}_{k}\}$. A contribution to the mixed anomaly of the gauge algebra $\mathfrak{g}_{I}$ with flavor symmetry $\mathfrak{f}_{k}$, $\mathcal{I}^{(8)}_{\mathfrak{g}_{I}\mathfrak{f}_{k},\text{mixed}}$, is (up to numerical factors) given by~\cite{Cordova:2020tij}
\begin{align}
\label{mixed}
\mathcal{I}^{(8)}_{\mathfrak{g}_{I}\mathfrak{f}_{k},\text{mixed}}=c_{2}(f_{g_{I}})\bigg(\eta^{Ik}c_{2}(f_{\mathfrak{f}_{k}})+\check{h}_{\mathfrak{g}_{I}}c_{2}(f_{R}) + (\eta^{II}-2)p_{1}(T)\bigg)\,,
\end{align}
where $c_{2}(f)$ is the second Chern class of the associated vector bundle, $p_{1}(T)$ is the first Pontryagin class of the tangent bundle, and $\check{h}_{\mathfrak{g}_{I}}$ is dual Coxeter number of the gauge algebra $\mathfrak{g}_{I}$. This mixed anomaly indicates that the 6D effective action on the tensor branch is no longer invariant under background gauge transformations corresponding to the global symmetries, and that this non-invariance depends on the gauge algebra\cite{Cordova:2020tij}. However, since each gauge algebra factor has an associated BPS instanton string which couples to $b_{I}^{(2)}$, the Bianchi identity of the three-form field strength $H_{I}^{(3)}$ can be modified to cancel this non-invariance of the action,
\begin{align}
\frac{1}{2\pi}dH^{(3)}_{I}=-\frac{1}{4}\eta^{Ik}c_{2}(f_{\mathfrak{f}_{k}})+\check{h}_{\mathfrak{g}_{I}}c_{2}(f_{R})  - \frac{1}{4}(\eta^{II}-2)p_{1}(T)\,.
\end{align}
On the other hand, the background field $B^{(2)}_{\LST}$ modifies the Bianchi identity to
\begin{align}
\label{eq:dHLST}
\frac{1}{2\pi}dH^{(3)}_{\LST}=-\frac{1}{4}\sum_{k=1}^{n_{f}}\kappa_{\mathfrak{f}_{k}}c_{2}(f_{\mathfrak{f}_{k}})+\kappa_{R}c_{2}(f_{R}) - \frac{1}{4}\kappa_{P}p_{1}(T)\,.
\end{align}
This modification mixes the 1-form and 0-form symmetries of the theory, which is why the anomaly coefficients are also called the 2-group structure constants. Using relation~\eqref{tensorrelation}, we can compute these constants:
\begin{align}
\label{2group}
\kappa_{\mathfrak{f}_{k}}=-\sum_{I=1}^{n_{T}}l_{\LST,I}\eta^{Ik}\,,\qquad  
\kappa_{R}=-\sum_{I=1}^{n_{T}}l_{\LST,I}\check{h}_{\mathfrak{g}_{I}}\,,\qquad 
\kappa_{P}=-\sum_{I=1}^{n_{T}}l_{\LST,I}(\eta^{II}-2)\,.
\end{align}
In the following sections, we will review how to compute them geometrically from Calabi-Yau data using F-theory, focusing on the universal 2-group symmetry constants $\kappa_R$ and $\kappa_P$ \cite{DelZotto:2020sop}. Before that, we would like to remark that such continuous 2-group structure constants do not exist for 6D SCFTs, since the superconformal algebra in 6D prohibits the existence of a conserved 2-form current (coming from the 1-form symmetry) \cite{Cordova:2020tij}, and so terms arising in \eqref{mixed} would have to be cancelled by a dynamical Green--Schwarz mechanism in place of the non-dynamical version in equation~\eqref{eq:dHLST}.

\subsection{Heterotic LSTs}
\label{sec:Horwit}
We focus our discussion on Heterotic LSTs, of which there are two kinds, associated to the $\mathfrak{e}_{8} \times \mathfrak{e}_{8}$ and the $\mathfrak{so}_{32}$ Heterotic strings. We only review the essential ingredients of these theories; a detailed description can be found in \cite{DelZotto:2022ohj}. NS5 branes in the $\mathfrak{so}_{32}$ LST are perturbative and have been described in~\cite{Witten:1995gx}. A rank $N$ LST with minimal SUSY arises as the worldvolume theory of $N$ NS5 branes probing an ALE singularity $\mathbbm{C}^{2}/\Gamma_{\mathfrak{g}}$ in the Heterotic string context, where $\Gamma_{\mathfrak{g}} \subset SU(2)$ and $\mathfrak{g}$ labels the resulting Lie algebra according to the McKay correspondence \cite{Intriligator:1997dh,Blum:1997mm,Blum:1997fw}. From the point of view of the worldvolume theory, the $\mathfrak{so}_{32}$ algebra corresponds to the flavor symmetry of the LST.
Since $\pi_{1}(S^{3}/\Gamma_{\mathfrak{g}}) \cong \Gamma_{\mathfrak{g}}$ (where the lens space $S^{3}/\Gamma_{\fg}$ is the boundary of the ALE space), one also has to specify a choice of flat connection at infinity, which is encoded in the embedding morphism $\mathfrak{\lambda}$:
\begin{align}
\label{eq:flatcons}
\lambda: \pi_{1}(S^{3}/\Gamma_{\mathfrak{g}}) \cong \Gamma_{\mathfrak{g}} \hookrightarrow \text{Spin}(32)/\mathbbm{Z}_{2}\,,
\end{align}
where $\text{Spin}(32)/\mathbbm{Z}_{2}$ is the actual gauge group of the Heterotic string. We will denote the corresponding LST by $\mathcal{\Tilde{K}}_{N}(\lambda;\mathfrak{g})$. Furthermore, a non-trivial $\lambda$ means that the actual flavor symmetry of this LST will be the commutant of $\lambda(\Gamma_{\mathfrak{g}})$ in $\text{Spin}(32)/\mathbbm{Z}_{2}$~\cite{Blum:1997fw,Blum:1997mm,Intriligator:1997dh,Intriligator:1997pq,Brunner:1997gf}. 

The $\mathfrak{e}_{8} \times \mathfrak{e}_{8}$ LST on the other hand can be described in M-theory via the Horava-Witten setup \cite{Horava:1995qa}. In this picture, we have the two M9 branes as end-of-the-world branes of the  11D M-theory interval, each of which carries an $\mathfrak{e}_{8}$ gauge algebra. The LST (of rank $N$) arises as the worldvolume theory of $N$ parallel M5 branes probing a $\mathbbm{C}^{2}/\Gamma$ singularity transverse to the M9 branes \cite{Ganor:1996mu}. Again, the $\mathfrak{e}_{8} \times \mathfrak{e}_{8}$ corresponds to the flavor symmetry of the LST. Due to the presence of the singularity, the M5 branes fractionate. Each fraction is characterized by (discrete) three-form flux, which breaks the gauge algebra $\mathfrak{g}$ to a subalgebra. These fractions were determined via F-theory in \cite{DelZotto:2014hpa}. For this class of LSTs, we have two morphisms $\mu_{a}$, $a=1,2$, that describe the flat connections at infinity for to the two $\mathfrak{e}_{8}$ factors as 
\begin{align}
\label{eq:flatcone}
\mu_{a}: \pi_{1}(S^{3}/\Gamma_{\mathfrak{g}}) \cong \Gamma_{\mathfrak{g}} \hookrightarrow \text{E}_{8}\,. 
\end{align}
We denote this LST by $\mathcal{K}_{N}(\mu_{1},\mu_{2},\mathfrak{g})$.

LSTs in 6D have $\mathcal{N}=(1,0)$ tensor multiplets whose scalar vevs parameterize the tensor branch and whose two-form fields couple to BPS strings. In the Horava-Witten setup for $\mathfrak{e}_{8} \times \mathfrak{e}_{8}$ LSTs, these BPS strings are stretched between M9-M5 (Dirac self-pairing 1) and M5-M5 (Dirac self-pairing 2) branes. For a minimal 6D theory with $N$ M5 branes, we get the quiver
\begin{align}
\lbrack  \mathfrak{e}_{8}\rbrack     \, \,
{\overset{\mathfrak{g}}{1}}\, \,
\underbrace{{\overset{\mathfrak{g}}{2}}\, \, \, \,
{\overset{\mathfrak{g}}{2}}\, \,\, \, 
. \, \,
. \, \,
. \, \,
{\overset{\mathfrak{g}}{2}}\, \, \, \,
{\overset{\mathfrak{g}}{2}} \, \,}_{N-2}
{\overset{\mathfrak{g}}{1}}\, \,
\lbrack  \mathfrak{e}_{8}\rbrack     \,.
\end{align}
In the quiver, each node corresponds to an M5 brane and the number on the node is the Dirac self-pairing of the BPS string associated to it. The gauge algebra $\mathfrak{g}$ is associated to the singularity $\Gamma_{\fg}$. This is a special case of the quiver we introduced in~\eqref{eq:quiverF}, known as the partial tensor branch. One obtains the full tensor branch by taking into account the fractionalization of the respective branes, which can be determined via F-theory.

An $\mathfrak{e}_{8} \times \mathfrak{e}_{8}$ LST can also be obtained from fusing two orbi-instanton theories $\mathcal{T}(\mu_{a},\mathfrak{g})$, which are 6D SCFTs corresponding to an M9-M5 brane system in the presence of an ALE singularity $\mathbbm{C}^{2}/\Gamma_{\mathfrak{g}}$~\cite{DelZotto:2014hpa,Heckman:2015bfa,Mekareeya:2017jgc,Frey:2018vpw} and flat $\fe_8$ connections $\mu_a$.
For $N>2$, one has in addition superconformal matter theories $\mathcal{T}_{N-2}(\mathfrak{g},\mathfrak{g})$ associated to the $(N-2)$ M5 branes probing the singularity. The fusion of these theories can be thought of as a gauging of the diagonal global symmetry algebras $\mathfrak{g}$ of the two SCFTs (which we represent by $\mathrel{\stackon[-1pt]{{-}\mkern-5mu{-}\mkern-5mu{-}\mkern-5mu{-}}{\mathfrak{g}}}$), and then coupling this (now local) algebra $\mathfrak{g}$ to a tensor multiplet. This can be succinctly written as
\begin{align}
\mathcal{K}_{N}(\mu_{1},\mu_{2},\mathfrak{g})=\mathcal{T}(\mu_{1},\mathfrak{g}) \mathrel{\stackon[-1pt]{{-}\mkern-5mu{-}\mkern-5mu{-}\mkern-5mu{-}}{\mathfrak{g}}} \mathcal{T}_{N-2}(\mathfrak{g},\mathfrak{g}) \mathrel{\stackon[-1pt]{{-}\mkern-5mu{-}\mkern-5mu{-}\mkern-5mu{-}}{\mathfrak{g}}} \mathcal{T}(\mu_{2},\mathfrak{g})\,.
\end{align}
This picture is particularly useful since orbi-instanton quivers $\mathcal{T}(\mu_{a},\mathfrak{g})$) have been classified in~\cite{Frey:2018vpw} for any $\mathfrak{g}$.

\begin{figure}[t!]
  \centering
  \begin{tikzpicture}[scale=1.3]
    \draw[very thick, ->] (-4.5,2.5) -- (-2.3,0.8);  
  	\node (3) at (-3.1,1.8) [ text width=3cm,align=center ] {\textbf{ $C_{\mathfrak{e}_{8}\times \mathfrak{e}_{8}}$ }}; 
 	\draw[very thick, ->] (2.5,2.5) -- (0.3,0.8);  
  	\node (3) at (1.0,1.8) [ text width=3cm,align=center ] {\textbf{ $C_{\fso_{32} }$ }}; 
 	\node (1) at (-1,0.8) [draw,rounded corners,very thick,text width=2.8cm,align=center ] {{\bf \textcolor{black}{5D SQFT}}};
	\node (2) at (-4,3) [draw,rounded corners,very thick,text width=3.4cm,align=center ] {{\bf \textcolor{black}{\underline{HET $\text{E}_8 \times \text{E}_8$}\\  LST $\mathcal{K}_{N} (\mu_{1},\mu_{2},\mathfrak{g}) $ \\}}};
	\node (A) at (2,3) [draw,rounded corners,very thick,text width=3.4cm,align=center ] {{\bf \textcolor{black}{\underline{HET $\text{Spin}(32)/\mathbbm{Z}_2$}\\  LST $\mathcal{\Tilde{K}}_{N'}(\lambda;\mathfrak{g}) $ \\}}};
  \end{tikzpicture}
\caption{T-duality of an $\text{E}_8\times \text{E}_8$ and an \text{Spin}(32)$/\mathbbm{Z}_2$ Heterotic LST: The maps $C$ denote respective circle reductions, followed by a flow in the Coulomb branch to a point where the become the identical 5D SQFT.}
\label{fig:Duality}
\end{figure}
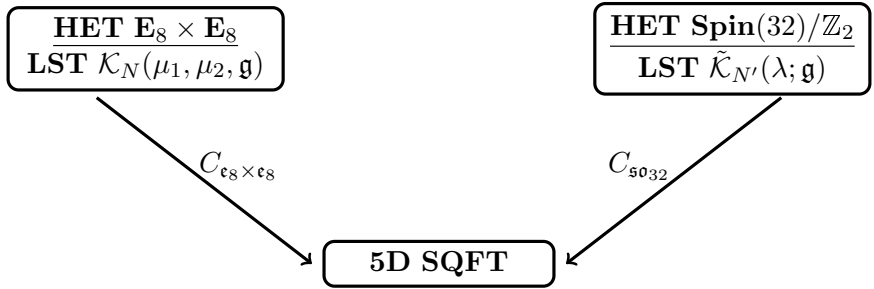

A notion of T-duality for Heterotic LSTs was introduced in~\cite{Intriligator:1997dh}. A theory $\mathcal{K}_{N} (\mu_{1},\mu_{2},\mathfrak{g})$ may have a corresponding T-dual theory $\mathcal{\Tilde{K}}_{N'}(\lambda;\mathfrak{g})$ inherited from the usual T-duality between the $\mathfrak{e}_{8} \times \mathfrak{e_{8}}$ and $\mathfrak{so}_{32}$ 10D Heterotic string compactified on a circle. Wilson lines along the circle can break each of the algebras to the common $\mathfrak{so}_{16} \times \mathfrak{so}_{16}$ subalgebra, and hence reach a common theory at that point in moduli space. This duality extends to Heterotic LSTs, where one compactifies the corresponding LSTs on a circle, and moves in the Coulomb branch of the resulting 5D theories until the same theory is reached. This means that for a given choice of flat connection on the $\mathfrak{e}_{8}\times \mathfrak{e}_{8}$ side~\eqref{eq:flatcone}, one expects a corresponding morphism \eqref{eq:flatcons} on the $\mathfrak{so}_{32}$ side. It is one of the main aims of this work to map out these dual theories. More precisely, we can define the maps $C_{\mathfrak{e}_{8}\times \mathfrak{e}_{8}} $ and $C_{\mathfrak{so}_{32}}$ such that
\begin{align}
C_{\mathfrak{e}_{8}\times \mathfrak{e}_{8}} (\mathcal{K}_{N} (\mu_{1},\mu_{2},\mathfrak{g}))=C_{\mathfrak{so}_{32}}(\mathcal{\Tilde{K}}_{N'}(\lambda;\mathfrak{g}))
\end{align}
i.e., these maps describe the respective circle reductions and flows of the corresponding LSTs in the Coulomb branch to the point where they match, see Figure~\ref{fig:Duality}. From this, one could be tempted to define the T-duality map
\begin{align}
T_{D}: \mathcal{K}_{N} (\mu_{1},\mu_{2},\mathfrak{g}) \rightarrow \mathcal{\Tilde{K}}_{N'}(\lambda;\mathfrak{g})\,,\qquad
T_{D}=C_{\mathfrak{so}_{32}}^{-1}\circ C_{\mathfrak{e}_{8}\times \mathfrak{e}_{8}}\,.
\end{align}
There are, however subtleties in defining this map. First, it was oberved in~\cite{DelZotto:2022xrh} that $T_D$ is one-to-many. Secondly, in this paper, we find examples presented in Section~\ref{sec:exotic} for which $C_{\mathfrak{so}_{32}}$ cannot be uniquely inverted. Nevertheless, there are sets of necessary conditions that can be checked to ensure that two given LSTs are T-dual. 

First, the dimension of the 5D Coulomb branch, dim(CB), along with the rank of the flavor algebra of the LST, should match on field-theoretic grounds. These are given by
\begin{align}
\label{eq:coulFlav}
\Dim(\text{CB})=n_{T} + \text{rk}(\fg), \qquad \Dim(\text{WL})=\rk(\fg_{F})\,,
\end{align}
where $\fg$ and $\fg_{F}$ are the 6D gauge and flavor algebras respectively, and dim(WL) denotes the number of Wilson line parameters of the flavor algebra $\fg_F$ that one can turn on along the circle. The expression for dim(CB) arises because the rank of the gauge algebra in 5D is at least as large as the rank in 6D, since one gets extra Abelian gauge fields from reducing the 2-form fields $b_{I}^{(2)}$ on a circle. Finally, it was conjectured in \cite{DelZotto:2020sop} that T-dual LSTs have the same universal 2-group structure constants defined in~\eqref{2group}. 

\subsection{F-theory construction and T-duality}
\label{sec:revFtheory}
The F-theory construction of 6D LSTs proceeds by compactifying F-Theory on a CY threefold $X$, which is an elliptic fibration over a non-compact K{\"a}hler surface,
\begin{align*}
\begin{array}{c}
    T^{2} \rightarrow X\\
    \;\;\;\;\;\;\;\;\;\;\;\downarrow^{\hspace{0.1cm}\pi}\\
    \qquad \quad B_{2}\,.
\end{array}
\end{align*}
Elliptic fibrations for which the fiber has complex dimension one are called flat; otherwise, they are called non-flat (see e.g.\ \cite{DelZotto:2022xrh} for a discussion in the context of LSTs). We will assume the existence of a section, which implies that $X$ is birational to a Weierstrass model of the form
\begin{align}
    p=Y^{2}+X^{3}+fXZ^{4}+gZ^{6}=0\,.
\end{align}
Here, $X,Y,Z$ are projective coordinates of $\mathbbm{P}^{2}_{2,3,1}$, the holomorphic section is
\begin{align}
s_{0}:~[X:Y:Z]=[1:-1:0]\,,
\end{align}
and $f,g$ are sections of line bundles of the anticanonical class of the base of degrees 4 and 6, respectively. If the discriminant $\Delta=4f^{3}+27g^{2}$ vanishes over some locus in the base, the model becomes singular. The vanishing order of $\{f,g,\Delta\}$ over a base divisor encodes the gauge algebra of the 6D theory according to Kodaira's classification (the gauge theory arises from D7 branes wrapping these divisors in the type~IIB picture). In the base, one has both compact and non-compact divisors, which gives rise to gauge and flavor symmetries, respectively. Matter arises at codimension two in the base where two singular divisors intersect. At these intersection points, one typically obtains bifundamental matter. If matter appears to be charged under a single gauge algebra only, this signals field-theoretically that there is additional matter that is charged under a flavor algebra, which may not realized explicitly in the geometry. 

For Heterotic LSTs, the base $B_2$ is birational~\cite{Bhardwaj:2015oru} to $B_{\LST} =\mathbbm{P}^{1} \times \mathbbm{C}$. Moreover, Heterotic/F-theory duality imposes that the CY threefold also has a K3 fibration
\begin{align}
\label{eq:K3fib}
\begin{split}
    K3 \rightarrow &X\\
    &\downarrow\\
    &\mathbbm{C}\,.
\end{split}
\end{align}
Equipped with this, we can discuss how the LST data is encoded in the threefold geometry. The BPS strings arise from D3 branes wrapping compact curves in the base, and the Dirac pairing $\eta^{IJ}$ is the (negative) intersection matrix of these curves,
\begin{align}
\eta^{IJ}=-(\Sigma^{I},\Sigma^{J})\,,
\end{align}
where $\Sigma^{I}$ is a basis of $H_{2}(B_{2},\mathbbm{Z})$. For an LST we have the \textit{LST curve}, which is the unique curve class in the base with self-intersection~0. This corresponds to the 0 eigenvalue of the matrix $\eta^{IJ}$. After circle compactification of the 6D LST as described in Section~\ref{sec:Horwit}, the dimension of the 5D Coulomb branch is given in terms of the 6D geometric data by
\begin{align}
\label{eq:CBdim}
\text{dim(CB)}=h^{1,1}(X) \, .
\end{align}
The 2-group structure constants can be computed from the quiver data via~\eqref{2group}. Finally, the (geometrically realized) rank of the flavor algebra is just the number of non-compact divisors in the geometry.

If two different 6D LSTs are related by T-duality, they correspond to the same gauge theory upon circle compactification to 5D. From a field theory point of view, the 5D theory UV-completes to two different 6D LSTs. We construct such T-dual theories by identifying different elliptic fibrations within the same CY threefold, which will give rise to different 6D LSTs. This fits in nicely with M/F-theory duality: F-theory on a CY threefold $X$ (times an additional $S^1$) is dual to M-theory on $X$. Hence, if we compactify the two different 6D LSTs on a circle, integrate out the massive KK modes, and move on the Coulomb branch, we should obtain the same 5D theory, corresponding to the 5D theory obtained from compactifying M-theory on $X$. 

To check whether two theories are T-dual, their data~\eqref{2group} and~\eqref{eq:coulFlav} has to match. Since the Coulomb branch dimension is related to $h^{1,1}$, it cannot change under T-duality, which simply corresponds to choosing different elliptic fibrations within the same CY. The matching of the two-group structure constants is less obvious from the geometry, but we will see that they indeed match. Since these invariants already match in 6D, one can view them as necessary conditions for two LSTs to be T-dual. For Heterotic LSTs, $\kappa_{P}$ is always 2, so the non-trivial match will come from $\kappa_{R}$. Finally, since the number of non-compact curves also does not change upon choosing different elliptic fibrations, the rank of the geometric realized flavor algebra should be preserved as well. However, it is known that the full field-theoretic flavor symmetry $\fg_{F}$ is not always realized in F-theory~\cite{Bertolini:2015bwa}. To establish a match of the flavor ranks across T-duals, we carry out a careful field theory analysis that takes non-toric flavor symmetries and ABJ anomalies into account.

\subsection{Toric construction}
\label{sec:torcon}
In this section, we discuss the toric construction of CY threefolds underlying the engineering of Heterotic LSTs in F-theory. We start by describing how to construct a compact threefold with an elliptic fibration structure first, before moving on to the non-compact threefold $X$ that describes the LST. We then elaborate on the K3 fibration \eqref{eq:K3fib}, as this is where the flavor data of the LST is encoded in.
\subsubsection{General construction}
\label{sec:toriccons}
We construct a compact Calabi-Yau $3$-fold $X$ as the anti-canonical hypersurface in an ambient toric variety $\mathcal{A}$ that is defined via a reflexive 4-dimensional lattice polytope $\Delta_4\subset\mathbbm{Z}^4$~\cite{Batyrev:1994pg}. Reflexive polytopes can only have one internal point, which is conventionally taken to be the origin in $\mathbbm{Z}^{4}$. Every point $v_{p} \in \Delta_{4}$, $p=1,\ldots,N_P$, is associated to a complex coordinate $x_{p}$. The toric variety given by the points $v_p$ is in general singular, and one needs to pick a (fine, regular, star) triangulation to fully resolve the singularities while keeping the manifold K\"ahler. The existence of such a triangulation is guaranteed in our case, see for example~\cite{Huang:2019pne}.

We also require that $X$ has an elliptic fibration structure. A sufficient (but very strong and far from necessary~\cite{Anderson:2017aux, Grimm:2019bey}) condition for this is that the ambient variety $\mathcal{A}$ admits a fibration structure, from which the elliptic fibration of the Calabi-Yau hypersurface is inherited,
\begin{align}
\begin{split}
    F \rightarrow &~\mathcal{A}\\ 
    &~\downarrow\\
    &~B_{2}\,.
\end{split}
\end{align}
Here, $B_{2}$ is the base and $F$ is a compact complex two-dimensional weak Fano variety whose anti-canonical hypersurface is a CY 1-fold, i.e., a torus. If we arrange the lattice polytope points $v_p$ in a $4\times N_P$ matrix, the fibration will descend from the ambient space to the CY if this matrix is of the form
\begin{align}
\Delta_{4}\sim
\begin{pmatrix}
  F & 0\\ 
  T & B_{2}
\end{pmatrix}\,.
\end{align}
Since there are 16 reflexive 2D polytopes $F$, there are 16 different toric ambient spaces or fiber types. Since T-dual LSTs arise from different elliptic fibrations of the same CY threefold, T-duality correspond to a choice of which two coordinates we take as the fiber and which two we take as the base. If this can be done in multiple ways such that the the fiber coordinates over the generic point form a 2D reflexive polytope, there are multiple elliptic fibration structures.

To obtain the hypersurface equation for $X$ in $\mathcal{A}$, we follow Batyrev's construction~\cite{Batyrev:1994pg}, which also requires the dual polytope $\Delta^{*}_{4}$
\begin{align}
\Delta^{*}_{4}=\{m \in M_{M_{R}}~|~\langle v, m \rangle \geq -1 \text{ for all } v\in \Delta_{4}\}\,.
\end{align}
If $\Delta_{4}$ is reflexive, so is $\Delta^{*}_{4}$. The hypersurface describing a \textit{compact} CY is then
\begin{align}
\label{CY}
p=\sum_{m \in \Delta^{*}_{4}}\prod_{v_{p} \in \Delta_{4}} a_{m}\; x_{p}^{\langle v_{p}, m \rangle +1}\,.
\end{align}
Note that for non-compact LST bases, the polytope is only semi-convex, i.e.,  the origin $(0,0,0,0)\in\mathbbm{Z}^4$ is the unique point interior in a codimension-one face of $\Delta_4$. Due to the non-compactness, some of the coordinates are invariant under the toric $\mathbbm{C}^{*}$ scalings. Hence, they can appear with arbitrary powers in \eqref{CY}, which means that $\Delta^{*}_{4}$ is an infinite prism~\cite{Bouchard:2003bu,DelZotto:2022xrh}. In practice, one truncates the polynomials at some order. 

We consider Heterotic LSTs for which the base is birational to $\mathbbm{P}^{1} \times \mathbbm{C}$. We denote the affine $\mathbbm{P}^{1}$ coordinates by $v_{w_{0}}$ and $v_{w_{1}}$, and the $\mathbbm{C}$ coordinate by $v_{y_{0}}$. The toric representation of the base is, up to a SL(2,$\mathbbm{Z}$) transformation,
\begin{align}
\label{eq:ToricBase}
\begin{split}
\begin{tikzpicture}
\draw[->]  (9,-5) -- (10,-5) node (mynode) at (10.4,-5){$v_{w_{1}}$};
\draw[<-]  (8,-5) -- (9,-5) node (mynode) at (7.6,-5){$v_{w_{0}}$};
\draw[->]  (9,-5) -- (9,-4) node (mynode) at (9,-3.6){$v_{y_{0}}$};
\filldraw[black] (9,-5) circle (2pt);
\end{tikzpicture}\,.
\end{split}
\end{align}
Since the vectors in the toric diagram for the base~\eqref{eq:ToricBase} do not span $\mathbbm{Z}^2$, it is no longer compact, leading to non-compact Calabi-Yau manifolds. In practice, we may always construct the compact CY threefold first, and then take a decompactification limit which would operationally correspond to removing a base ray in the toric variety; we will call the resulting semi-convex polytope $\hat{\Delta}_{4}$. The singularity structure of the fibers over the base would still be encoded in the CY hypersurface equation~\eqref{CY}, with the singularities over the curves $v_{w_{0}}$ and $v_{w_{1}}$ encoding flavor algebras instead of gauge algebras. If the hypersurface equation is in Tate form, one can directly read off the singularities. Otherwise, one can map it to the birational Weiersstrass form and then use Kodaira's classification to read off the singularity type. Equivalently, one can identify the toric tops~\cite{Bouchard:2003bu} over the base rays of the lattice polytope $\hat{\Delta}_4$ and read off the singularities from them. This makes it also easy to geometrically engineer LSTs with a given gauge and flavor algebra.

\subsubsection{K3 fibration}
The (geometric) flavor algebra of the Heterotic LST is encoded in a K3 fibration $S$ of the elliptically fibered CY threefold $X$. All non-compact fibral divisors of the threefold $X$ pull back to divisors of the K3 fiber $S$ and therefore embed into the Picard lattice \Pic$(S)$. Since the K3 fiber $S$ of $X$ is elliptically fibered, the Picard lattice of $S$ is 
\begin{align}
    \Pic(S)=U \oplus W   \, ,
\end{align}
where $U$ is the 2D hyperbolic lattice corresponding to the divisor class of the elliptic curve and the base, and $W$ is the so-called frame lattice. The latter is given in terms of divisors that do not intersect the zero-section of the elliptic fibration and are hence shrinkable. According to the Shioda-Tate formula, the frame lattice $W$ splits into free Mordell-Weil generators of the elliptic fiber and ADE divisors,
\begin{align}
\label{eq:FlavorK3}
    W = \text{MW}(S) \oplus \sum_{i} \fg_{\text{ADE},i}\,.
\end{align}
These translate to Abelian and non-Abelian flavor algebra factors, respectively. We therefore have $\rk(W)=\rk(\fg_F)$. Similar to the K\"ahler moduli of compact surfaces, volumes of the curves in $W$ correspond to Wilson lines in the flavor algebra of the 5D circle reductions.

K3 fibrations of $X$ can be obtained in a similar way to elliptic fibrations, this time ensuring that the reflexive polytope $\Delta_{4}$ contains a 3D reflexive sub-polytope $\Delta_{3}$. This means that the K3s we construct will be toric hypersurfaces in one of the 4319 reflexive 3D polytopes $\Delta_3$, which is completed to a semi-convex 4D polytope $\hat{\Delta}_4$. We can determine the the Picard lattice Pic$(S)$ of the K3 fiber from $\Delta_3$ and its dual $\Delta_3^{*}$. A subtlety is that there are in general \textit{non-toric} contributions to the Picard lattice, which come from ambient space divisors that intersect the K3 hypersurface more than once, and hence contribute a correction term to the Picard rank. We split the total Picard lattice into a toric and a non-toric part, 
\begin{align}
    \Pic(S)=\Pic(S)_\text{tor} + \Pic(S)_\text{cor}\,. 
\end{align}
The toric contribution comes from the one-dimensional rays and has rank
\begin{align}
\label{eq:Pictoric}
    \rk(\Pic(S)_\text{tor})= l(\Delta_3)-3\,,
\end{align}
where $l$ are points of the polytope $\Delta_3$. The contribution of the non-toric part to the Picard rank is
\begin{align}
\label{eq:Picnontoric}
\rk(\Pic(S)_\text{cor})=\sum_{\theta_1, \theta_1^\circ} l(\theta_1)l(\theta^{\circ}_1) \, ,
\end{align}
where $\theta_1$ and $\theta^\circ_1$ are codimension 1 faces of $\Delta_3$ and their duals in $\Delta_3^{*}$, respectively. The presence of non-toric divisors means that a single ambient divisor class, whose volumes is given by an ambient K\"ahler space modulus $t_i$, descends to multiple independent divisor classes on the K3 fiber $S$, whose volumes are all forced to be equal. In Section~\ref{sec:nsl}
we will exploit this construction to our advantage and use it to construct non-simply laced flavor algebras, even though K3s only have an ADE classification.

%%%%%%%%%%%%%%%%%%%%%%%%%%%%%%%%%%%%%%%%%%%%%%%%%%%%%%%%%%%%%%%%%%%%%%%%%%%%%
\section{Flavor rank matching}
\label{sec:flavmatch}
For any choice of ADE singularity $\Gamma_\mathfrak{g}$ together with a trivial choice for the flat connection at infinity, one finds the following form for the T-dual pair of LSTs~\cite{Aspinwall:1996vc}
\begin{align}
\lbrack  \mathfrak{e}_{8}\rbrack     \, \
\overset{\fg}{1} \, \,
\overset{\fg}{2}\, \,\overset{\fg}{2}\, \,
\ldots\, \, \overset{\fg}{2}\, \, \overset{\fg}{1} \, \, 
\lbrack \mathfrak{e}_{8} \rbrack
\qquad\xleftrightarrow{\text{T-duality}}\qquad
\lbrack  \mathfrak{so}_{32}\rbrack     \, \
{\overset{\mathfrak{sp}_{n}}{1}}  \, \,
{\overset{\mathfrak{so}_{m}}{4}}  \, \,
\dots
\myoverset{{\overset{\mathfrak{sp}_{l}}{1}}}
{\overset{\mathfrak{so}_{k}}{4}}  \, \,
\dots
\end{align}
where the quiver on the $\mathfrak{so}_{32}$ side is made up of $\mathfrak{so}$, $\mathfrak{sp}$ and $\fsu$ gauge algebras, arranged such that the base satisfies fiber-base duality, i.e., the topology is that of the Dynkin diagram of $\mathfrak{g}$, or foldings thereof \cite{Intriligator:1997dh} for choices of $\fg=\{ \fsu_n, \fso_{4n+2}, \fe_6 \}$. In such cases, the duality map $T_D$ appears to be an isomorphism between the two theories and all data, such as the flavor rank, matches straight-forwardly.\footnote{In fact, for $\fg=\fsu_n$, we obtain an additional flavor contribution, which makes the match less trivial than expected.} Once we turn on non-trivial holonomies for the flavor group, the situation however will become much less trivial and we will start finding more than one dual. 

While the non-Abelian flavor algebras correspond to the degeneration of the fiber over non-compact divisors in the base, $\fu_1$ flavor factors arise from the free part of the Mordell–Weil group of the torus fiber~\cite{Lee:2018ihr}. This flavor data is encoded in the Picard lattice of the K3 fibration and hence invariant under T-duality. However, since the K3 does not necessarily capture the full field-theoretic (FT) flavor algebra, the question remains whether the rank of the full flavor algebra actually matches across the T-duality. Field-theoretically, one would expect this to be the case. A subtlety in determining the rank of $\fg_F$ is that $\fu_1$ flavor symmetries can have mixed anomalies with a gauge symmetry. Such a mixed anomaly does not render the theory inconsistent, but breaks the $\fu_1$ flavor symmetry. In the presence of multiple $\fu_1$ factors, all of which have mixed anomalies, some linear combinations might actually be non-anomalous and survive. Once this is properly taken into account, we find that the flavor ranks match for T-dual models. In fact, one may regard the match of flavor ranks across duals as a consistency check of the methods developed in~\cite{Fazzi:2020}, which we review next.

\subsection[Origins of \texorpdfstring{$\fu_1$}{u1} flavor symmetries in 6D theories]{Origins of \texorpdfstring{$\boldsymbol{\fu_1}$}{u1} symmetries in 6D theories}
There are many sources for $\fu_1$ flavor symmetries, arising from the various kinds of matter representations in the 6D quiver theory~\cite{Lee:2018ihr,Fazzi:2020}. In particular, if one has matter in a complex representation, the flavor symmetry is $\fu_{n} \sim \fsu_{n} \times \fu_{1}$.
For example, an $\fsu_n$ on a $(-2)$-curve contributes $2n$ hypermultiplets~\cite{Johnson:2016qar} in the fundamental representation $\boldsymbol{n}$ of $\fsu_n$, and hence leads to a $\fu_{2n}$ flavor symmetry. 

Most of the $\fu_1$ symmetries in our theories will arise in this way, or in a similar way in the case of bifundamental hypermultiplets or other complex representations of $\fsu_n$. Another source for $\fu_1$ flavor factors are single hypermultiplets in pseudo-real representations, since these can be split into two half-hypermultiplets that transform under an $\fso_{2} \sim \fu_1$ flavor symmetry. We summarize the possible representations that can contribute Abelian flavor factors in Table~\ref{tab:flavs1}.

\begin{table}[t]
    \centering
    \begin{tabular}{|c|c|c|} \hline
        $\fg$ &  Matter & Generic Flavor Symmetry\\ \hline
        $\mathfrak{su}_{n}$  & $m \times \boldsymbol{n}$ ($n\geq 3$) & $\mathfrak{u}_{m}$ \\ 
        & $k \times \boldsymbol{\frac{n(n-1)}{2}}$ ($n\geq 5$) & $\mathfrak{u}_{k}$ \\ \hline
      $\mathfrak{sp}_{n/2}$ & $1 \times \boldsymbol{n}$ & $\mathfrak{so}_{2}$ \\ \hline
      $\mathfrak{so}_{10}$ & $m \times \boldsymbol{16}$ & $\mathfrak{u}_{m}$ \\ \hline
       $\mathfrak{so}_{11}$ & $1 \times \boldsymbol{32}$ & $\mathfrak{so}_{2}$ \\ \hline
        $\mathfrak{so}_{12}$ & $1 \times \boldsymbol{32}$ & $\mathfrak{so}_{2}$ \\ \hline
      $\mathfrak{e}_{6}$ & $m \times \boldsymbol{27}$ & $\mathfrak{u}_{m}$ \\ \hline
      $\mathfrak{e}_{7}$ & $1 \times \boldsymbol{56}$ & $\mathfrak{so}_{2}$ \\ \hline \hline
        $\mathfrak{su}_{n \geq 3}$ $\times$ $\mathfrak{su}_{k \geq 3}$  & $m \times \boldsymbol{(n,k)}$ & $\mathfrak{u}_{m}$  \\ \hline
        $\mathfrak{su}_{n \geq 3}$ $\times$ $\mathfrak{sp}_{k/2 \geq 1}$ & $m \times \boldsymbol{(n,k)}$ & $\mathfrak{u}_{m}$  \\ \hline
      \hline
    \end{tabular}
    \caption{Flavor symmetries with $U(1)$ contributions for matter charged under one and two gauge algebra factors. All hypermultiplets are full hypers.}
    \label{tab:flavs1}
\end{table}

Lastly, one can also obtain $\fu_1$ flavor factors from E-string theories, i.e., from a $(-1)$-curve with no gauge algebra. This theory has a generic $\mathfrak{e}_8$ flavor symmetry that can be partially gauged,
\begin{align}
\cdots \, \,
{\overset{\mathfrak{g}_L}{m}} \, \,
{\overset{{\left[\mathfrak{g}_{F}\right]}}{1}} \, \, 
{\overset{\mathfrak{g}_R}{n}} \, \,
\cdots \, \,
\end{align}
where $g_{L} \times g_{R} \subset \mathfrak{e}_{8}$. The ungauged part, given by the commutant, contributes a global symmetry $\mathfrak{g}_{F}=[\mathfrak{g}_{L}\times \mathfrak{g}_{R}, \mathfrak{e}_{8}]$. This commutant can contain $\fu_1$ factors, for example when $\fg_L=\fe_6$ and $\fg_R=\fsu_2$, we get $\fg_F=\mathfrak{u}_1$.

\subsection{ABJ anomalies}
\label{sec:ABJAnomalies}
The naive Abelian flavor contributions discussed above are can be anomalous and hence may not survive in the EFT. A hypermuliplet which transforms in a representation $\boldsymbol{\rho}$ of a non-Abelian gauge algebra $\mathfrak{g}$ and with charge $q$ under a global $\fu_1$ corresponds to a mixed anomaly in the 6D anomaly polynomial $\mathcal{I}^{(8)}$ of the form
\begin{align}
\label{ABJf}
\mathcal{I}^{(8)}_{\text{ABJ}} \sim \frac{1}{6}qf_{\fu_1}\text{Tr}_{\boldsymbol{\rho}}f_{\fg}^{3}\,.
\end{align}
This can be non-vanishing if the gauge algebra $\fg$  has a non-trivial cubic Casimir, which only occurs for $\mathfrak{g}=\fsu_{n}$ with $n > 2$. Each such $\fsu_n$ algebra imposes a non-trivial constraint for the global $\fu_1$ flavor factors, such that typically only a few linear combinations of $\fu_1$ factors remain non-anomalous and hence unbroken. Generically, the number of non-anomalous $\fu_1$ factors is given by
\begin{align}
\label{condABJ}
\#(\text{$\fu_1$ factors})=\#(\text{naive $\fu_1$ factors}) - \#(\mathfrak{su}_{\text{n}}~\text{flavor algebras with}~n>2)\,,
\end{align}
where the naive number of $\fu_1$ factors can be read off from Table~\ref{tab:flavs1}.

Let us illustrate this in an example, where the LST is given by $M$ M5 branes probing a $\mathbbm{C}^{2}/\mathbbm{Z}_{N}$ singularity with $N>2$ in the $\text{E}_8 \times \text{E}_8$ Heterotic string. This theory is described by the quiver 
\begin{align}
\label{ABJex} 
\begin{split}
~~~~~~~\overbrace{~\hspace{3.4cm}}^{\text{left ramp}}~\overbrace{\hspace{4.0cm}}^{\text{plateau }(M\times\;\mathfrak{su}_{N})}~\overbrace{~\hspace{3.2cm}}^{\text{right ramp}}~\hspace{2.2cm}\\[-2pt]
\begin{array}{lllllllllllllllllllllll}
 \lbrack  \mathfrak{e}_{8}\rbrack  \, \, 
1 \, 
\,  {\overset{\mathfrak{su}_{1}}{2}}   
& &  \hspace{-0.4cm} {\overset{\mathfrak{su}_2}{2}}& & \hspace{-0.5cm} 
{\overset{\mathfrak{su}_3}{2}} &    \ldots  
&   {\overset{\mathfrak{su}_{N}}{2}}&  \cdots    
&  \hspace{-0.4cm}  {\overset{\mathfrak{su}_{N}}{2}}  & \cdots &  {\overset{\mathfrak{su}_{N}}{2}} &  \cdots 
& \hspace{-0.4cm} {\overset{\mathfrak{su}_3}{2}}&
& \hspace{-0.4cm}  {\overset{\mathfrak{su}_{2}}{2}}  && \hspace{-0.4cm} 
{\overset{\mathfrak{su}_{1}}{2}}   \, \,  1 \, \, 
\lbrack \mathfrak{e}_{8} \rbrack      \\   
  & \hspace{-0.2cm}\scaleto{{(\overline{\textbf{1}},\textbf{2})_{0}}}{8pt} \hspace{-0.5cm}&&\hspace{-0.2cm}\scaleto{{(\overline{\textbf{2}},\textbf{3})_{0}}}{8pt}\hspace{-0.5cm}    & &\hspace{-0.1cm}\scaleto{{(\overline{\textbf{N-1}},\textbf{N})_{0}}}{8pt}\hspace{-0.1cm} & \scaleto{{ (\textbf{N})_{N}}}{8pt}& \hspace{-0.1cm}\scaleto{{(\overline{\textbf{N}},\textbf{N})_{-1}}}{8pt}\hspace{-0.1cm}&&\scaleto{{(\overline{\textbf{N}},\textbf{N})_{-1}}}{8pt}&\scaleto{{ (\textbf{N})_{N}}}{8pt}&\hspace{-0.1cm}\scaleto{{(\overline{\textbf{N}},\textbf{N-1})_{0}}}{8pt}\hspace{-0.1cm}&&\hspace{-0.2cm}\scaleto{{(\overline{\textbf{3}},\textbf{2})_{0}}}{8pt}\hspace{-0.5cm}&&\hspace{-0.2cm}\scaleto{{(\overline{\textbf{2}},\textbf{1})_{0}}}{8pt}\hspace{-0.5cm}& 
\end{array}\,,\hspace{1cm}
\end{split}
\end{align} 
where we also indicate the various matter representations and for later convenience already their charge under the (single non-anomalous) $\fu_1$ flavor symmetry. We use a convention where $\fsu_0=\fsu_1=\emptyset$. Let us explain how to compute these charges. The number of $\fu_1$ flavor symmetry factors can be counted as follows:
\begin{itemize}
\item There are $(N-3)$ flavor $\fu_1$ factors on the left ramp, which we label by $\ell=4,\ldots, N$, associated with the $(N-3)$ bifundamentals $(\ell-1,\ell)$. Note that the bifundamentals $(1,2)$ and $(2,3)$ are uncharged under flavor $\fu_1$ factors. Furthermore, we label the $(N-1)$ gauge factors by $\ell=1,\ldots, N-1$.
\item On the middle plateau, there are $M$ $\fsu_N$ factors which we label by $p=N,\ldots, N + M$. The ones on the left and right ``edge'' comes with a single fundamental in addition to the bifundamentals, signaling the presence of an additional $\mathfrak{u}_1$ flavor factor. We thus get $(M-1)$ flavor $\fu_1$ factors from the plateau, with two extra contributions from the edges.
\item The right ramp is identical to the left ramp and hence also contributes $(N-3)$ flavor $\fu_1$ factors, which we label by $r=N+M+1,\ldots, 2N+M-3$; we label the $(N-1)$ gauge factors by $r=1,\ldots, 2N+M-1$.
\end{itemize}
The total number of naive $\fu_1$ factors from the quiver is then
\begin{align}
(N-3) + 1 + (M-1) + 1 + (N-3)=2N+M-5\,.
\end{align}
Using the prescription in \eqref{condABJ}, the number of non-anomalous $\fu_1$ factors is 
\begin{align}
(2N+M-5)-(M+2(N-3))=1\,.
\end{align}
Thus, a single $\fu_1$ flavor factor is expected to be non-anomalous and survive. 

The $\fu_1$ charges $q$ in such a linear quiver of $\mathfrak{su}_{n}$ factors is fixed by the adjacent $\mathfrak{su}_{n}$ groups. Choosing a normalization such that the smallest $\fu_1$ charge is 1, we can write the ABJ anomaly~\eqref{ABJf} as
\begin{align}
\label{ABJanom}
\begin{split}
6\mathcal{I}^{(8)}_\text{ABJ}&=
\sum_{\stackrel{\text{\tiny left}}{\text{ \tiny ramp}}}q_\ell f_{\fu_{1_\ell}}\text{Tr}(f^{3}_{\ell}) +
\sum_{\text{\tiny plateau}} q_p f_{\fu_{1_p}}\text{Tr}(f^{3}_{p}) +
\sum_{\stackrel{\text{\tiny right}}{\text{ \tiny ramp}}} q_r f_{\fu_{1_r}}\text{Tr}(f^{3}_{r})\\
&= \sum_{\ell=3}^{N-1} \text{Tr}(f^{3}_{\fsu_{\ell}})\left[-\ell f_{\fu_{1_\ell}}+(\ell+1) f_{\fu_{1_{\ell+1}}}\right] - \text{Tr}(f^{3}_{\mathfrak{su}_{N}}) f_{\fu_{1_N}} \\ 
&+ \sum_{p=N}^{N+M}\text{Tr}(f^{3}_{\fsu_{N_{p}}})\left[-N f_{\fu_{1_{p}}}+ N f_{\fu_1{_{p+1}}})\right] \\ 
&+ \sum_{r=N+M+1}^{2N+M-3} \text{Tr}(f^{3}_{\fsu_{r}})\left[r f_{\fu_{1_{r}}}-(r+1) f_{\fu_{{r+1}}}\right] - \text{Tr}(f^{3}_{\fsu_{N+M}}) f_{\fu_{1_{N+M}}} \, .\\ 
\end{split}
\end{align}
This anomaly is non-vanishing. However, using a different $\fu_1$ basis, one can find exactly one linear combination for which the anomaly vanishes, since the matrix of $\fu_1$ flavor charges is a $(M+2N-6) \times (M+2N-5)$ matrix, where the rows refer to the number of $\fsu_{N}$ algebras with $N>2$ and the columns are the number naive $\fu_1$ factors in~\eqref{ABJanom}. This matrix has a one-dimensional kernel
\begin{align}
U(1)_\text{non-anomalous}=\{\underbrace{0,0, \cdots, 0}_{\text{left ramp}}, \underbrace{N, -1,-1, \cdots ,-1,-1,N}_{\text{plateau}},\underbrace{0,\cdots,0,0}_{\text{right ramp}} \}
\end{align}
As expected~\cite{Fazzi:2020}, all $\fu_1$ charges in the ramp regions are zero and only the plateau region has matter charged under the flavor $\fu_1$; we give the $\fu_1$ charges of this surviving flavor symmetry in~\eqref{ABJex}.

Let us now consider the $\text{Spin}(32)/\mathbbm{Z}_2$ T-dual of the above LST. It turns out that the dual quivers depend on whether wee have an $\fsu_{2N}$ or $\fsu_{2N+1}$ singularity. For $\fsu_{2N}$ with $M$ M5 branes probing the singularity, the T-dual LST quiver is
\begin{align}
\label{eq:ExDualQ1}
\lbrack  \mathfrak{so}_{32}\rbrack    \, \,   
{\overset{\mathfrak{sp}_{4N+M-1}}{1}}   \, \,   
\underbrace{{\overset{\mathfrak{su}_{8N+2M-10}}{2}} \, \, 
{\overset{\mathfrak{su}_{8N+2M-18}}{2}} \, \,
\cdots \, \,
{\overset{\mathfrak{su}_{2M+6}}{2}}}_{(N-1) \text{~nodes}} \, \,
{\overset{\mathfrak{sp}_{M-1}}{1}} \,.
\end{align}
As expected from the Heterotic picture, we get an $\mathfrak{so}_{32}$ flavor symmetry. Using Table~\ref{tab:flavs1} and equation~\eqref{condABJ}, we find that such a quiver has a single non-anomalous $\fu_1$, and all bifundamentals are charged under it. This shows that the total flavor rank is 17, matching the flavor rank from the $\mathfrak{e}_{8} \times \mathfrak{e}_{8}$ dual. Following the procedure outlined above, we can compute the the non-anomalous $\fu_1$ flavor charges from the kernel of the associated charge matrix,
\begin{align}
\begin{split}
q_{1}&=(-4)^{N-2}\;(M-1)\;\left(\frac{M+3}{4}\right)_{N-2}\,, \\[10pt]
q_{\alpha}&=\frac{(4N+M-1)(4N+M-5)}{(4N-4\alpha+M-1)(4N-4\alpha+M+3)}\;q_{1}, \hspace{0.5cm} 2 \leq \alpha \leq N-1  \\[12pt]
q_{N}&=\frac{(M+7)(4N+M-1)(4N+M-5)}{(M+3)(M-1)^{2}}\;q_{1}\,,
\end{split}
\end{align}
for the $N$ bifundamentals appearing in~\eqref{eq:ExDualQ1}. In these expressions, $(a)_{n}$ is the Pochhammer symbol
\begin{align}
    (a)_{n}=\prod_{k=1}^{n}(a+k-1) \, .
\end{align}
Similarly, for  $\mathfrak{su}_{2N+1}$ with $M$ M5 branes probing the singularity, the T-dual LST quiver is
\begin{align}
\lbrack  \mathfrak{so}_{32}\rbrack    \, \,   
{\overset{\mathfrak{sp}_{4N+M+1}}{1}}   \, \,   
\underbrace{{\overset{\mathfrak{su}_{8N+2M-6}}{2}} \, \, 
{\overset{\mathfrak{su}_{8N+2M-14}}{2}} \, \,
\cdots \, \,
{\overset{\mathfrak{su}_{2M+10}}{2}}}_{(N-1)~\text{nodes}} \, \,
\underset{{\left[N_{A}=1\right]}}
{\overset{\mathfrak{su}_{2M+2}}{1}} \,.
\end{align}
The last curve has matter transforming in the two-fold antisymmetric irrep, which contributes to the naive $\fu_1$ counting according to Table~\ref{tab:flavs1}. So we have a total of $(N+1)$ $\fu_1$ factors, out of which a single non-anomalous $\fu_1$ combination survives. The charges can again be written in terms of Pochhammer symbols, 
\begin{align}
\begin{split}
q_{1}&=(-1)^{N+1}4^{N-2}(M+1)\bigg(\frac{M+5}{4}\bigg)_{N-2}\,, \\[10pt]
q_{\alpha}&=\frac{(4N+M-3)(4N+M+1)}{(4N+M-4\alpha+1)(4N+M-4\alpha+5)}~q_{1}, \hspace{0.5cm} 2 \leq \alpha<N\,, \\[12pt]
q_{A}&=\frac{(2M+10)(4N+M-3)(4N+M+1)}{(M+1)(M+5)}~q_{1}\,,
\end{split}
\end{align}
where $q_{A}$ is the charge of the anti-symmetric matter irrep.  Hence, $M$ M5 branes probing a $\mathbbm{Z}_N$ singularity in either $\text{E}_8\times \text{E}_8$ or $\text{Spin}(32)/
\mathbbm{Z}_2$ Heterotic theory give rise to additional $\fu_1$ flavor symmetry contributions. The flavor rank matches when ABJ flavor anomalies are taken into account.

\subsection[Flavor rank matching in \texorpdfstring{$D_{n}$}{Dn} models]{Flavor rank matching in \texorpdfstring{$\boldsymbol{D_{n}}$}{Dn} models}
In this section, we illustrate the construction of models with $D_{n}$ singularities and non-trivial choices of flat connection at infinity (i.e., different Higgs branches of the $\mathfrak{e}_{8} \times \mathfrak{e}_{8}$ theory). To construct these theories, we can start with a theory with some gauge algebra on an isolated curve. The generic FT flavor algebra of such theories has been computed in~\cite{Bertolini:2015bwa,Johnson:2016qar}. Upon gauging a part of the flavor algebra, it becomes a gauge algebra on adjacent curves in the quiver, and we can read off the remaining FT algebra from the commutant. If this commutant contains $\fu_1$ factors, we compute the ABJ anomaly and find the non-anomalous linear combinations of $\fu_1$ generators. The examples we describe here have more than one dual theory whose flavor ranks match in a non-trivial way. A more exhaustive list of models together with their field-theoretic flavor symmetries can be found in Appendix~\ref{appendix:tables}.

\subsubsection*{Examples for a \texorpdfstring{$\boldsymbol{D_{4}}$}{D4} singularity}\vspace{-12pt}
Consider the M NS5 branes in the $\text{E}_8 \times \text{E}_8$ Heterotic theory with a $D_{4}$ singularity and holonomies $(\mu_{1},\mu_{2})$ that break the flavor algebra to $\mathfrak{e}_{6}$. This theory may be represented as a fusion of two orbi-instanton theories $\mathcal{T}(\fe_{6},\fso_8)$, with an $\mathcal{T}_{M-2}(\fso_8, \fso_8)$ conformal matter theory,
\begin{align}
\label{eq:fuse}
\lbrack  \mathfrak{e}_{6}\rbrack    
{\overset{{{{\left[\mathfrak{u}_{1}\right]}}}}{1}} \, \,
{\overset{\mathfrak{su}_2}{2}} \, \,
\underset{{\left[\mathfrak{su}_{2}\right]}}
{\overset{\mathfrak{so}_7}{3}} \, \, 
1 \, \, 
\textcolor{blue}{[ \mathfrak{so}_8 ]}
\mathrel{\stackon[-1pt]{{-}\mkern-5mu{-}\mkern-5mu{-}\mkern-5mu{-}}{\textcolor{blue}{\mathfrak{so}_8}}}
\textcolor{blue}{[\mathfrak{so}_8]} \underbrace{ 
1\,
{\overset{\mathfrak{so}_8}{4} \, 1\,  
{\overset{\mathfrak{so}_8}{4}} \, 1\,
{\overset{\mathfrak{so}_8}{4}}  \,
 \ldots  1\, }}_{M-2}
\textcolor{blue}{[\fso_{8}]}
\mathrel{\stackon[-1pt]{{-}\mkern-5mu{-}\mkern-5mu{-}\mkern-5mu{-}}{  \textcolor{blue}{\mathfrak{so}_{8}}}} 
\textcolor{blue}{[\mathfrak{so}_8]} \ldots
\lbrack \mathfrak{e}_{6} \rbrack\,,
\end{align}
where the blue flavor factors are not realized in the toric geometry construction. This fusion results in an LST with quiver
\begin{align}
\label{eq:d4ex}
\lbrack  \mathfrak{e}_{6}\rbrack     \, \,    
{\overset{{\textcolor{blue}{{\left[\mathfrak{u}_{1}\right]}}}}{1}} \, \,
{\overset{\mathfrak{su}_2}{2}} \, \,
\underset{\textcolor{blue}{\left[\mathfrak{su}_{2}\right]}}
{\overset{\mathfrak{so}_7}{3}} \, \, 
1 \, \, 
{\overset{\mathfrak{so}_8}{4}} \, \,
\underbrace{1 \, \,
{\overset{\mathfrak{so}_8}{4}} \, \,}_{\times (M-3)} \, \,
1 \, \, 
\underset{\textcolor{blue}{\left[\mathfrak{su}_{2}\right]}}
{\overset{\mathfrak{so}_7}{3}} \, \, 
{\overset{\mathfrak{su}_{2}}{2}} \, \,
{\overset{{\textcolor{blue}{{\left[\mathfrak{u}_{1}\right]}}}}{1}} \, \,
\lbrack \mathfrak{e}_{6} \rbrack   \, ,   \qquad 
\end{align}
The little string charge vector $\vec{l}_{\LST}$, defined in \eqref{tensorrelation}, is
\begin{align}
\vec{l}_{\LST}=(1,1,1,2,1 ,\underbrace{2,1}_{\times (M-3)},2,1,1,1) \, .
\end{align}
The torically realized flavor is $\mathfrak{e}_{6}^{2} \times \mathfrak{u}_{1}$. The $\mathfrak{u}_{1}$ appears since we have realized the quiver with an elliptic fibration given by the fiber polytope $F_{11}$, which has a non-trivial MW group~\cite{Klevers:2014bqa}. The FT flavor is $\mathfrak{e}_{6}^{2} \times \mathfrak{u}_{1}^{2} \times \mathfrak{su}_{2}^{2}$, as indicated in the quiver, which means that the rank of all FT flavor algebras is 16. This theory has two T-duals. 

\begin{figure}[t!]
\begin{picture}(0,60)
\put(70,0){\includegraphics[scale=0.5]{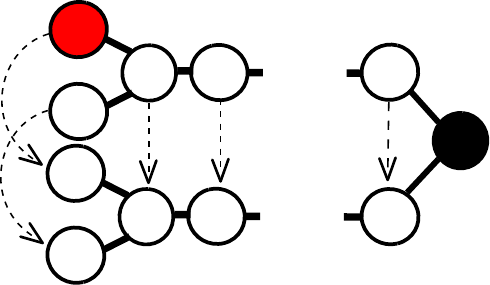}} 
\put(290,15){\includegraphics[scale=0.5]{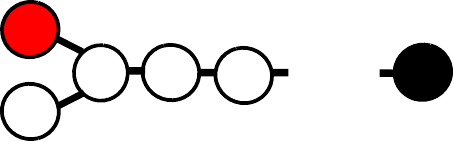}} 
\put(137,16){$\ldots$}
\put(137,50){$\ldots$}
\put(364,31){$\ldots$}
\put(220,30){\Large $\xrightarrow{\text{T-dual}}$}
\end{picture}
\caption{\label{fig:SO4nFolding} Folding of the base quiver in~\eqref{eq:ex1DQ1}. The affine node is in red, and the black node is the central node that is invariant under the folding.}
\end{figure}
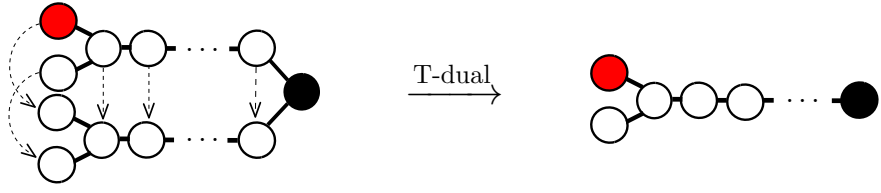

\textbf{First dual}\\
The first dual is given by the quiver
\begin{align}
\label{eq:ex1DQ1}
\textcolor{blue}{\lbrack  \mathfrak{u}_{1}\rbrack } \times \left( 
{\overset{\mathfrak{su}_{2M-2}}{2}}  \, \,   
{\overset{\mathfrak{sp}_{2M-2}}{1}}  \, \,     
{\overset{\mathfrak{su}_{2M+6}}{2}} \, \,  
\textcolor{blue}{\lbrack \mathfrak{su}_{16} \rbrack }\right)  \, ,   \qquad \vec{l}_{\LST}=(1, 2,1) \,,
\end{align}
where the $\fu_1$ factor on the right indicates a (non-toric) Abelian flavor factor under which all bifundamental matter of the theory is charged. The elliptic fibration descends from an $F_{13}$ fiber polytope, which comes with $\mathbbm{Z}_{2}$ MW torsion. The topology of the base is that of an $\mathfrak{so}_{8}$ Dynkin diagram folded along the central node as depicted in Figure~\ref{fig:SO4nFolding}. We will frequently encounter folded Dynkin diagrams under fiber-base-duality in models with MW torsion (see Appendix~\ref{appendix:tables}), in accordance with the discussion in~\cite{DelZotto:2022ohj}.

The torically realized part of flavor algebra is $\mathfrak{su}_{12} \times \mathfrak{su}_{2}^2$, and all intersect the rightmost $\fsu_{2M+6}$ node. The rank of the torically realized flavor algebras matches the torically realized flavor rank of the T-dual~\eqref{eq:d4ex}. However, in both cases we also have flavor algebra factors that are not torically realized. These include $\fu_1$ factors coming from the bifundamentals in the quiver, cf.\ Table~\ref{tab:flavs1}.

To find the FT flavor, we first reconstruct the matter required on each curve by anomaly freedom~\cite{Johnson:2016qar}. From these matter irreps, we can determine their flavor algebras. The gauge algebras on the neighboring curves can be thought of as arising from gauging a part of this flavor algebra, such that the remnant flavor algebra is given by the commutant of the full flavor algebra and the adjacent gauge subalgebras. For the quiver~\eqref{eq:ex1DQ1}, this procedure plays out as follows:
\begin{itemize}
    \item The curvess ${\overset{\mathfrak{su}_{2M+6}}{2}}$ require $ (4M+12)$ hypermultiplets in the fundamental representation, which transform in a $\mathfrak{u}_{4M+12} \sim \mathfrak{su}_{4M+12} \times \mathfrak{u}_{1,A}$ flavor algebra. Upon gauging the $\mathfrak{sp}_{2M-2}$ subalgebra of the adjacent curve,\footnote{We indicate which subalgebra we gauged by an asterisk.} we obtain $\mathfrak{su}_{4M+12} \times \mathfrak{u}_{1,A}\supset\mathfrak{su}_{16} \times \mathfrak{sp}_{2M-2}^{*} \times \mathfrak{u}_{1,A} \times \mathfrak{u}_{1,B}$.
    \item The curves ${\overset{\mathfrak{sp}_{2M-2}}{1}}$ require ($4M+4$) hypermultiplets in the fundamental representation, which have an $\mathfrak{so}_{8M+8}$ flavor algebra. Gauging the $\mathfrak{su}_{2M+6}$ and the $\mathfrak{su}_{2M-2}$ subalgebras on the adjacent curves gives $\mathfrak{so}_{8M+8}\supset\mathfrak{su}_{2M+6}^{*} \times \mathfrak{su}_{2M-2}^{*} \times \mathfrak{u}_{1,B}  \times \mathfrak{u}_{1,C} $.
\end{itemize}
Hence, the total naive flavor symmetry algebra is $\mathfrak{su}_{16} \times  \mathfrak{u}_{1,A}  \times \mathfrak{u}_{1,B}  \times \mathfrak{u}_{1,C} $. Carrying out the ABJ anomaly analysis described in Section~\ref{sec:ABJAnomalies}, we find that only a single linear combination of $\fu_1$ factors is anomaly-free. Hence, the actual FT flavor symmetry is $\mathfrak{su}_{16} \times \mathfrak{u}_{1}$ with rank 16, matching that of the original model.

\textbf{Second dual}\\
The quiver of the other T-dual model is
\begin{align}
\textcolor{blue}{\lbrack  \mathfrak{so}_{24}\rbrack}     \, \,   
{\overset{\mathfrak{sp}_{M+3}}{1}}   \, \,  
\myoverset{\overset{\mathfrak{sp}_{M-3}}{1}}
{\myunderset{\overset{\mathfrak{sp}_{M-3}}{1}}
{\overset{\mathfrak{so}_{4M+4}}{4}}}  \, \, 
{\overset{\mathfrak{sp}_{M-1}}{1}} \, \,
\textcolor{blue}{\lbrack \mathfrak{so}_{8} \rbrack}   \, ,   \qquad 
\vec{l}_{\LST}=(1,1,1,1,1) \, .
\end{align}
The elliptic fibration in this T-dual comes from an $F_{9}$ fiber polytope, which has MW rank 2, as well as an $\fsu_2$ and $\fso_{20}$ fiber. The rank of the torically realized flavor algebra is thus 13, just as the toric flavor ranks of the other two T-duals. To get the full flavor algebra, we proceed as before:
\begin{itemize}
    \item The curve ${\overset{\mathfrak{sp}_{M-1}}{1}}$ requires ($2M+6$) hypers, which have an $\mathfrak{so}_{4M+12}$ flavor algebra. Gauging the $\mathfrak{so}_{4M+4}$ subalgebra of the adjacent node gives $ \mathfrak{so}_{4M+12}\supset\mathfrak{so}_{4M+4}^{*}\times \mathfrak{so}_{8}$.
    \item The curve ${\overset{\mathfrak{sp}_{M+3}}{1}}$ requires ($2M+14$) hypers, which have an $\mathfrak{so}_{4M+28}$ flavor algebra. Gauging the $\mathfrak{so}_{4M+4}$ subalgebra of the adjacent nodes gives $\mathfrak{so}_{4M+28}\supset\mathfrak{so}_{24} \times \mathfrak{so}_{4M+4}^{*}$.
\end{itemize}
Hence, the total flavor symmetry is $\mathfrak{so}_{8} \times \mathfrak{so}_{24}$, which has again rank 16. This shows that the FT flavor rank matches across all T-dual models. Besides the flavor rank, we can also compute the Coulomb branch dimension and 2-group structure constants for $M>2$, which also match for all three duals,
\begin{align}
\text{Dim(CB)}=6M +2 \, , \qquad\qquad \kappa_{R}=8M+2\, .
\end{align} 

\subsubsection*{Examples for a \texorpdfstring{$\boldsymbol{D_{8}}$}{D8} singularity}\vspace{-12pt}
The other example that we will discuss in detail has $M$ M5 branes probing a $D_{8}$ singularity. We choose flavor holonomies $(\mu_{1},\mu_{2})$ that leave an $\mathfrak{so}_{12}\times \fsu_2$ flavor symmetries on both $\fe_8$ sides. Similar to~\eqref{eq:fuse}, we can obtain this model by fusing SCFTs,
\begin{align}
\label{eq:fuse2}
\lbrack  \mathfrak{so}_{12}\rbrack    
{\overset{\mathfrak{sp}_{3}}{1}}   \, \, 
\textcolor{blue}{[ \mathfrak{so}_{16} ]}
\mathrel{\stackon[-1pt]{{-}\mkern-5mu{-}\mkern-5mu{-}\mkern-5mu{-}}{\textcolor{blue}{\mathfrak{so}_{16}}}}
\textcolor{blue}{[\mathfrak{so}_{16}]} \underbrace{ 
\overset{\mathfrak{sp}_{4}}{1}\,
{\overset{\mathfrak{so}_{16}}{4} \, 
\overset{\mathfrak{sp}_{4}}{1} \,
{\overset{\mathfrak{so}_{16}}{4}} \, 
\overset{\mathfrak{sp}_{4}}{1} \,
{\overset{\mathfrak{so}_{16}}{4}}  \,
 \ldots  \overset{\mathfrak{sp}_{4}}{1}\, }}_{M-2}
\textcolor{blue}{[\fso_{16}]}
\mathrel{\stackon[-1pt]{{-}\mkern-5mu{-}\mkern-5mu{-}\mkern-5mu{-}}{  \textcolor{blue}{\mathfrak{so}_{16}}}} 
\textcolor{blue}{[\mathfrak{so}_{16}]} {\overset{\mathfrak{sp}_{3}}{1}}   \, \, 
\lbrack \mathfrak{so}_{12} \rbrack  \, .
\end{align}
This results in an LST with quiver
\begin{align}
\label{exflav1}
\lbrack  \mathfrak{so}_{12}\rbrack     \, \,   
{\overset{\mathfrak{sp}_{3}}{1}}   \, \,  
\myoverset{1}
{\underset{\left[ \mathfrak{sp}_{1}\right]}
{\overset{\mathfrak{so}_{16}}{4}}}  \, \,
{\overset{\mathfrak{sp}_4}{1}} \, \, 
\underbrace{{\overset{\mathfrak{so}_{16}}{4}} \, \,
{\overset{\mathfrak{sp}_4}{1}}}_{\times (M-2)}  \, \,
\myoverset{1}
{\underset{\left[ \mathfrak{sp}_{1}\right]}
{\overset{\mathfrak{so}_{16}}{4}}} \, \,
{\overset{\mathfrak{sp}_3}{1}}  \, \,
\lbrack \mathfrak{so}_{12} \rbrack   \, ,   \qquad 
\vec{l}_{\LST}=(1, 1,1, 2, \underbrace{1,2}_{\times (M-2)},1,1,1) \, .
\end{align}
In the toric description, we find that the elliptic fibration descends from the 2D toric $F_{13}$. The torically realized flavor algebra is $\mathfrak{so}_{12}^{2} \times \mathfrak{sp}_{1}$. The non-compact divisor over which we have an $\mathfrak{sp}_{1}$ flavor algebra intersects two compact curves in the base, yielding the two $\mathfrak{sp}_{1}$ flavor factors required by field theory on these quiver nodes. Hence, the full FT flavor algebra is $\mathfrak{so}_{12}^{2} \times \mathfrak{sp}_{1}^{2}$ of rank 14. This model has again two T-duals. 

\textbf{First dual}\\
We find the first dual with an $F_{9}$ fiber type, whose free part of the MW group has rank 2 and an $\fsu_{12}$ fiber. The quiver is given by
\begin{align}
\label{exflav1dual}
\textcolor{blue}{\lbrack  \mathfrak{u}_{1,D_{1}} \times  \mathfrak{u}_{1,D_{2}} \rbrack } \times \bigg(
{\overset{\mathfrak{su}_{2M+2}}{2}}  \, \,   
\myoverset{\myoverset{\overset{\mathfrak{sp}_{2M-6}}{~1^{\star}}}{\overset{\mathfrak{su}_{4M-4}}{~2^{\star}}}}
{\underset{\textcolor{blue}{\left[\mathfrak{su}_{2}\right]}}{\overset{\mathfrak{su}_{4M+4}}{2}}} \, \,     
{\overset{\mathfrak{su}_{2M+8}}{2}} \, \,  
\lbrack \mathfrak{su}_{12} \rbrack \bigg)  \, ,   \qquad \vec{l}_{\LST}=(1, 2^\star,2^\star,2, 1) \, , \qquad (\star : \text{for } M>2)
\end{align}
which has the topology of a folded $\mathfrak{so}_{16}$ Dynkin diagram, cf.\ Figure~\ref{fig:SO4nFolding}. The torically realized flavor algebra has rank $13$ . To find the full FT flavor, we take into account that the presence of bifundamentals of $\mathfrak{su}_k$ algebras leads to $\fu_1$ factors and reconstruct the other algebra factors as above from the matter content:
\begin{itemize}
    \item The curve ${\overset{\mathfrak{su}_{2M+8}}{2}}$ requires $(4M+16)$ hypers in the fundamental representation, leading to a $\mathfrak{u}_{4M+16}\sim\mathfrak{su}_{4M+16} \times \mathfrak{u}_{1,A}$ flavor algebra. Gauging the subalgebra of the adjacent node induces a breaking $\mathfrak{u}_{4M+16}\supset\mathfrak{su}_{12} \times \mathfrak{su}_{4M+4}^{*} \times\mathfrak{u}_{1,{A}} \times \mathfrak{u}_{1,B_1}$.
    \item  The curve ${\overset{\mathfrak{su}_{4M+4}}{2}}$ requires $(8M+8)$ hypers in the fundamental representation, leading to a\linebreak $\mathfrak{u}_{8M+8}\sim\mathfrak{su}_{8M+8} \times \mathfrak{u}_{1,C}$ flavor algebra. Gauging the subalgebras of the adjacent nodes induces a breaking $\mathfrak{u}_{8M+8}\supset \mathfrak{su}_{2M+2}^{*} \times \mathfrak{su}_{4M-4}^{*} \times \mathfrak{su}_{2M+8}^{*} \times \mathfrak{su}_{2} \times \mathfrak{u}_{1,B_1} \times \mathfrak{u}_{1,B_2} \times\mathfrak{u}_{1,B_3} \times \times \mathfrak{u}_{1,C}$.
     \item The curve ${\overset{\mathfrak{su}_{4M-4}}{2}}$ requires $(8M-8)$ hypermultiplets in the fundamental representation, leading to a\linebreak  $\mathfrak{u}_{8M-8}\sim\mathfrak{su}_{8M-8} \times \mathfrak{u}_{1,D}$ flavor algebra. Gauging the subalgebra of the adjacent node induces a breaking $\mathfrak{u}_{8M-8}\supset\mathfrak{su}_{4M+4}^{*} \times \mathfrak{sp}_{2M-6}^{*} \times \mathfrak{u}_{1,B_{2}} \times\mathfrak{u}_{1,E}$.
\end{itemize}
Out of these 6 naive $\fu_1$ factors, 2 linear combinations are non-anomalous. Hence, the FT flavor is $\mathfrak{su}_{2} \times \mathfrak{su}_{12} \times \mathfrak{u}_{1,D_{1}} \times \mathfrak{u}_{1,D_{2}}$, which has rank 14, as expected from T-duality.

\textbf{Second dual}\\
The second dual is given by the quiver
\begin{align}
\begin{split}
 \textcolor{blue}{\lbrack \mathfrak{so}_{8}\rbrack }    \, \,   
{\overset{\mathfrak{sp}_{M-1}}{1}}   \, \,  
\myoverset{\overset{\mathfrak{sp}_{M-3}}{1^{\star}}}
{\underset{\textcolor{blue}{\left[\mathfrak{sp}_{2}\right]}}{\overset{\mathfrak{so}_{4M+4}}{4^{\star}}}} \, \, 
{\overset{\mathfrak{sp}_{2M-2}}{1}}  \, \,
{\overset{\mathfrak{so}_{4M+4}}{4}} \, \, 
{\overset{\mathfrak{sp}_{2M-2}}{1}} \, \, 
\myoverset{\overset{\mathfrak{sp}_{M-3}}{1^{\star}}}
{\overset{\mathfrak{so}_{4M+4}}{4^{\star}}} \, \, 
{\overset{\mathfrak{sp}_{M+1}}{1}} \, \,
\lbrack \mathfrak{so}_{16} \rbrack   \, ,   \qquad 
&\vec{l}_{\LST}=(1, 1^\star,1, 2,1,2,1^\star,1,1)\,. \\
&(\star : \text{for } M>2)
\end{split}
\end{align}
Since we only have $\mathfrak{so}$ and $\mathfrak{sp}$ gauge factors, there are no $\fu_1$ flavor factors. This is also true from the geometric engineering perspective, since we have an $F_{13}$ fiber polytope, which has trivial free MW rank. The toric flavor rank 13 matches the toric flavor ranks of the other duals. The FT flavor is $\mathfrak{so}_{8} \times \mathfrak{sp}_{2} \times \mathfrak{so}_{16}$, which has rank 14, and hence the full FT flavor rank of all duals matches as well. The Coulomb branch dimension and 2-group structure constants also match, and are given for $M>2$ by
\begin{align}
\text{Dim(CB)}=14M +4 \, , \qquad \kappa_{R}=24M\, .
\end{align}

\subsection{Flavor and K3 polarizations in Heterotic LSTs}
A feature of Heterotic LSTs is that their flavor group is described by the polarization of an elliptic K3 fiber, which allows for a simple description of flavor enhancements. Consider for example the affinization of the E-string theory, which results in a Heterotic LST given by the quiver 
\begin{align}
\label{eq:EStringMinimal}
    1\, \,  1 \, .
\end{align}
The minimally polarized K3 over this base has at least one elliptic fibration, leading to $\Pic(S_0)=U$, but misses other flavor factors completely. At the same time, the transcendental lattice is maximal, $T(S_0) = U \oplus \text{E}_8 \oplus \text{E}_8$.  This allows to describe more of the field-theoretic flavor group in terms of the K3 geometry, as long as we do not change the LST quiver. In terms of geometric realization of the flavor groups, we can rotate components from $T(S_0)$ into the Picard lattice. This is possible if:
\begin{enumerate}
    \item All non-compact divisors/singularities pull back to the flavor K3 and intersect the compact curves, i.e., no flavor group decouples.
    \item The enhancement does not introduce new singularities that would require additional compact divisors to fully resolve the threefold and hence modify the quiver.
\end{enumerate}
This way, one can define a filtration of polarized K3 surfaces $S_i$ with filtration parameter the Picard rank. Along this filtration, the rank of the transcendental lattice decreases until a maximal endpoint $S_\text{max}$ is reached. For the example in~\eqref{eq:EStringMinimal}, the endpoints are two $\fe_8$ components that yield the familiar quiver
\begin{align}
  [\fe_8] \, \,  1 \, \, 1 \, \, [\fe_8] \, .
\end{align}
This choice of polarization describes the full field-theoretic flavor group geometrically, which is reached by rotating the K3 lattice to obtain a polarization with $\Pic(S_{\text{max}})=U \oplus \text{E}_8 \oplus \text{E}_8$. 
As the rank of $\Pic(S)$ enhances, it becomes more and more likely to find extra elliptic fibration structures\footnote{For elliptic threefolds in toric ambient spaces, the number of elliptic fibrations increases with the number of K\"ahler moduli \cite{Anderson:2016cdu,Huang:2019pne}.} and consequently more T-dual models. In particular, for the familiar case where $\Pic(S_{UV})=U\oplus \text{E}_8 \oplus \text{E}_8$, there exists a second fibration with $\Pic(S)= U \oplus \text{Spin}(32)$ and LST quiver 
\begin{align}
    [ \fso_{32} ]~~\overset{\fsu_2}{0} \, ,
\end{align}
which realizes the $\text{E}_8\times \text{E}_8 \leftrightarrow \text{Spin}(32)/\mathbbm{Z}_2$ Heterotic duality.

%%%%%%%%%%%%%%%%%%%%%%%%%%%%%%%%%%%%%%%%%%%%%%%%%%%%%%%%%%%%%%%%%%%%%%%%%%%%%
\section{Engineering non-simply laced flavor groups}
\label{sec:nsl}
6D SCFTs and LSTs can have non-simply laced flavor groups. Typical examples are the "frozen" conformal matter theories discussed in \cite{Mekareeya:2017jgc}. 
A simple example is the SCFT quiver
\begin{align}
    [\ff_4]\,\, 1 \, \, \overset{\fg_2}{3} \, \, \overset{\fsu_2}{2} \, ,
\end{align}
which is an E-string attached to the ``$3 ~~ 2$'' non-Higgsable cluster theory. The vanilla E-string comes with an $\fe_8$ flavor symmetry, but the adjacent $\overset{\fg_2}{3}$ gauges a $\fg_2$ subgroup, which leaves a maximal $\ff_4 = [\fg_2,\fe_8]$ flavor subalgebra.

This raises the question of how these flavor symmetries are realized in IIB or F-theory geometry. In the classical D7 brane picture, non-simply laced symmetries arise from a monodromy action which acts as an outer automorphism on the world-volume gauge algebra $\fg$, reducing it to a non-simply laced version. However, flavor groups can be thought of as D7 branes that wrap non-compact divisors and thus naively do not experience such monodromy effects. Hence this could be another case where D7 branes or the F-theory geometry might not be able to give the full field theory flavor algebra 
\cite{Bertolini:2015bwa}. However, in the 6D SCFTs limit, non-simply laced flavor algebras can be obtained from 3-form flux as part of a triple \cite{deBoer:2001wca} that freezes the flavor group to a non-simply laced one, and in the case of orbi-instanton theories, as part of the holonomies acting at infinity \cite{Frey:2018vpw}. As LSTs can be fused from such SCFTs, such an effect should carry over, and one might ask whether such fluxes are mapped across dualities, and how they are encoded in the K3 geometry. Geometrically, non-simply laced subgroups can occur at special loci in moduli space where the volume of curves that resolve an ADE resolution are related such that they can be folded on top of each other. Such loci in moduli space can be enforced by choosing a description where some divisors are non-toric and hence descend from the same ambient space divisor.

\subsection{Non-simply laced flavor from freezing fluxes in 6D SCFTs}
In the construction of 6D SCFTs, $M$ M5 branes can fractionalize in the presence of a $\fg$-type transverse singularity~\cite{DelZotto:2014hpa} given by the quiver
\begin{align}
[\fg ] \, \, \overset{\fg}{2}\, \, \overset{\fg}{2}\, \, \ldots \overset{\fg}{2}\,\, [\fg ]\, .
\end{align}
The $\fg$-singularity leads to a fractionalization of the M5 brane into $p(\fg)$ fractions which can be computed from F-theory \cite{DelZotto:2014hpa} in terms of conformal matter insertions. Each M5 fraction carries a unit of $1/p(\fg)$ M-theory three-form flux in the transverse M-theory direction $\mathbbm{C}^2/\Gamma_{\fg}$,
\begin{align}
    \int_{S^3/\Gamma_{\fg}}  C_3 = \frac{n}{d} \text{ mod } 1\, .
\end{align}
The fluxes freeze the original $\fg$-type singularity to a subgroup $\mathfrak{h}$ \cite{deBoer:2001wca}. This way, non-simply laced groups naturally appear in superconformal matter chains probed by an ADE singularity. For example, if $\fg=\fe_8$, the M5 breaks into $p(\fe_8)=12$ fractions, described by the quiver
\begin{align}
\label{eq:E8E8Conf}
\begin{array}{c@{~\;}c@{~\;}c@{~\;}c@{~\;}c@{~\;}c@{~\;}c@{~\;}c@{~\;}c@{~\;}c@{~\;}c@{~\;}c@{~\;}c}
    [\fe_8] & 1 &2 &\overset{\fsu_2}{2} &\overset{\fg_2}{3} &\overset{}{1} &\overset{\ff_4}{5} &\overset{}{1} &\overset{\fg_2}{3}&\overset{\fsu_2}{2} &2 &1 &[\fe_8]  \,,\\ 
  &  \frac{1}{12}& \frac16 & \frac{1}{4} & \frac{1}{3} & \frac{5}{12} & \frac{1}{2} & \frac{7}{12} & \frac{2}{3} & \frac{3}{4} &   \frac{5}{6}  & \frac{11}{12} &  
    \end{array}
\end{align}
where the bottom line gives the number of fractional flux quanta. In Table~\ref{tab:frozen} we list the flux fractions $\frac{d}{n}$ that can lead to non-simply laced gauge algebras. In the above quiver, we can take one of the tensor vevs to infinity, which effectively decompactifies the respective curve and turns its gauge group into a flavor group. After such a decoupling, there might be some left-over residual flux that ``freezes'' the flavor algebra to a non-simply group~\cite{Mekareeya:2017sqh}. For example, upon decoupling~\eqref{eq:E8E8Conf} at the $(-5)$-curve, one is left with half a unit of flux, which leads to 
\begin{align}
\label{eq:F4E8Frozen}
     [\ff_4]   \, \,\overset{}{1} \, \,\overset{\fg_2}{3}\, \, \overset{\fsu_2}{2} \, \,2 \, \,1 \, \,[\fe_8]  \, ,
\end{align}
realizing a non-simply laced flavor algebra. 
Similarly, when decoupling~\eqref{eq:E8E8Conf} at the $(-3)$-curve, one obtains the quiver 
\begin{align} 
   [\fg_2] \, \,  \overset{}{1} \, \,\overset{\ff_4}{5}  \, \,\overset{}{1} \, \,\overset{\fg_2}{3}\, \, \overset{\fsu_2}{2} \, \,2 \, \,1 \, \,[\fe_8]  \, ,
\end{align} 
which leaves a residual flux of $1/3$ that freezes the $\fg_2$ flavor group. 

\begin{table}[t]
    \centering
    \begin{tabular}{|c|c|c|c|c|} \hline
        $\fg$ &  $\fso_{2k+8}$ & $\fe_6$ & $\fe_7$  &$ \fe_8$\\ \hline
      $d/n$  & $2$ & $2$ & $2$\qquad $3$     & $ 2\qquad 3\qquad4$   \\ 
      $\mathfrak{h}$ & $\fsp_k$ & $\fsu_3$& \, \, $\fso_7$ \,    \, $\fsu_2$\, \,&  $\ff_4$ \,\,\, $\fg_2$   \quad  $\fsu_2$  \\ \hline
    \end{tabular}
    \caption{The unbroken subalgebra $\mathfrak{h}$ of $\fg$ upon turning on $d/n$ fractional freezing flux units~\cite{Tachikawa:2015wka}.}
    \label{tab:frozen}
\end{table}

As can be seen from Table~\ref{tab:frozen}, this freezing mechanism does not work for constructing $\fso_{2k+1}$ flavor symmetries with $k>3$. Instead, these can be obtained by starting with a $\mathcal{T}(\fe_k, \fso_{2k+8})$ matter theory~\cite{DelZotto:2014hpa} ,
\begin{align}
\begin{split}
\mathcal{T}(\fe_6, \fso_{20})&: \qquad [\fe_6 ] \, \, \overset{}{1} \, \,  \overset{\fsu_3} {3} \,\,\overset{}{1}   \overset{\fso_9}{4} \, \, \overset{\fsp_1}{1}  \, \, \overset{\fso_{11}}{4}\, \,   \overset{\fsp_2}{1} \, \, \overset{\fso_{13}}{4} ~~~\ldots~~~  \overset{\fso_{19}}{4} \, \, \overset{\fsp_{6}}{1}\, \, [\fso_{20}] \,,\\
\mathcal{T}(\fe_7, \fso_{22})&: \qquad  \lbrack  \fe_7 \rbrack \, \,  \overset{}{1}\,\,  \overset{\fsu_2}{2} \, \, \overset{\fso_7} {3} \,\,\overset{\fsp_1}{1} \, \, \overset{\fso_{11}}{4}\, \,   \overset{\fsp_2}{1} \, \, \overset{\fso_{13}}{4} ~~~ ~\,\ldots~~~   \overset{\fso_{21}}{4} \, \, \overset{\fsp_{7}}{1}\, \, [\fso_{22}] \,,  \\
\mathcal{T}(\fe_8, \fso_{22})&: \qquad  \lbrack  \fe_8  \rbrack \, \, 1\,\, 2  \,\, \overset{\fsu_2}{2} \,\, \overset{\fg_2}  {3}\,\, 1 \overset{\fso_9}{4} \, \, \overset{\fsp_1}{1} \, \, \overset{\fso_{11}}{4}\, \,   \overset{\fsp_2}{1} \, \, \overset{\fso_{13}}{4} \ldots   \overset{\fso_{23}}{4} \, \, \overset{\fsp_{8}}{1}\, \, [\fso_{24}] \,,
\end{split}
\end{align}
and decoupling the chain at an appropriate $\fso$ or $\fsp$ flavor point.

\subsection{Non-simply laced LSTs from fusion}
Using the SCFT building blocks discussed above, we can start with an $\mathcal{K}_N(\text{E}_8,\text{E}_8,\fe_8)$ theory~\cite{DelZotto:2022ohj,DelZotto:2022xrh,DelZotto:2023ahf} obtained by fusing two $\mathcal{T}(\fe_8,\fe_8)$ theories. The associated quiver is
\begin{align}
\lbrack  \fe_8  \rbrack \, \, 1\, \, 2 \, \,\overset{\fsu_2}{2}  \, \,\overset{\fg_2}{3}\, \, \overset{}{1} \, \,\textcolor{red}{\overset{\ff_4}{5}} \, \,\overset{}{1} \, \,\overset{\fg_2}{3}\, \, \overset{\fsu_2}{2} \, \,2 \, \,1 \, \, 
\textcolor{blue}{[\fe_8]}
\mathrel{\stackon[-1pt]{{-}\mkern-5mu{-}\mkern-5mu{-}\mkern-5mu{-}}{\textcolor{blue}{\mathfrak{e}_{8}}}}
\textcolor{blue}{[\fe_8]} \underbrace{ \ldots  
\overset{\fe_8}{12}
\ldots }_{N-2}
\textcolor{blue}{[\fe_8]}
\mathrel{\stackon[-1pt]{{-}\mkern-5mu{-}\mkern-5mu{-}\mkern-5mu{-}}{  \textcolor{blue}{\mathfrak{e}_{8}}}} 
\textcolor{blue}{[\fe_8]}
\, \, 1\, \, 2 \, \,\overset{\fsu_2}{2}  \, \,\overset{\fg_2}{3}\, \, \overset{}{1} \, \,\overset{\ff_4}{5} \, \,\overset{}{1} \, \,\overset{\fg_2}{3}\, \, \overset{\fsu_2}{2} \, \,2 \, \,1 \, \, 
[\fe_{8}]\,,
\end{align}
where $N$ is the number of $\fe_8$ gauge algebra factors. Subsequently, we decompactify the $(-5)$-curve in the left SCFT block (highlighted in red), resulting in a frozen conformal matter theory.\footnote{Note that frozen conformal matter theories can also be obtained from orbi-instanton theories with non-trivial holonomy~\cite{Frey:2018vpw}.} Upon adding E-string matter at the gluing $\fe_8$ algebras, one obtains the LST quiver  
\begin{align}
\label{eq:QuiverF4E8}
    [\ff_4] \, \, 1 \, \,\overset{\fg_2}{3}\, \, \overset{\fsu_2}{2} \, \,2 \, \,1 \, \,
\myoverset{\displaystyle 1}{\overset{\fe_8}{\myunderset{\vphantom{3^{3^{3^{3^{3}}}}}\displaystyle 1}{12}}} %
\underbrace{\, \, 1\, \, 2 \, \,\overset{\fsu_2}{2}  \, \,\overset{\fg_2}{3}\, \, \overset{}{1} \, \,\overset{\ff_4}{5} \, \,\overset{}{1} \, \,\overset{\fg_2}{3}\, \, \overset{\fsu_2}{2} \, \,2 \, \,1 \, \,\overset{\fe_8} {12}\ldots}_{(N-2)}%
\myoverset{\displaystyle 1}{\overset{\fe_8} {12}}  \, \, 1\, \, 2 \, \,\overset{\fsu_2}{2}  \, \,\overset{\fg_2}{3}\, \, \overset{}{1} \, \,\overset{\ff_4}{5} \, \,\overset{}{1} \, \,\overset{\fg_2}{3}\, \, \overset{\fsu_2}{2} \, \,2 \, \,1 \, \, [\fe_8] \,,
%\ldots \overset{\fe_8} {12} \ldots \overset{\fe_8}{12} \ldots \myoverset{\displaystyle 1}{\overset{\fe_8} {12}}  \ldots [\fe_8] \,.
\end{align}
Note that the rank of the flavor groups is $\rk(G_F)=12$. The T-dual quiver is
\begin{align}
\label{eq:NonSimplyDual}
    [\fso_{25}] \, \, \overset{\fsp_{N+6}}{1} \, \, \overset{\fso_{4N+15}}{4} \, \,
\overset{\fsp_{3N+1}}{1}\,\, \overset{\fso_{8N+5}}{4} \,\,
\overset{\fsp_{5N-1}}{1}\,\, \overset{  \overset{ \overset{[N_F=\frac12]}{ \fsp_{3N-5}}}{\displaystyle 1}  }{\overset{\fso_{12N-5}}{4}}\,\,
\underset{[N_F=\frac12]}{\overset{\fsp_{4N-4}}{1}}\,\, \overset{\fso_{4N+4}}{4}  \, .  
\end{align}
The quiver has the expected shape of an affine $\fe_8$ algebra, as well as a flavor algebra that is broken from the primordial $\fso_{32}$ algebra to the non-simply laced rank $12$ subalgebra $\fso_{25}$.

\subsection{Non-simply laced LSTs from non-favorable K3s}
\begin{figure}[t]
\centering
\begin{picture}(0,200)
\put(150,150){\Large $\mathcal{A}$}
\put(40,50){$B_1$}
\put(-80,140){$D_{\alpha_1}$}
\put(-90,130){$D_{\alpha_3}$}
\put(-110,115){$D_{\alpha_2}$}
\put(-80,105){$D_{\alpha_4}$}
\put(-120,20){\includegraphics[scale=0.4]{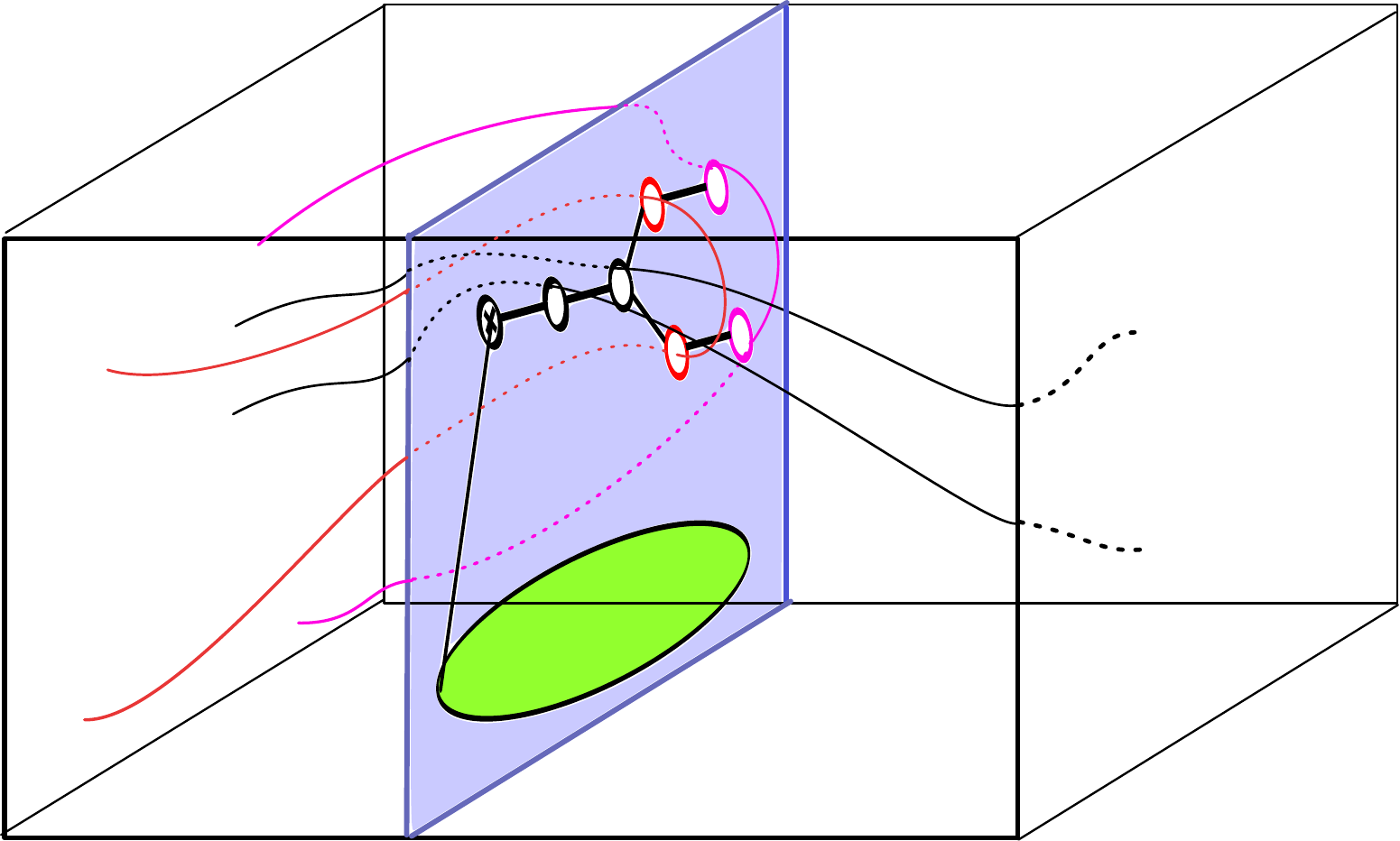}}
\end{picture}
\caption{A K3 hypersurface (blue) with an $\fe_6$ fiber in a 3D ambient space $\mathcal{A}$. The two ambient divisors $D_{\alpha_1}$ an $D_{\alpha_2}$ (red and pink) intersect the K3 twice, each providing two resolution curves of the $\fe_6$ singularity.}
\label{fig:K3Ambient}
\end{figure}
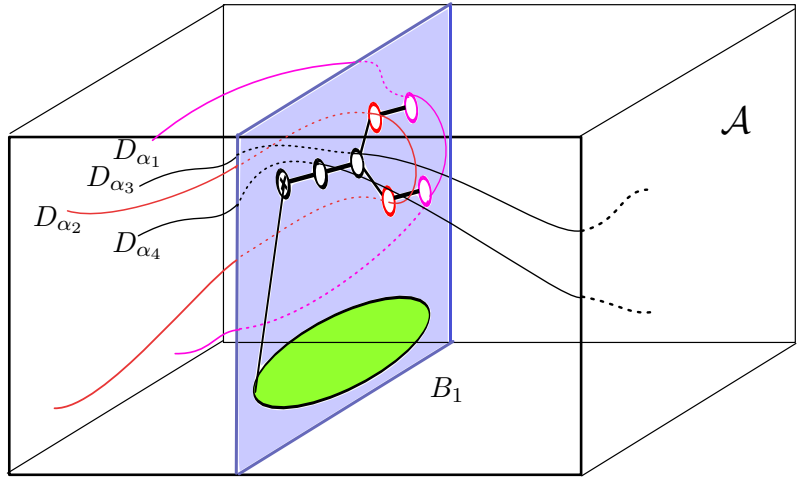

To obtain non-simply laced flavor algebras from folding ADE algebras, we need a symmetry group $\sigma_n$ corresponding to the corresponding outer automorphism of the Dynkin diagram. Curves that correspond to nodes that are folded onto one another need to have the same volume, which requires restricting the K3 moduli space to loci that are compatible with $\sigma_n$. We can enforce this relation among divisor volumes by taking a toric ambient space divisor that intersects the K3 hypersurface $n$ times.\footnote{For compact CY threefolds, such contributions are called non-K\"ahler favorable~\cite{Anderson:2016cdu}.}  This is illustrated for a K3 inside a toric ambient space~$\mathcal{A}$, which is a $T^2$ fibration over $B_1$ with a Kodaira singularity of type $\textit{IV}^*$ in Figure~\ref{fig:K3Ambient}.
 
\begin{table}[t!]
    \centering
    \begin{align*}
    \begin{array}{ccccc}
\begin{array}{|c|c|}
\hline
\multicolumn{2}{|c|}{\text{Generic fiber}}\\ \hline
    X &  (1,0,0)\\
    Y & (0,1,0)\\
    Z &  (-2,-3,0)  \\
    \hline
\end{array}
&\oplus&
\begin{array}{|c|c|}
\hline
\multicolumn{2}{|c|}{\text{"$\ff_4$" fiber}}\\ \hline
    \alpha_0 & (-2,-3,-1)  \\
    \alpha_1 & (0-1,-1)\\
    \alpha_2 & (-1,-2,-2) \\
    \alpha_3 & (-2,-3,-3) \\
    \alpha_4 & (-2,-3,-2)\\
    \hline
\end{array}
&\oplus&
\begin{array}{|c|c|}
\hline
\multicolumn{2}{|c|}{\text{$\fe_8$ Fiber}}\\ \hline 
\beta_1 & (-2,-3,1)\\
\beta_2 & (-2,-3,2)\\
\beta_3 & (-2,-3,3)\\
\beta_4 & (-2,-3,4)\\
\beta_5 & (-2,-3,5)\\
\beta_6 &   (-2,-3,6)\\ 
\hat{\beta}_1 &
(0,-1,2)\\
\hat{\beta}_2 &
(-1,-1,3)\\
\hat{\beta}_3 &
(-1,-2,4)\\
\hline
\end{array}
\end{array}
\end{align*}
\caption{Toric rays of the K3 fiber for the quiver~\eqref{eq:QuiverF4E8}. The ``$\ff_4$'' rays $\alpha_1$ and $\alpha_2$ intersect the K3 hypersurface twice, splitting the $\ff_4$ into an $\fe_6$ fiber.}
\label{tab:K3Rays}
\end{table}
We demonstrate this construction for the flavor K3 of the LST given in~\eqref{eq:QuiverF4E8}. The toric rays of the K3 surface $S$ with $\ff_4 \times \fe_8$ flavor group are given in Table~\ref{tab:K3Rays}. Using \eqref{eq:Pictoric} and \eqref{eq:Picnontoric}, the dimension of the toric and non-toric Picard lattices of $S$ are
\begin{align}
    \rk(\Pic(S))=\rk(\Pic(S)_\text{tor})+\rk(\Pic(S)_\text{cor})=14+2 \, . 
\end{align} 
The K3 has two reducible fibers at the two poles of the base $\mathbbm{P}^1_{x_0,x_1}$. Upon resolution, the base divisors are replaced by
\begin{align}
    (\alpha_i) \rightarrow x_0 = \prod_i \alpha_i^{-v(\alpha_i)_3}\,, \qquad  (\beta_j,\hat{\beta}_k) \rightarrow x_1 = \prod_j \beta^{v(\beta_j)_3} \prod_k \hat{\beta}^{v(\hat{\beta}_k)_3} \,, 
\end{align}
where $v(\alpha_i)_3$ is the third component of the vector that defines to toric ambient space divisor $\alpha_i$, and similarly for $v(\beta)_3$ and $v(\hat{\beta}_i)_3$. Over the divisor $D_{x_1}=\{x_1=0\}$, we find a Kodaira type $II^*$ fiber corresponding to an $\fe_8$ algebra. Over $D_{x_0}=\{x_0=0\}$, we expect a type $IV^*$ fiber, i.e., an $\fe_6$ algebra. However, since there are only $14-2=12$ toric resolution  divisors (we need to subtract one for the fiber and one for the base), eight of which correspond to resolution divisors for $\fe_8$, only 4 are left for the type $IV^*$ singularity. The missing two are the non-toric divisors. They descend from $D_{\alpha_1}=\{\alpha_1=0\}$ and $D_{\alpha_0}=\{\alpha_2=0\}$, which split into two on the K3. We can see this from the hypersurface equation
\begin{align}
\begin{split}
p&= \hat{\beta}_2 Y^2  + \alpha_1^2 \alpha_2 \hat{\beta}_1^2 \hat{\beta}_3 X^3 + \prod_i \alpha_i \prod_j \beta_j \prod_k \hat{\beta}_k~ X Y Z + \hat{\beta}_1\hat{\beta}_3\prod_i \alpha_i^2 \prod_j \beta_j^2 \prod_k \hat{\beta}_k~X^2 Z^2\\ 
& + \hat{c}_1(x_0,x_1)\alpha_0^2 \alpha_4\hat{\beta}_2\hat{\beta}_3 \prod_j \beta_j^3 \prod_k \hat{\beta}_k~Y Z^3 + c_1(x_0,x_1)\alpha_0^2\alpha_ 4\hat{\beta}_3 \prod_i \alpha_i \prod_j \beta_j^4 \prod_k \hat{\beta}_k^2~X Z^4   \\ 
& + c_3(x_0,x_1)\alpha_0^4 \alpha_4^2 \beta_0^5 \beta_2^4 \beta_3^3 \beta_4^2 \beta_5~Z^6 \,,
\end{split}
\end{align}
where the $c_i$ are homogeneous polynomials of degree $i$ in $[x_0:x_1]$. We can use this to show that the divisors $D_{\alpha_1}$ and $D_{\alpha_2}$ both descend from the ambient space divisor $D_{x_0}$, along which the hypersurface becomes reducible,
\begin{align}
\label{eq:SplitDivisors}
    {D_{\alpha_{1}}}\cap p={D_{\alpha_{2}}}\cap p=Y^2 +   c_1 Y  + c_3  = \left[Y- \frac12 \left(  c_3 + \sqrt{D_{2,X}} \right)\right]\left[Y- \frac12 \left(  c_3 - \sqrt{D_{2,X}} \right)\right]\,,
\end{align}
where we have used the toric scalings to set coordinates to one, and we introduced $D_{2,X}:=c_3^2(x_0,x_1)+ 4 c_1(x_0,x_1)$, which is some complex number. Hence, the toric realization of the $\fe_6$ fiber forces the two outer $\fe_6$ fibral curves to have exactly the same volume, as expected for $\ff_4$.

\subsection{Non-simply laced LSTs from decoupling gravity}
\label{ssec:gravdecouple}
In this section we discuss the appearance of non-simply laced flavor group from a gravity decoupling perspective. We take the flavor K3 and fiber it over a compact $\mathbbm{P}^1$ base to obtain a compact threefold $X$. To keep the discussion as simple as possible, we will have no additional curves that stay compact in the LST limit. As an example, we take an elliptic fibration over a Hirzebruch $\mathbbm{F}_{12}$ base. This means that we add the rays
\begin{align}
\label{eq:compactRays}
    v_1=(-2,-3,12,1) \, , \qquad v_2=(-2,-3,0,-1)
\end{align}
to the K3 polytope $\Delta_3$ in Table~\ref{tab:K3Rays} which is schematically completed to the 4D rays $(\Delta_3,0)$,
resulting in a CY threefold with Hodge numbers
\begin{align}
h^{1,1}(X)=15\,,\qquad h^{2,1}(X)=135\,.
\end{align}
The corresponding quiver is
\begin{align}
\begin{array}{ccc}
&0&\\
\overset{~~\ff_4}{-12}&&\overset{\fe_8}{12}\\
&0&
\end{array}
\end{align} 
The $0$ curve is in the class of the $\mathbbm{P}^1$ fiber $[F]$ and the $\ff_4$ and $\fe_8$ sit over the classes $[H]$ and $[E]=[H-12F]$, respectively. The intersection ring is
\begin{align}
    H \cdot F=E \cdot F=1\, , \quad  H^2=-E^2=12\, , \quad F^2=E \cdot H=0 \, .
\end{align}
The K\"ahler form of the base $\mathbbm{F}_{12}$ is
\begin{align}
    J=v_1 F+v_2 H\,,
\end{align}
where $v_i$ are K\"ahler moduli. With this, the overall volume of the $\mathbbm{F}_{12}$ base is
\begin{align}
    \text{Vol}(B)= \int_{\mathbbm{F}_{12}} J \wedge J = 2 v_1 v_2+12 v_2^2 \, ,
\end{align}
and the volumes of the respective curves are
\begin{align}
    \text{Vol}(F)= \int_F J= v_2 \, , \qquad
    \text{Vol}(H)= \int_H J=v_1 +12v_2 \, , \qquad
    \text{Vol}(E)= \int_E J= v_1 \,.
\end{align} 

From this, we see that there are two gravity decoupling limits: We can take $v_2\to\infty$, which sends the fiber $F$ to infinite volume and only leaves the shrinkable $(-12)$-curve that flows to an SCFT, or we can take $v_1 \rightarrow \infty$ while keeping the $\mathbbm{P}^1$ with volume $v_2$ finite, which is the little string limit \cite{Hayashi:2023hqa}. In the latter case, we find that both $H$ and $E$ decompactify and become flavor divisors that intersect the finite curve $F$. From this perspective, we may also discuss the role of the monodromy divisor $D_{2,X}$, 
\begin{align}
\label{eq:HomD2X}
    [D_{2,X}]=6(2 F+H+E) \,,
\end{align} 
which is the branching locus in the base $B_2$ of the two-section $\{X=0\}$ in the Tate model. This divisor interchanges the two fibral $\mathbbm{P}^1$'s of the local $\fe_6$ resolutions, which results in an $\ff_4$ resolution. This means that the two divisors $D_{\alpha_1}$ and $D_{\alpha_2}$ are reducible on the K3 but not in the compact threefold. From its homology class~\eqref{eq:HomD2X}, we see that the divisor $D_{2,X}$ intersects $D_{x_0}$ in $84$ points. In the gravity decoupling limit where the divisor $D_{x_0}$ becomes non-compact, these intersections are send to infinity. While the monodromy is technically gone, the K3 itself is still forced to be at a special locus in moduli space where only the toric Picard group is realized as flavor holonomies, giving rise to the $\ff_4$ subgroup of the $\fe_6$ singularity.

\subsection{Non-simply laced flavor algebras and T-duality}
In the previous sections we have argued that the Picard group, if non-K\"ahler favorable, can enforce the appearance of non-simply laced subgroups $\widehat{\fg}\subset \fg$ of the geometric K3 singularity. This is not changed by choosing different (toric) fibrations and should hence be preserved in M-Theory, where we identify the toric Picard group of the K3 with the flavor holonomies/Wilson lines under circle compactification. 
 
Let us exemplify this using the T-dual of the quiver~\eqref{eq:F4E8Frozen} given by~\eqref{eq:NonSimplyDual}. The flavor group is given by the K3 in Table~\ref{tab:K3Rays}. The base of the second fibration is specified by projecting the toric ambient rays $v=(x,y,z)$ onto the $x$ coordinate. The generic fiber ambient space is given by the rays $v_i=(0,y_i,z_i)$, corresponding to the coordinates $Y, \alpha_1, \hat{\beta}_2$ in Table~\ref{tab:K3Rays}. It is easy to see that this toric ambient space is $\mathbbm{P}^2_{1,2,3}$. We can identify the tops that restrict to $D_{x_0}$ and $D_{x_1}$ of the $\mathbbm{P}^1_{x_0,x_1}$ base. While $D_{x_1}$ has a trivial top, we find for $D_{x_0}$
\begin{align}
    x_0 =  \alpha_2 \hat{\beta}_4 \hat{\beta}_3 \left(Z \alpha_0  \alpha_3 \alpha_4 \prod_{j} \beta_j\right)^2 \, .
\end{align} 
Mapping this to the singular Weierstrass model reveals an $I_9^*$ fiber corresponding to an $\fso_{26}$ algebra. However, in the toric description we only have 13 independent fibral divisors in $D_{x_0}$, and in particular only three multiplicity one nodes ($\alpha_2$, $\hat{\beta}_3 $, $\hat{\beta}_4$). The divisor $D_{\alpha_2}$ intersects the K3 hypersurface twice and resolved
both the spinor and co-spinor root of $\fso_{26}$, which allows to identify the two and results in an $\fso_{25}\subset \fso_{26}$ subalgebra. The second non-toric divisor is $D_{\alpha_1}$, which maps to the two sections $X=0$ in the typical Tate-model nomenclature. Since $D_{\alpha_1}$ intersects the K3 twice, the two-section splits into two rational sections, which enhances the Mordell-Weil group of the toric fiber of the K3. However, this enhanced $\fu_1$ flavor algebra is absent in the 6D LST (both in this model and its T-dual). As for the divisor $D_{\alpha_2}$, the volume of the additional section that would give rise to the flavor $\fu_1$ is linked to that of the generic elliptic fiber, which in field theory terms is the inverse radius of the 6D circle that is decompactified.   

Finally we remark that all T-duality considerations respect the 6D SUGRA compactification discussed in Section~\ref{ssec:gravdecouple}. When embedding the K3 in an elliptic threefold by adding the rays~\eqref{eq:compactRays}, the T-dual configuration has an $\mathbbm{F}_4$ base with quiver 
\begin{align}
\begin{array}{ccc}
&0&\\
\overset{  }{-4}&&\overset{\fso_{25}}{4}\\
&0&
\end{array}\, .
\end{align} 
In this compact model, the gauge group is indeed just $\fso_{25}$. The enhancement to $\fso_{26}$ is absent due to the monodromy around the additional compact $\mathbbm{P}^1$ direction.

\subsection{Examples of non-simply laced flavor constructions}
Consider an LST that is obtained by fusing two orbi-instanton theories $\mathcal{T}(\fso_{13},\fe_7)$ with an $\mathcal{T}_{M-2}(\fe_7, \fe_7)$ conformal matter theory~\cite{Frey:2018vpw},
\begin{align}
\label{eq:so132Fusion}
\lbrack  \mathfrak{so}_{13}\rbrack    
{\overset{\mathfrak{sp}_2}{1}}  \,
{\overset{\mathfrak{so }_{11}}{4}} 
\underset{\left[N_f=1 / 2\right]} 
{\overset{\mathfrak{sp}_1}{1}} 
{\overset{\mathfrak{so}_7}{3}} \, 
{\overset{\mathfrak{su}_2}{2}} 
1   
\textcolor{blue}{[ \mathfrak{e}_7 ]}
\mathrel{\stackon[-1pt]{{-}\mkern-5mu{-}\mkern-5mu{-}\mkern-5mu{-}}{\textcolor{blue}{\mathfrak{e}_7}}}
\textcolor{blue}{[\mathfrak{e}_7]} \underbrace{ \ldots 
{\overset{\mathfrak{e}_7}{8} \, 1\,  
{\overset{\mathfrak{su}_2}{2}} \,
{\overset{\mathfrak{so}_7}{3}}  \,
 {\overset{\mathfrak{su}_2}{2}}  
1\, {\overset{\mathfrak{e}_7}{8}} \ldots   }}_{M-2}
\textcolor{blue}{[\fe_{7}]}
\mathrel{\stackon[-1pt]{{-}\mkern-5mu{-}\mkern-5mu{-}\mkern-5mu{-}}{  \textcolor{blue}{\mathfrak{e}_{7}}}} 
\textcolor{blue}{[\mathfrak{e}_7]} \ldots
\lbrack \mathfrak{so}_{13} \rbrack  
\end{align}
which results in the LST quiver 
\begin{align}
\label{eq:so1312Quiver}
\lbrack  \mathfrak{so}_{13}\rbrack \, \,     
{\overset{\mathfrak{sp}_2}{1}}  \, \,   
{\overset{\mathfrak{so }_{11}}{4}} \, \,   
\underset{\left[N_f=1 / 2\right]} 
{\overset{\mathfrak{sp}_1}{1}} \, \, 
{\overset{\mathfrak{so}_7}{3}} \, \,
{\overset{\mathfrak{su}_2}{2}} \, \,
1 \, \,
{\overset{\mathfrak{e}_7}{8}} \, \,
\underbrace{1 \, \,
{\overset{\mathfrak{su}_2}{2}} \, \,
{\overset{\mathfrak{so}_7}{3}} \, \,
 {\overset{\mathfrak{su}_2}{2}} \, \,
1\, \,
{\overset{\mathfrak{e}_7}{8}}}_{\times (M-1)}\, \,
1 \, \,
{\overset{\mathfrak{su}_2}{2}} \, \,
{\overset{\mathfrak{so}_7}{3}} \, \,
\underset{\left[N_f=1 / 2\right]}
{\overset{\mathfrak{sp}_1}{1}} \, \,
{\overset{\mathfrak{so}_{11}}{4}} \, \,
{\overset{\mathfrak{sp}_2}{1}} \, \,
\lbrack \mathfrak{so}_{13} \rbrack   \, ,
\end{align}
The flavor K3 is given by the vertices
\begin{align}
    \Delta_3= \left(\begin{array}{r@{~}r@{~}r} (-2, &-3, &-2)\\ (1, &0, &2)\\ (1, &0, &-2)\\(0, &1, &0)\\(-2, &-3, &2)
\end{array}\right)\, , \qquad \rk(\Pic(S)_\text{tor})=14\,,\qquad \rk(\Pic(S)_\text{cor})=3 \, .
\end{align}
The toric part of the Picard group is 14, which matches the rank $12$ flavor group. There is again a correction term which is attributed to the toric rays $x_i: (1,0,-1),(1,0,0),(1,0,1)$ for $i=1,2,3$. Each of these ambient divisors intersects the K3 hypersurface $S$ twice. 

The LST with quiver~\eqref{eq:so1312Quiver} is realized for a choice of elliptic fibration where the K3 base is given by projection onto the last coordinate. The elliptic fiber is a Tate model with $I_3^*$ fibers at north and south pole of the $\mathbbm{P}^1$ base. Within those fibers, $x_1$ and $x_3$ resolve both the spinor and co-spinor node, fixing their volumes to the same values, such that only an $\fso_{13} \subset \fso_{14}$ subalgebra is realized. The other toric divisor $x_2$ is identified with the two-section $X=0$ of the Tate model. On the K3, this two-section intersects the hypersurface twice, resulting in another section with a non-toric Mordell-Weil contribution. As the class of this $\fu_1$ is the same as of the generic fiber, it does not give rise to an additional flavor holonomy or a $\fu_1$ flavor factor in the quiver.

The polytope $\Delta_3$ has a second toric fibration structure given by projecting onto the first coordinate. The elliptic fiber descends from an $F_{13}$ fiber polytope with $\mathbbm{Z}_2$ MW torsion. In addition, one finds an $I_4^*$ and $I_8$ fiber over the base rays. Three of the eight divisors in the $I_8$ fiber are given by the double intersection of the $x_i$, meaning that the fiber only realizes an $\fsp_4 \subset \fsu_8$ flavor subalgebra. The full T-dual LST is then
\begin{align}
\label{eq:so13dual}
\lbrack  \mathfrak{sp}_{4}\rbrack     \, \, 
\myoverset{{\overset{\mathfrak{sp}_{M-1}}{1}}}
{\overset{\mathfrak{so}_{4M+12}}{4}}
{\overset{\mathfrak{sp}_{3M+1}}{1}}  \, \,     
\myoverset{{\overset{\mathfrak{sp}_{2M-2}}{1}}}
{\overset{\mathfrak{so}_{8M+8}}{4}} \, \,   
{\overset{\mathfrak{sp}_{3M+1}}{1}}  \, \,
{\overset{\mathfrak{so}_{4M+12}}{4}} \, \,
{\overset{\mathfrak{sp}_{M+3}}{1}} \, \,
\lbrack \mathfrak{so}_{16} \rbrack   \, ,   
\end{align} 
Both models have the same Coulomb branch dimension and two-group structure constants, which for $M>2$ read
\begin{align}
\text{Dim(CB)}=18M +25 \, , \qquad \kappa_{R}=48M+46\, .
\end{align} 

Note that by choosing the singularity type to be $\fg=\fso_8$ we can construct another LST with the same flavor group. This amounts to fusing two $\mathcal{T}(\fso_{13},\fso_8)$~\cite{Frey:2018vpw} with $\mathcal{T}_{M-3}(\fso_8,\fso_8)$ in~\eqref{eq:so132Fusion}, which gives
\begin{align}
\lbrack  \mathfrak{so}_{13}\rbrack     \, \,    
{\overset{\mathfrak{sp}_1}{1}} \, \,
{\overset{\mathfrak{g}_2}{3}} \, \,  
1 \, \, 
{\overset{\mathfrak{so}_8}{4}} \, \,
\underbrace{1 \, \,
{\overset{\mathfrak{so}_8}{4}} \, \,}_{\times (M-3)} \, \,
1 \, \,
{\overset{\mathfrak{g}_2}{3}} \, \, 
{\overset{\mathfrak{sp}_{1}}{1}} \, \,
\lbrack \mathfrak{so}_{13} \rbrack   \, .
\end{align}
Since we only changed the gauged part of the quiver~\eqref{eq:so132Fusion} but not the $\fso_{13}\times\fso_{13}$ flavor algebras, the K3 remains the same. In particular, we already know that this K3 has a second fibration with flavor group $\fso_{16}\times\fsp_4$ as in~\eqref{eq:so13dual}. The corresponding quiver is
\begin{align}
\begin{array}{ccc}
\overset{\mathfrak{sp}_{M-3}}{1} && \overset{\mathfrak{sp}_{M-3}}{1}  \\
 & \underset{\lbrack  \mathfrak{sp}_{4}\rbrack }{{\overset{\mathfrak{so}_{4M+4}}{4}}} & \\
\overset{\mathfrak{sp}_{M-3}}{1} && \underset{\lbrack \mathfrak{so}_{16} \rbrack}{\overset{\mathfrak{sp}_{M+1}}{1}}
\end{array}
\end{align}
This means that we have engineered two morphisms 
\begin{align}
    \mu_1 \times \mu_2: \Gamma_\fg \hookrightarrow \text{E}_8 \times \text{E}_8 \qquad\text{ with }\qquad \fg_F= \fso_{13}\times \fso_{13} \,,
\end{align}
for $\fg=\fe_7$ and $\fso_8$ that lead to the same flavor group breaking. These must then have two identical dual morphisms 
 \begin{align}
    \lambda: \Gamma_\fg \hookrightarrow \text{Spin(32)}/\mathbbm{Z}_2 \qquad\text{ with }\qquad  \fg_F= \fso_{16}\times \fsp_{4} \, ,
\end{align}
Geometrically, this arises since the Picard group of the flavor K3 is identical in both models.

%%%%%%%%%%%%%%%%%%%%%%%%%%%%%%%%%%%%%%%%%%%%%%%%%%%%%%%%%%%%%%%%%%%%%%%%%%%%%
\section{Exotic Models}
\label{sec:exotic}
In this section we will discuss certain curiosities arising for LSTs and their duals. The first one includes a model that has two distinct \text{Spin}(32)$/\mathbbm{Z}_2$
duals with the same base quiver shape, which occurs for $\fg=\fe_6$ singularities. This model is a counterexample to a conjecture made in~\cite{DelZotto:2022ohj} that the base quiver shape, the 2-group structure constants, and the Coulomb branch data \text{uniquely} determines a pair of $\text{E}_8\times \text{E}_8 \leftrightarrow \text{Spin}(32)/\mathbbm{Z}_2$ duals. As a second curiosity, we discuss an example where exceptional gauge algebras appear in both Heterotic models, which obscures a type $I$ dual picture. As a third curiosity, we construct a family of self-dual models.

\subsection{Different duals with same base topology} 
\label{sec:DifferentDuals}
\begin{figure}[t!]
\begin{picture}(0,60)
\put(70,0){\includegraphics[scale=0.5]{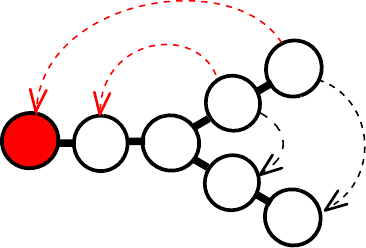}} 
\put(290,15){\includegraphics[scale=0.5]{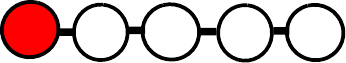}} 
\put(200,25){\Large $\xrightarrow{\text{T-dual}}$}
\end{picture}
\caption{\label{fig:E6Syms} Two $\mathbbm{Z}_2$ automorphisms of affine $\fe_6$ highlighted with black and red arrows respectively. The former one is only present in the affine extension. Both lead to an affine $\ff_4$-shaped quiver with inequivalent $\fso_{32}$ models.}
\end{figure}
In all our examples, the $\mathfrak{e}_{8} \times \mathfrak{e}_{8}$ LST with an $\mathfrak{ e}_{6}$ singularity always has an $\mathfrak{so}_{32}$ dual where the topology of the base is that of an affine $\ff_{4}$ Dynkin diagram, i.e., a $\mathfrak{e}_{6}$ folded by its $\mathbbm{Z}_2$ outer automorphism. This was already predicted in \cite{Intriligator:1997dh} and explicit models were constructed in \cite{DelZotto:2022ohj} due to properties of the representation theory of $\Gamma_{\fe_6}$. However, in all cases where we find a third dual, this is due to the affine extension having an additional $\mathbbm{Z}_2$ automorphism that acts non-trivial on the affine node. Indeed, affine $\fe_6$ also admits such an enhanced automorphism, but due to its triality symmetry, the action of this $\mathbbm{Z}_2$ is equivalent to the $\mathbbm{Z}_2$ symmetry that does not involve the affine node, cf.\ Figure~\ref{fig:E6Syms}. Hence, although both $\mathbbm{Z}_2$ actions on affine $\fe_6$ lead to an equivalent affine $\ff_4$ shape, they give rise to inequivalent models. 
 
Let us illustrate this explicitly in the model $\mathcal{K}_{M}(\mu_{1},\mu_{2},\mathfrak{e}_{6})$ with $M>2$ and holonomies $(\mu_{1},\mu_{2})$ that lead to a flavor algebra of $\mathfrak{so}_{12}^{2} \times \fu_1^{2}$,
\begin{align}
\lbrack  \mathfrak{so}_{12}\rbrack     \, \,   
{\overset{\mathfrak{sp}_{1}}{1}}   \, \,  
{\overset{\mathfrak{so}_{7}}{3}}  \, \,
{\overset{\mathfrak{su}_2}{2}} \, \,     
{\overset{{{{\left[\mathfrak{u}_{1}\right]}}}}{1}} \, \,
{\overset{\mathfrak{e}_{6}}{6}}   \, \,
\underbrace{{1 \, \,
\overset{\mathfrak{su}_{3}}{3}} \, \,
1 \, \,
{\overset{\mathfrak{e}_6}{6}}}_{\times (M-1)}  \, \,    
{\overset{{{{\left[\mathfrak{u}_{1}\right]}}}}{1}} \, \,
{\overset{\mathfrak{su}_{2}}{2}}   \, \, 
{\overset{\mathfrak{so}_{7}}{3}}   \, \, 
{\overset{\mathfrak{sp}_{1}}{1}}   \, \, 
\lbrack \mathfrak{so}_{12} \rbrack   \, ,   \qquad 
\nonumber
\vec{l}_{\LST}=(1, 1, 2,3,1 \underbrace{3,2,3,1}_{\times (M-1)},3,2,1,1) \, .
\end{align}
We find the following two duals:
\begin{align}     
\label{eq:dual1}
\textcolor{blue}{\lbrack  \mathfrak{u}_{1,D_{1}} \times \mathfrak{u}_{1,D_{2}}\rbrack } \times \bigg( 
{\overset{\mathfrak{sp}_{M-3}}{1}} \, \,
{\overset{\mathfrak{so}_{4M+4}}{4}} \, \,  
{\overset{\mathfrak{sp}_{3M-1}}{1}}  \, \, 
\underset{\left[\mathfrak{su}_{2}\right]}
{\overset{\mathfrak{su}_{4M+4}}{2}} \, \,
{\overset{\mathfrak{su}_{2M+8}}{2}}  \, \,     
\lbrack \mathfrak{su}_{12} \rbrack \bigg)   \, ,   \qquad \vec{l}_{\LST}=(1,1,3, 2,1) \, .
\end{align}

\begin{align}
\textcolor{blue}{\lbrack  \mathfrak{u}_{1,D}\rbrack } \times \left( 
\lbrack  \mathfrak{so}_{16}\rbrack     \, \,   
{\overset{\mathfrak{sp}_{M+3}}{1}}   \, \,  
{\underset{\left[\mathfrak{so}_{4}\right]}{\overset{\mathfrak{so}_{4M+12}}{4}}} \, \, 
{\overset{\mathfrak{sp}_{3M-1}}{1}}  \, \,
{\overset{\mathfrak{su}_{4M}}{2}} \, \, 
{\overset{\mathfrak{su}_{2M+2}}{2}} \, \, 
\lbrack \mathfrak{su}_{4} \rbrack \right)  \, ,   \qquad 
\vec{l}_{\LST}=(1, 1,3,2,1) \, 
\end{align}
Indeed, in both cases, the base topology is that of affine $\mathfrak{f}_{4}$, but the number of 5-branes and the flavor algebra are different. Thus, the base topology, 2-group structures and Coulomb branch dimension are not enough to determine an LST. Indeed, this data is for all three models (for $M>2$)
\begin{align}
\text{rk}(G_F)=14\, , \qquad     \text{Dim(CB)}=12M +12 \, , \qquad \kappa_{R}=24M+16\, .
\end{align}
Thus, from the perspective of the $\fso_{32}$ Heterotic string on an $\fe_6$ singularity, there are two models 
\begin{align}
    \tilde{\mathcal{K}}_N ( \lambda, \fe_6) \qquad \text{ and } \qquad \tilde{\mathcal{K}}_{N^\prime}(\lambda^\prime, \fe_6) \, ,
\end{align}
with different $\lambda$ and number of NS5 branes, both of which are dual to the same Heterotic $\fe_8 \times \fe_8$ theory.

\subsection{Duals with exceptional gauge algebras}
Instantons or NS5 branes in the $\fso_{32}$ Heterotic theory are typically perturbative and can be dualized to type $I$ theories with orientifolds (see e.g.~\cite{DelZotto:2022ohj}). Thus the appearing gauge algebra factors on such theories are of ortho-symplectic type such $\mathfrak{sp}_n$, $\mathfrak{so}_n$, and in special cases $\mathfrak{su}_n$ gauge algebras while exceptional algebras are absent. However, we find a class of isolated models (with $M=1$) that seem to have an $\mathfrak{e}_{7}$ gauge algebra in the dual.
Consider for example the theory $\mathcal{K}_{1}(\mu_{1},\mu_{2},\mathfrak{e}_{8})$ with flavor holonomies $(\mu_{1},\mu_{2})$ that lead to flavor algebras $\mathfrak{e}_{8} \times \mathfrak{so}_{15}$. The quiver of this theory is
\begin{align}
\begin{split}
\lbrack  \mathfrak{e}_{8}\rbrack     \, \,   
1 \, \,   
2 \, \,  
{\overset{\mathfrak{su}_2}{2}} \, \,
{\overset{\mathfrak{g}_2}{3}} \, \,
1 \, \,
{\overset{\mathfrak{f}_4}{5}} \, \
1 \, \,
{\overset{\mathfrak{g}_2}{3}} \, \,
{\overset{\mathfrak{su}_2}{2}} \, \,
2 \, \,
1 \, \,
{\overset{\mathfrak{e}_8}{12}} \, \,
1 \, \,
2 \, \,
{\overset{\mathfrak{su}_{2}}{2}}  \, \,
{\overset{\mathfrak{g}_2}{3}} \, \, 
1 \, \,
{\overset{\mathfrak{so}_{9}}{4}} \, \,
{\overset{\mathfrak{sp}_1}{1}} \, \,
{\overset{\mathfrak{so}_{11}}{4}} \, \,
{\overset{\mathfrak{sp}_2}{1}} \, \,
{\overset{\mathfrak{so}_{13}}{4}} \, \,
{\overset{\mathfrak{sp}_3}{1}} \, \,
\lbrack \mathfrak{so}_{15} \rbrack   \, ,   \qquad \\
\vec{l}_{\LST}=(1,1,1,1,2,1,3,2,3,4,5,1,7,6,5,4,7,3,5,2,3,1,1) ]\,.\qquad\qquad~
\end{split}
\end{align}
The dual theory is given by $\mathcal{\Tilde{K}}_{1}(\lambda;\mathfrak{g})$ with flavor holonomy $\lambda$ that leads to an $\mathfrak{so}_{7}\times \mathfrak{so}_{24}$ flavor algebra with the quiver
\begin{align}
\lbrack  \mathfrak{so}_{7}\rbrack     \, \,   
{\overset{\mathfrak{su}_{2}}{2}}   \, \,  
1 \, \,
{\overset{\mathfrak{e}_7}{8}} \, \,
1 \, \,
{\overset{\mathfrak{su}_{2}}{2}} \, \, 
{\overset{\mathfrak{so}_{7}}{3}}  \, \,
{\overset{\mathfrak{sp}_{1}}{1}} \, \, 
{\overset{\mathfrak{so}_{12}}{4}} \, \, 
{\overset{\mathfrak{sp}_{3}}{1}} \, \, 
{\overset{\mathfrak{so}_{16}}{4}} \, \,
{\overset{\mathfrak{sp}_5}{1}} \, \,
{\overset{\mathfrak{so}_{20}}{4}} \, \,
{\overset{\mathfrak{sp}_7}{1}} \, \,
\lbrack \mathfrak{so}_{24} \rbrack   \, ,   \qquad 
\vec{l}_{\LST}=(1,2,1,6,5,4,7,3,5,2,3,1,1) \, . \nonumber
\end{align}
The 6D anomalies cancel in both models, so these are consistent 6D gauge theories. Furthermore, the dimension of the Coulomb branch, the flavor rank, and $\kappa_{R}$ match, 
\begin{align}
\text{rk}(G_F)=15\, , \qquad     \text{Dim(CB)}=64 \, , \qquad \kappa_{R}=194\,,
\end{align} 
indicating that both theories are indeed dual  LSTs. As mentioned earlier, the appearance of an $\mathfrak{e}_{7}$ singularity seems puzzling from the perspective of the perturbative $\mathfrak{so}_{32}$ Heterotic string. It would be interesting to understand the origin of such a singularity.

\subsection{A self-dual model}
In this section we construct a self-dual model, i,.e., an LST such that
\begin{align}
   T_D(\mathcal{K}_N (\mu_1, \mu_2, \fg)) = \tilde{\mathcal{K}}_{N^\prime} (\lambda, \fg) = \mathcal{K}_N (\mu_1, \mu_2, \fg) \, ,
\end{align}
i.e. the $\text{E}_8 \times \text{E}_8$ model is \textit{exactly} the same as its $\text{Spin}(32)/\mathbbm{Z}_2$ counterpart, even though the duality map $T_D$ is non-trivial (the theories sit at different loci in the 5D moduli space). 

In order to do so, we choose the flavor holonomies such that we break $\fe_8\times\fe_8$ and $\fso_{32}$ to the maximal $\fso_{16}\times \fso_{16}$ subalgebra.\footnote{The full flavor group is in fact $(\text{Spin}(16) \times \text{Spin}(16))/\mathbbm{Z}_2$.} Next, we have to choose some singularity $\fg$ that has the chance to lead to the same quiver and gauge algebras in both Heterotic models. As there are typically no exceptional gauge algebras on the $\fso_{32}$ side, we restrict our search to $A$ and $D$ type singularities.
Since $A_r$ and $D_{2m+1}$ singularities usually lead to a folding on the $\text{Spin}(32)/\mathbbm{Z}_2$ side, we are left with $D_{2N}$. Indeed, for a $\fg=\fso_{4N}$ singularity (with $N>2$) and $M$ NS5 branes on the $\text{E}_8\times \text{E}_8$ side
we obtain 
\begin{align}
\label{ogsd}
    \lbrack  \mathfrak{so}_{16}\rbrack     \, \,   
{\overset{\mathfrak{sp}_{N}}{1}}   \, \,  
\myoverset{{\overset{\mathfrak{sp}_{N-4}}{1}}}
{\overset{\mathfrak{so}_{4N}}{4}}  \, \,
{\overset{\mathfrak{sp}_{2N-4}}{1}} \, \, 
\underbrace{{\overset{\mathfrak{so}_{4N}}{4}} \, \,
{\overset{\mathfrak{sp}_{2N-4}}{1}}}_{\times (M-2)}  \, \,
\myoverset{{\overset{\mathfrak{sp}_{N-4}}{1}}}
{\overset{\mathfrak{so}_{4N}}{4}} \, \,
{\overset{\mathfrak{sp}_{N}}{1}}  \, \,
\lbrack \mathfrak{so}_{16} \rbrack 
\end{align}
which dualizes to the $\text{Spin}(32)/\mathbbm{Z}_2$ quiver
\begin{align}
\label{eq:selfudal}
    \lbrack  \mathfrak{so}_{16}\rbrack     \, \,   
{\overset{\mathfrak{sp}_{M+1}}{1}}   \, \,  
\myoverset{\overset{\mathfrak{sp}_{M-3}}{1}}
{\overset{\mathfrak{so}_{4M+4}}{4}} \, \, 
{\overset{\mathfrak{sp}_{2M-2}}{1}}  \, \,
\underbrace{{\overset{\mathfrak{so}_{4M+4}}{4}} \, \, 
{\overset{\mathfrak{sp}_{2M-2}}{1}}}_{\times N-3} \, \, 
\myoverset{\overset{\mathfrak{sp}_{M-3}}{1}}
{\overset{\mathfrak{so}_{4M+4}}{4}} \, \, 
{\overset{\mathfrak{sp}_{M+1}}{1}} \, \,
\lbrack \mathfrak{so}_{16} \rbrack
\end{align}
There also exists a third $\text{Spin}(32)/\mathbbm{Z}_2$ dual, which is not of relevance here but can be found in Appendix~\ref{appendix:tables}. 
Evidently, upon choosing $M=N-1$, the two quivers~\eqref{eq:selfudal} and~\eqref{ogsd} become identical theories and their invariants are
\begin{align}
 \text{rk}(G_F)=16\, , \qquad     \text{Dim(CB)}= 4 N^2-4\, , \qquad \kappa_{R}= 8 N^2 - 16 N+ 10  \, .
\end{align} 
While the two models are identical as six-dimensional theories, it is important remark that their 10D origin as well as the T-duality, and hence the duality map $T_{D}$, between the two is non-trivial. The two theories are not just trivially the same upon circle reductions, but sit at different points of their 5D Coulomb branch moduli space. Phrased differently, when starting from either quiver and compactifying on a circle, one can move in the 5D moduli space to a different locus and find the exact same theory there.\footnote{We expect that this symmetry is reflected in the 5D moduli space of these theories.} This can also be seen from the underlying fiber/base duality: Upon compactifying LST~\eqref{ogsd}, all gauge algebra factors -- including the $\fso_{4N}$ factors on the $-4$ curves -- are shrunk to a singular point. However, these curves are precisely those that assemble the dual base quiver on the \text{Spin}(32)$/\mathbbm{Z}_2$ side. Thus, from the perspective of the $\text{E}_8 \times \text{E}_8$ LST reduced on a circle, the Spin$(32)/\mathbbm{Z}_2$ appears to have a shrunken $\fso_{4N}$ base and vice versa. The most generic point at which both theories are identical as 5D theories is on their maximal Coulomb branch, where all singularities are resolved.

%%%%%%%%%%%%%%%%%%%%%%%%%%%%%%%%%%%%%%%%%%%%%%%%%%%%%%%%%%%%%%%%%%%%%%%%%%%%
\section{Conclusions and Outlook}
\label{sec:conclusion}
In this work, we construct Heterotic LSTs with more general flavor breaking on the $\text{E}_8 \times \text{E}_8$ side, and construct their T-duals on the $\text{Spin}(32)/\mathbbm{Z}_2$ side. We determine the 5D Coulomb branch dimensions, universal 2-group structure constants and full flavor algebra. In addition to the former two, we show that the rank of the flavor algebra also always matches, providing another non-trivial invariant of LSTs that related by T-duality. This requires to carefully take into account $\fu_1$ ABJ flavor anomalies and finding non-anomalous linear combinations.

We discuss in detail the construction of Heterotic LSTs with $\fso$ and non-simply laced flavor symmetries. For the latter, we observe that T-duality preserves non-simply laced flavor algebras, suggesting a type of conserved freezing flux contribution in the M-theory description. Geometrically, T-dual heterotic LSTs can be constructed from non-compact K3-fibered CY threefolds with multiple elliptic fibrations. Different choices of elliptic fibrations correspond to different 6D LSTs, which become the same model in the 5D M-theory picture. The K3 fibers encode the flavor algebras. By exploiting non-K\"ahler favorable divisors, we engineer models that are forced to sit at special loci in the K3 moduli space, at which the volumes of several curves that resolve the ADE fiber singularity have the same volume, which allows modding out an outer automorphism to obtain non-simply laced subgroups. This procedure is inherent to a choice of polarization for the flavor K3 Picard lattice and preserved upon choosing different elliptic fibration structures, which means it is invariant under T-duality. This explains why the flavor algebra is also always reduced in T-dual theories of such models.

We use the geometric engineering procedure to construct LSTs with three T-dual theories. For these to appear, the flavor group on the $\text{E}_8\times \text{E}_8$ side must be broken. Moreover, a singularity $\fg$ whose affine extension has an additional $\mathbbm{Z}_2$ automorphism that involves the affine node is needed, which happens for $\fg=\{\fso_{2N}, \fe_6,\fe_7\}$. When these conditions are met, one obtains two Heterotic $\text{Spin}(32)/\mathbbm{Z}_2$ duals where the base quivers of one theory is in the shape of the affine $\fg$ algebra and the other is of the same shape with the outer automorphism modded out. We highlight this for the special case of an $\fe_6$ singularity, whose affinization has two inequivalent $\mathbbm{Z}_2$ symmetries, both of which fold the Dynkin diagram to an affine $\ff_4$-shaped quiver, but with different flavor and gauge algebra factors. The T-duality map $T_D$ thus provides multiple $\text{Spin}(32)/\mathbbm{Z}_2$ group homomorphisms for a single $\text{E}_8\times \text{E}_8$ theory and can therefore not be an isomorphism. 

Finally, we construct a family of $\text{E}_8\times \text{E}_8 \leftrightarrow \text{Spin}(32)/\mathbbm{Z}_2$ self-T-dual models. These two LSTs have exactly the same $D_{2N}$ type of singularity and Spin$(16)^2/\mathbbm{Z}_2$ flavor group. However, these theories are at different loci in the 5D Coulomb branch moduli space upon circle compactification to 5D. This suggests that the 5D moduli space has a $\mathbbm{Z}_2$ symmetry.

While this work continues to carve out the large landscape of Heterotic little string theories, there are still several open problems. First, it would be interesting to understand how the duality map $T_D$ acts on the homomorphisms $(\mu_1,\mu_2)$ and $\lambda$ from first principles. For this, one should study the map of the nilpotent deformations in the $\text{E}_8\times\text{E}_8$ theory to the $\text{Spin}(32)/\mathbbm{Z}_2$ theory, akin to the analysis of~\cite{Frey:2018vpw} for the orbi-instanton theories.

There are also two more LST type of structures that have not yet been discussed: The first ones are LSTs with twisted T-duals that are related by discrete holonomy reductions~\cite{Bhardwaj:2019fzv,DelZotto:2020sop,Bhardwaj:2022ekc}. Geometrically, these can be engineered from genus-one fibered Calabi-Yau threefolds~\cite{Oehlmann:2019ohh,Anderson:2023wkr,Anderson:2023tfy}. Second, it would be interesting to extend our explorations to IIB compactifications with $O7^+$ planes, which correspond to frozen phases of F-theory~\cite{deBoer:2001wca,Tachikawa:2015wka,Bhardwaj:2018jgp, Morrison:2023hqx}. 

Lastly, the perturbative Heterotic string is a part of the string spectrum of any 6D SUGRA theory with $n_T>0$ tensor multiplets, which are precisely the LST sectors we studied here. A general perspective of T-duality for Heterotic LSTs could therefore potentially elucidate properties of BPS strings across dualities when gravity is included. In particular, one may ask whether T-dualities of such SUGRA theories are captured completely by the T-dualities of their respective LST subsectors.

%%%%%%%%%%%%%%%%%%%%%%%%%%%%%%%%%%%%%%%%
\section*{Acknowledgements}
We thank Florent Baume and Craig Lawrie for discussions, and Florent Baume, Michele del Zotto and Muyang Liu for collaborations on related topics. The work of FR is supported by the NSF grants PHY-2210333 and PHY-2019786 (The NSF AI Institute for Artificial Intelligence and Fundamental Interactions). The work of HA, PKO, and FR is also supported by startup funding from Northeastern University.%%%%%%%%%%%%%%%%%%%%%%%%%%%%%%%%%%%%
 
\appendix
\clearpage
\section{List of T-dual quivers and their invariants}
\label{appendix:tables}
    \centering
    \resizebox{1\textwidth}{!}{
    \begin{tabular}{|p{1300pt}|c|}  \hline
       \resizebox{2.7\textwidth}{!}{
           ($\mathfrak{g}$,
            $\text{rk}(G_{F})$, 
            $\kappa_{R}$,
            $\text{Dim(CB)})^{T}$  
              }
      & \resizebox{2.7\textwidth}{!}{\Huge T-dual theory description} \\ \hline
      \resizebox{2.2\columnwidth}{!}{
  \begin{minipage}{0.1\textwidth}\Huge
     \begin{align*}
       &\mathfrak{so}_{4N}^{M} \\ \\
&16 \\ \\
       & 8NM-8M+2 \\ \\
       & 4NM-2M+2N-2
       \end{align*}
  \end{minipage}
}
        &  \resizebox{16\textwidth}{!}{
  \begin{minipage}{0.1\textwidth}\Huge
   \begin{align*}
    \lbrack  \mathfrak{so}_{16}\rbrack     \, \,   
{\overset{\mathfrak{sp}_{N}}{1}}   \, \,  
\myoverset{{\overset{\mathfrak{sp}_{N-4}}{1}}}
{\overset{\mathfrak{so}_{4N}}{4}}  \, \,
{\overset{\mathfrak{sp}_{2N-4}}{1}} \, \, 
\underbrace{{\overset{\mathfrak{so}_{4N}}{4}} \, \,
{\overset{\mathfrak{sp}_{2N-4}}{1}}}_{\times (M-2)}  \, \,
\myoverset{{\overset{\mathfrak{sp}_{N-4}}{1}}}
{\overset{\mathfrak{so}_{4N}}{4}} \, \,
{\overset{\mathfrak{sp}_{N}}{1}}  \, \,
\lbrack \mathfrak{so}_{16} \rbrack   \, ,   \qquad \qquad 
\textcolor{blue}{\lbrack  \mathfrak{u}_{1,D} \rbrack } \times \bigg(\myoverset{\overset{\myoverset{\left[\mathfrak{su}_{16} \right]}{\mathfrak{su}_{2M+2N+2}}}{2}}
{\myunderset{\overset{\mathfrak{su}_{2M+2N-6}}{2}}
{\overset{\mathfrak{su}_{4M+4N-12}}{2}}}  \, \,   
\underbrace{{\overset{\mathfrak{su}_{4M+4N-20}}{2}}  \, \,     
{\overset{\mathfrak{su}_{4M+4N-28}}{2}} \, \,
\dots \dots}_{\times (N-3)}
{\overset{\mathfrak{sp}_{2M-2N+2}}{1}} \, \, \bigg),  \qquad \qquad \lbrack  \mathfrak{so}_{16}\rbrack     \, \,   
{\overset{\mathfrak{sp}_{M+1}}{1}}   \, \,  
\myoverset{\overset{\mathfrak{sp}_{M-3}}{1}}
{\overset{\mathfrak{so}_{4M+4}}{4}} \, \, 
{\overset{\mathfrak{sp}_{2M-2}}{1}}  \, \,
\underbrace{{\overset{\mathfrak{so}_{4M+4}}{4}} \, \, 
{\overset{\mathfrak{sp}_{2M-2}}{1}}}_{\times N-3} \, \, 
\myoverset{\overset{\mathfrak{sp}_{M-3}}{1}}
{\overset{\mathfrak{so}_{4M+4}}{4}} \, \, 
{\overset{\mathfrak{sp}_{M+1}}{1}} \, \,
\lbrack \mathfrak{so}_{16} \rbrack
      \end{align*}
  \end{minipage}
} \\ \hline

              \resizebox{1.6\columnwidth}{!}{
  \begin{minipage}{0.05\textwidth}\Huge
     \begin{align*}
       &\mathfrak{so}_{4N+2}^{M} \\ \\
&14 \\ \\
       & 8NM-4M \\ \\
       & 4NM+2N-3
       \end{align*}
  \end{minipage}
}
        &  \resizebox{18\textwidth}{!}{
  \begin{minipage}{0.1\textwidth}\Huge
   \begin{align*}
\lbrack  \mathfrak{so}_{10}\rbrack     \, \,   
{\overset{\mathfrak{sp}_{N+1}}{1}}  \, \,
\myoverset{\overset{\overset{\left[\mathfrak{u}_{1} \right]}{\mathfrak{sp}_{N-3}}}{1}}
{\underset{\left[\mathfrak{sp}_1\right]}
{\overset{\mathfrak{so}_{4N+2}}{4}}} \, \,
{\overset{\mathfrak{sp}_{2N-3}}{1}}  \, \,
\underbrace{{
\overset{\mathfrak{so}_{4N+2}}{4}} \, \,
{\overset{\mathfrak{sp}_{2N-3}}{1}} }_{\times (M-2)}  \, \,
\myoverset{\overset{\overset{\left[\mathfrak{u}_{1} \right]}{\mathfrak{sp}_{N-3}}}{1}}
{\underset{\left[\mathfrak{sp}_1\right]}
{\overset{\mathfrak{so}_{4N+2}}{4}}} \, \,
{\overset{\mathfrak{sp}_{N+1}}{1}}  \, \,
\lbrack \mathfrak{so}_{10} \rbrack   \, ,   \qquad \qquad
\textcolor{blue}{\lbrack  \mathfrak{u}_{1,D_{1}} \times  \mathfrak{u}_{1,D_{2}} \times \mathfrak{u}_{1,D_{3}} \rbrack } \times \bigg(\left[\mathfrak{su}_{2} \right]     \, \,  
\myoverset{\overset{\myoverset{\left[\mathfrak{su}_{2} \right]}{\mathfrak{su}_{2M+2N-4}}}{2}}
{\myunderset{\overset{\myunderset{2}{\mathfrak{su}_{2M+2N}}}{\left[\mathfrak{su}_{10} \right]}}
{\overset{\mathfrak{su}_{4M+4N-10}}{2}}}  \, \,   
\underbrace{{\overset{\mathfrak{su}_{4M+4N-18}}{2}}  \, \,     
{\overset{\mathfrak{su}_{4M+4N-26}}{2}} \, \,
\dots \dots}_{\times (N-3)}
{\overset{\mathfrak{su}_{4M-4N+6}}{1}} \, \,\bigg),  \qquad \qquad \textcolor{blue}{\lbrack  \mathfrak{u}_{1,D_{1}} \times  \mathfrak{u}_{1,D_{2}} \rbrack } \times \bigg(\lbrack  \mathfrak{su}_{3} \rbrack     \, \,   
{\overset{\mathfrak{su}_{2M}}{2}}   \, \,  
\underset{\left[\mathfrak{so}_{4}\right]}
{\overset{\mathfrak{sp}_{2M-2}}{1}}  \, \,
\underbrace{{\overset{\mathfrak{so}_{4M+4}}{4}} \, \, 
{\overset{\mathfrak{sp}_{2M-2}}{1}}}_{\times N-2} \, \, 
\myoverset{\overset{\mathfrak{sp}_{M-3}}{1}}
{\overset{\mathfrak{so}_{4M+4}}{4}} \, \, 
{\overset{\mathfrak{sp}_{M+1}}{1}} \, \,
\lbrack \mathfrak{so}_{16} \rbrack \bigg)
   \end{align*}
  \end{minipage}
}
         \\ \hline  
           
              \resizebox{2.6\columnwidth}{!}{
  \begin{minipage}{0.05\textwidth}\Huge
     \begin{align*}
       &\mathfrak{so}_{4N}^{M} \\ \\
&16 \\ \\
       &4N^{2}-8N+8NM-8M+6 \\ \\
       & 2N^{2}-N+4NM-2M-1
   \end{align*}
  \end{minipage}
}
        &   
         \resizebox{18\textwidth}{!}{
  \begin{minipage}{0.1\textwidth}\Huge
   \begin{align*}
    \textcolor{blue}{\lbrack  \mathfrak{u}_{1,D} \rbrack } \times \bigg(
\lbrack  \mathfrak{su}_{8}\rbrack     \, \,   
{\overset{\mathfrak{su}_{2N}}{2}}   \, \,  
{\overset{\mathfrak{sp}_{2N-4}}{1}}   \, \,  
{\overset{\mathfrak{so}_{4N}}{4}}  \, \,
\underbrace{{\overset{\mathfrak{sp}_{2N-4}}{1}} \, \,
\underset{\left[\mathfrak{sp}_1\right]}
{\overset{\mathfrak{so}_{4N}}{4}}}_{\times (M-1)}  \, \,
{\overset{\mathfrak{sp}_{2N-5}}{1}} \, \,
{\overset{\mathfrak{so}_{4N-4}}{4}}  \, \,
{\overset{\mathfrak{sp}_{2N-7}}{4}}  \, \,
{\overset{\mathfrak{so}_{4N-8}}{4}}  \, \,
{\overset{\mathfrak{sp}_{2N-9}}{4}}  \, \,
\cdots \, \,
{\overset{\mathfrak{so}_{12}}{4}}  \, \,
{\overset{\mathfrak{sp}_{1}}{1}}  \, \,
{\overset{\mathfrak{so}_{7}}{3}}  \, \,
{\overset{\mathfrak{su}_{2}}{2}}  \, \,
1 \, \,
\lbrack \mathfrak{e}_{7} \rbrack \bigg)  \, ,   \qquad \qquad
\textcolor{blue}{\lbrack  \mathfrak{u}_{1,D_{1}} \times  \mathfrak{u}_{1,D_{2}} \rbrack } \times \bigg(\myoverset{\overset{\myoverset{\left[\mathfrak{su}_{2} \right]}{\mathfrak{su}_{2M+3N-6}}}{2}}
{\myunderset{\overset{\myunderset{2}{\mathfrak{su}_{2M+3N}}}{\left[\mathfrak{su}_{14} \right]}}
{\overset{\mathfrak{su}_{4M+6N-14}}{2}}}  \, \,    
\underbrace{{\overset{\mathfrak{su}_{4M+6N-22}}{2}}  \, \,     
{\overset{\mathfrak{su}_{4M+6N-30}}{2}} \, \,
\dots \dots}_{\times (N-3)}
{\overset{\mathfrak{sp}_{2M-N+1}}{1}} \, \,\bigg),  \qquad \qquad \lbrack  \mathfrak{so}_{10}\rbrack     \, \,   
{\overset{\mathfrak{sp}_{M}}{1}}   \, \,  
\myoverset{\overset{\overset{\left[ \mathfrak{u}_{1} \right]}{\mathfrak{sp}_{M-2}}}{1}}
{\overset{\mathfrak{so}_{4M+6}}{4}} \, \, 
{\overset{\mathfrak{sp}_{2M}}{1}}  \, \,
\underbrace{\dots \dots 
 {\overset{\mathfrak{so}_{4M+4N-10}}{4}} \, \, 
{\overset{\mathfrak{sp}_{2M+2N-8}}{1}} \, \,
{\overset{\mathfrak{so}_{4M+4N-6}}{4}} \, \, 
{\overset{\mathfrak{sp}_{2M+2N-6}}{1}}}_{\times 2(N-3)} \, \, 
\myoverset{\overset{\overset{\left[\mathfrak{u}_{1} \right]}{\mathfrak{sp}_{M+N-4}}}{1}}
{\overset{\mathfrak{so}_{4M+4N-2}}{4}} \, \, 
{\overset{\mathfrak{sp}_{M+N}}{1}} \, \,
\lbrack \mathfrak{so}_{18} \rbrack
 \end{align*}
  \end{minipage}
} \\ \hline

              \resizebox{2.6\columnwidth}{!}{
  \begin{minipage}{0.05\textwidth}\Huge
     \begin{align*}
       &\mathfrak{so}_{4N+2}^{M} \\ \\
&16 \\ \\
       &4N^{2}-4N+4NM-1 \\ \\
       & 8N^{2}-16N+8NM-4M+8
\end{align*}
  \end{minipage}
}
        &   
         \resizebox{18\columnwidth}{!}{
  \begin{minipage}{0.01\textwidth}\Huge
   \begin{align*}
\lbrack  \mathfrak{e}_{6}\rbrack     \, \,   
1 \, \,
{\overset{\mathfrak{su}_{3}}{3}}  \, \,
1 \, \,
{\overset{\mathfrak{so}_{10}}{4}} \, \,
{\overset{\mathfrak{sp}_{2}}{1}} \, \,
{\overset{\mathfrak{so}_{14}}{4}} \, \,
{\overset{\mathfrak{sp}_{4}}{1}} \, \,
\cdots \, \,
{\overset{\mathfrak{sp}_{2N-4}}{1}} \, \,
\underset{\left[\mathfrak{sp}_{1}\right]}
{\overset{\mathfrak{so}_{4N+2}}{4}} \, \,
\underbrace{{{\overset{\mathfrak{sp}_{2N-3}}{1}}  \, \,
\overset{\mathfrak{so}_{4N+2}}{4}}}_{\times (M-2)}  \, \,
\overset{\mathfrak{sp}_{2N-3}}{1} \, \,
\underset{\left[\mathfrak{sp}_{1}\right]}
{\overset{\mathfrak{so}_{4N+2}}{4}} \, \,
\overset{\mathfrak{sp}_{2N-4}}{1} \, \,
\cdots \, \,
{\overset{\mathfrak{so}_{14}}{4}} \, \,
{\overset{\mathfrak{sp}_{4}}{1}} \, \,
\underset{\left[\mathfrak{sp}_{2}\right]}
{\overset{\mathfrak{so}_{10}}{4}} \, \,
1 \, \,
{\overset{\mathfrak{su}_{3}}{3}}  \, \,
1 \, \,
\lbrack \mathfrak{e}_{6} \rbrack   \, ,   \qquad  \qquad
\textcolor{blue}{\lbrack \mathfrak{u}_{1,D_{1}} \times  \mathfrak{u}_{1,D_{2}} \rbrack } \times \bigg(\myoverset{\overset{\myoverset{\left[\mathfrak{su}_{4} \right]}{\mathfrak{su}_{2M+4N-6}}}{2}}
{\myunderset{\overset{\myunderset{2}{\mathfrak{su}_{2M+4N-2}}}{\left[\mathfrak{su}_{12} \right]}}
{\overset{\mathfrak{su}_{4M+8N-16}}{2}}}  \, \,    
\underbrace{{\overset{\mathfrak{su}_{4M+8N-24}}{2}}  \, \,     
{\overset{\mathfrak{su}_{4M+8N-32}}{2}} \, \,
\dots \dots}_{\times (N-3)}
{\overset{\mathfrak{su}_{4M}}{1}} \, \, \bigg),   \qquad \qquad \textcolor{blue}{\lbrack \mathfrak{u}_{1,D} \rbrack } \times \bigg(\lbrack  \mathfrak{so}_{20}\rbrack     \, \,   
{\overset{\mathfrak{sp}_{M+2N-1}}{1}}   \, \,  
\myoverset{\overset{\overset{\left[ \mathfrak{so}_{4} \right]}{\mathfrak{sp}_{M+2N-5}}}{1}}
{\overset{\mathfrak{so}_{4M+8N-8}}{4}} \, \, 
\underbrace{\dots \dots 
 {\overset{\mathfrak{sp}_{2M+10}}{1}} \, \, 
{\overset{\mathfrak{so}_{4M+24}}{4}} \, \,
{\overset{\mathfrak{sp}_{2M+6}}{1}} \, \, 
{\overset{\mathfrak{so}_{4M+16}}{4}}}_{\times 2(N-3)} \, \, 
{\overset{\mathfrak{sp}_{2M+2}}{1}} \, \,
{\overset{\mathfrak{so}_{4M+8}}{4}} \, \,
{\overset{\mathfrak{sp}_{2M-2}}{1}} \, \, 
{\overset{\mathfrak{su}_{2M}}{2}} \, \,
\lbrack \mathfrak{su}_{4} \rbrack\bigg)
 \end{align*}
  \end{minipage}
} 
   \\ \hline

              \resizebox{1.2\columnwidth}{!}{
  \begin{minipage}{0.05\textwidth}\Huge
     \begin{align*}
       &\mathfrak{e}_{6}^{M} \\ \\
&	16 \\ \\
       & 24M+34 \\ \\
       & 	12M+22
   \end{align*}
  \end{minipage}
}
       &   
         \resizebox{18\columnwidth}{!}{
  \begin{minipage}{0.1\textwidth}\Huge
   \begin{align*}
     \lbrack  \mathfrak{so}_{14}
\rbrack     \, \,     {\overset{\mathfrak{sp}_2}{1}}  \, \,   {\overset{\mathfrak{so }_{10}}{4}}  \, \,   
{\overset{{\left[\mathfrak{u}_{1}
\right]}}{1}} \, \, 
{\overset{\mathfrak{su}_3}{3}} \, \,  
1 \, \, 
{\overset{\mathfrak{e}_6}{6}} \, \,
\underbrace{1 \, \,
{\overset{\mathfrak{su}_3}{3}} \, \,
1 \, \,
{\overset{\mathfrak{e}_6}{6}}}_{	\times (M-1)}\, \,
1 \, \,
{\overset{\mathfrak{su}_3}{3}} \, \,
{\overset{{\left[\mathfrak{u}_{1}
\right]}}{1}} \, \, 
{\overset{\mathfrak{so}_{10}}{4}}  \, \,
{\overset{\mathfrak{sp}_2}{1}} \, \,  
\lbrack \mathfrak{so}_{14} 
\rbrack   \, ,   \qquad \qquad
     	\textcolor{blue}{\lbrack  \mathfrak{u}_{1,D_{1}} 	\times  \mathfrak{u}_{1,D_{2}} 
\rbrack } 	\times \left( \lbrack  \mathfrak{su}_{2}
\rbrack     \, \, 
{\overset{\mathfrak{sp}_{M-1}}{1}}  \, \,   
{\overset{\mathfrak{so}_{4M+8}}{4}}  \, \,     
{\overset{\mathfrak{sp}_{3M+1}}{1}} \, \,   
{\overset{\mathfrak{su}_{4M+6}}{2}}  \, \,
{\overset{\mathfrak{su}_{2M+10}}{2}} \, \,
\lbrack \mathfrak{su}_{14} 
\rbrack 
\right),   \qquad \qquad
    	\textcolor{blue}{\lbrack  \mathfrak{u}_{1,D} 
\rbrack } 	\times \left(
\lbrack  \mathfrak{su}_{8}
\rbrack     \, \, 
{\overset{\mathfrak{su}_{2M+6}}{2}}  \, \,   
{\overset{\mathfrak{su}_{4M+4}}{2}}  \, \,     
{\overset{\mathfrak{sp}_{3M+1}}{1}} \, \,   
{\overset{\mathfrak{so}_{4M+12}}{4}}  \, \,
{\overset{\mathfrak{sp}_{M+3}}{1}} \, \,
\lbrack \mathfrak{so}_{16} 
\rbrack 
\right)  
  \end{align*}
  \end{minipage}
}
     \\ \hline
            
              \resizebox{1.2\columnwidth}{!}{
  \begin{minipage}{0.05\textwidth}\Huge
     \begin{align*}
       &\mathfrak{e}_{7}^{M} \\ \\
&	15 \\ \\
       & 48M+25 \\ \\
       & 18M+19
       \end{align*}
  \end{minipage}
}
        &   
         \resizebox{18\columnwidth}{!}{
  \begin{minipage}{0.1\textwidth}\Huge
   \begin{align*}
    \lbrack  \mathfrak{so}_{16}
\rbrack     \, \,   
{\overset{\mathfrak{sp}_{3}}{1}}  \, \,   
{\overset{\mathfrak{so}_{12}}{4}} \, \,  
{\overset{\mathfrak{sp}_1}{1}} \, \,
{\overset{\mathfrak{so}_7}{3}} \, \,
{\overset{\mathfrak{su}_2}{2}} \, \,
1 \, \,
\underbrace{{\overset{\mathfrak{e}_7}{8}} \, \,
1 \, \,
{\overset{\mathfrak{su}_2}{2}} \, \,
{\overset{\mathfrak{so}_7}{3}} \, \,
{\overset{\mathfrak{su}_2}{2}} \, \,
1}_{	\times (M-1)} \, \, 
\myoverset{{\overset{\left[\mathfrak{su}_2
\right]}{1}}}
{\overset{\mathfrak{e}_7}{8}} \, \,
1 \, \,
{\overset{\mathfrak{su}_{2}}{2}}  \, \,
{\overset{\mathfrak{so}_7}{3}} \, \, 
{\overset{\mathfrak{sp}_1}{1}} \, \, 
\lbrack \mathfrak{so}_{12} 
\rbrack   \, ,   \qquad \qquad
    	\textcolor{blue}{\lbrack  \mathfrak{u}_{1,D_{1}} 	\times  \mathfrak{u}_{1,D_{2}} 
\rbrack } 	\times \left(
{\overset{\mathfrak{so}_{4M+6}}{4}}  \, \,   
{\overset{\mathfrak{sp}_{4M-2}}{1}}  \, \,     
{\overset{\mathfrak{su}_{6M+1}}{2}} \, \,  
{\overset{\mathfrak{su}_{4M+6}}{2}}  \, \,
{\overset{\mathfrak{su}_{2M+10}}{2}} \, \,
\lbrack \mathfrak{su}_{14} 
\rbrack
\right),   
\qquad \qquad
   \lbrack  \mathfrak{so}_{12}
\rbrack     \, \, 
{\overset{\mathfrak{sp}_{M+1}}{1}}  \, \,   
{\overset{\mathfrak{so}_{4M+8}}{4}}  \, \,     
{\overset{\mathfrak{sp}_{3M-1}}{1}} \, \, 
\myoverset{{\overset{\mathfrak{sp}_{2M-3}}{1}}}
{\overset{\mathfrak{so}_{8M+4}}{4}}  \, \,
{\overset{\mathfrak{sp}_{3M}}{1}} \, \,
\underset{\left[\mathfrak{sp}_{1} 
\right]}
{\overset{\mathfrak{so}_{4M+12}}{4}}  \, \,
{\overset{\mathfrak{sp}_{M+3}}{1}}  \, \,
\lbrack \mathfrak{so}_{16} 
\rbrack  
 \end{align*}
  \end{minipage}
} 
      \\ \hline
    
              \resizebox{1.2\columnwidth}{!}{
  \begin{minipage}{0.05\textwidth}\Huge
     \begin{align*}
       &\mathfrak{e}_{7}^{M} \\ \\
&	16 \\ \\
       &48M+50 \\ \\
       & 18M+29
      \end{align*}
  \end{minipage}
}
       &   
         \resizebox{14\textwidth}{!}{
  \begin{minipage}{0.1\textwidth}\Huge
   \begin{align*}
   \lbrack  \mathfrak{e}_{8}
\rbrack     \, \,   
1 \, \,   
2 \, \,  
{\overset{\mathfrak{su}_2}{2}} \, \,
{\overset{\mathfrak{g}_2}{3}} \, \,
1 \, \,
{\overset{\mathfrak{f}_4}{5}} \, \
1 \, \,
{\overset{\mathfrak{g}_2}{3}} \, \,
{\overset{\mathfrak{su}_2}{2}} \, \,
1 \, \,
{\overset{\mathfrak{e}_7}{8}} \, \,
\underbrace{1 \, \,
{\overset{\mathfrak{su}_2}{2}} \, \,
{\overset{\mathfrak{so}_7}{3}} \, \,
{\overset{\mathfrak{su}_2}{2}} \, \,
1 \, \,
{\overset{\mathfrak{e}_7}{8}}}_{	\times (M-1)} \, \, 
1 \, \,
{\overset{\mathfrak{su}_{2}}{2}}  \, \,
{\overset{\mathfrak{so}_7}{3}} \, \, 
{\overset{\mathfrak{sp}_1}{1}} \, \,
{\overset{\mathfrak{so}_{12}}{4}} \, \, 
{\overset{\mathfrak{sp}_3}{1}} \, \,
\lbrack \mathfrak{so}_{16} 
\rbrack   \, ,   \qquad \qquad \lbrack  \mathfrak{so}_{8}
\rbrack     \, \,   
{\overset{\mathfrak{sp}_{M}}{1}}   \, \,  
{\overset{\mathfrak{so}_{4M+8}}{4}} \, \, 
{\overset{\mathfrak{sp}_{3M}}{1}}  \, \,
\myoverset{\overset{\mathfrak{sp}_{2M-2}}{1}}
{\overset{\mathfrak{so}_{8M+8}}{4}} \, \, 
{\overset{\mathfrak{sp}_{3M+2}}{1}} \, \, 
{\overset{\mathfrak{so}_{4M+16}}{4}} \, \, 
{\overset{\mathfrak{sp}_{M+6}}{1}} \, \,
\lbrack \mathfrak{so}_{24} 
\rbrack 
  \end{align*}
  \end{minipage}
} 
 \\ \hline

\resizebox{1.2\columnwidth}{!}{
  \begin{minipage}{0.1\textwidth}\Huge
     \begin{align*}
       &\mathfrak{e}_{7}^{M} \\ \\
&16 \\ \\
       & 48M+25 \\ \\
       & 18M+19
        \end{align*}
  \end{minipage}
}
        & \resizebox{18\textwidth}{!}{
  \begin{minipage}{0.1\textwidth}\Huge
   \begin{align*}
   \lbrack  \mathfrak{so}_{16}\rbrack     \, \,   
{\overset{\mathfrak{sp}_{3}}{1}}  \, \,   
{\overset{\mathfrak{so}_{12}}{4}} \, \,  
{\overset{\mathfrak{sp}_1}{1}} \, \,
{\overset{\mathfrak{so}_7}{3}} \, \,
{\overset{\mathfrak{su}_2}{2}} \, \,
1 \, \,
\underbrace{{\overset{\mathfrak{e}_7}{8}} \, \,
1 \, \,
{\overset{\mathfrak{su}_2}{2}} \, \,
{\overset{\mathfrak{so}_7}{3}} \, \,
{\overset{\mathfrak{su}_2}{2}} \, \,
1}_{\times (M-1)}  \, \,
\myoverset{{\overset{\left[\mathfrak{su}_2\right]}{1}}}
{\overset{\mathfrak{e}_7}{8}} \, \,
1 \, \,
{\overset{\mathfrak{su}_2}{2}} \, \,
{\overset{\mathfrak{so}_7}{3}} \, \,
{\overset{\mathfrak{su}_2}{2}} \, \,
1\, \,
\lbrack \mathfrak{e}_{7} \rbrack   \, ,   \qquad \qquad
   \textcolor{blue}{\lbrack \mathfrak{u}_{1,D} \rbrack } \times \left(
{\overset{\mathfrak{so}_{4M+6}}{4}}  \, \,   
{\overset{\mathfrak{sp}_{4M-2}}{1}}  \, \,     
{\overset{\mathfrak{su}_{6M+1}}{2}} \, \,  
{\overset{\mathfrak{su}_{4M+6}}{2}}  \, \,
{\overset{\mathfrak{su}_{2M+11}}{2}} \, \,
\lbrack \mathfrak{u}_{16} \rbrack \right),    \qquad \qquad \lbrack  \mathfrak{so}_{12}\rbrack     \, \,   
{\overset{\mathfrak{sp}_{M+1}}{1}}   \, \,  
{\overset{\mathfrak{so}_{4M+8}}{4}} \, \, 
{\overset{\mathfrak{sp}_{3M-1}}{1}}  \, \,
\myoverset{\overset{\mathfrak{sp}_{2M-3}}{1}}
{\overset{\mathfrak{so}_{8M+4}}{4}} \, \, 
{\overset{\mathfrak{sp}_{3M}}{1}} \, \, 
{\overset{\mathfrak{so}_{4M+12}}{4}} \, \, 
{\overset{\mathfrak{sp}_{M+4}}{1}} \, \,
\lbrack \mathfrak{so}_{20} \rbrack 
\end{align*}
  \end{minipage}
} 
     \\ \hline
   \end{tabular}}

% next page
\clearpage 

    \resizebox{1\textwidth}{!}{
    \begin{tabular}{|p{1300pt}|c|}  \hline
       \resizebox{2.7\textwidth}{!}{
           ($\mathfrak{g}$,
            $\text{rk}(G_{F})$, 
            $\kappa_{R}$,
            $\text{Dim(CB)})^{T}$  
              }
      & \resizebox{2.7\textwidth}{!}{\Huge T-dual theory description} \\ \hline

              \resizebox{1.3\columnwidth}{!}{
  \begin{minipage}{0.05\textwidth}\Huge
     \begin{align*}
       &\mathfrak{so}_{16}^{M} \\ \\
&16 \\ \\
       & 24M+74 \\ \\
       & 14M+48
      \end{align*}
  \end{minipage}
} 
      &  \resizebox{16\textwidth}{!}{
  \begin{minipage}{0.1\textwidth}\Huge
   \begin{align*}
\lbrack  \mathfrak{e}_{8}\rbrack     \, \,   
1  \, \,   
2 \, \,  
{\overset{\mathfrak{su}_2}{2}} \, \,
{\overset{\mathfrak{g}_2}{3}} \, \,
1 \, \,
{\overset{\mathfrak{so}_9}{4}} \, \,
{\overset{\mathfrak{sp}_1}{1}} \, \,
{\overset{\mathfrak{so}_{11}}{4}} \, \,
{\overset{\mathfrak{sp}_2}{1}} \, \,
{\overset{\mathfrak{so}_{13}}{4}} \, \,
{\overset{\mathfrak{sp}_3}{1}} \, \,
{\overset{\mathfrak{so}_{15}}{4}} \, \,
{\overset{\mathfrak{sp}_4}{1}} \, \,
\underbrace{{\overset{\mathfrak{so}_{16}}{4}} \, \,
{\overset{\mathfrak{sp}_4}{1}}}_{\times (M-1)}  \, \,
\myoverset{1}
{\overset{\mathfrak{so}_{16}}{4}} \, \,
{\overset{\mathfrak{sp}_4}{1}} \, \,
\lbrack \mathfrak{so}_{15} \rbrack   \, ,   \qquad \qquad 
\lbrack  \mathfrak{so}_{8}\rbrack     \, \,   
{\overset{\mathfrak{sp}_{M}}{1}}   \, \,  
\myoverset{\overset{\mathfrak{sp}_{M-2}}{1}}
{\overset{\mathfrak{so}_{4M+8}}{4}} \, \, 
{\overset{\mathfrak{sp}_{2M+2}}{1}}  \, \,
{\overset{\mathfrak{so}_{4M+16}}{4}} \, \, 
{\overset{\mathfrak{sp}_{2M+6}}{1}} \, \, 
\myoverset{\overset{\mathfrak{sp}_{M+2}}{1}}
{\overset{\mathfrak{so}_{4M+24}}{4}} \, \, 
{\overset{\mathfrak{sp}_{M+8}}{1}} \, \,
\lbrack \mathfrak{so}_{24} \rbrack
 \end{align*}
  \end{minipage}
} 
     \\ \hline

      \resizebox{1.3\columnwidth}{!}{
  \begin{minipage}{0.01\textwidth}\Huge
     \begin{align*}
       &\mathfrak{e}_{8}^{M} \\ \\
&14 \\ \\
       &120M+24 \\ \\
       & 30M+24
          \end{align*}
  \end{minipage}
}
         &   
         \resizebox{18\textwidth}{!}{
  \begin{minipage}{0.1\textwidth}\Huge
   \begin{align*}
\lbrack  \mathfrak{e}_{8}\rbrack     \, \,   
1  \, \,   
2 \, \,  
{\overset{\mathfrak{su}_2}{2}} \, \,
{\overset{\mathfrak{g}_2}{3}} \, \,
1 \, \,
{\overset{\mathfrak{f}_4}{5}} \, \,
1 \, \,
{\overset{\mathfrak{g}_{2}}{3}} \, \,
{\overset{\mathfrak{su}_2}{2}} \, \,
2 \, \,
1 \, \,
\myoverset{1}
{\overset{\mathfrak{e}_{8}}{12}} \, \,
\underbrace{1  \, \,   
2 \, \,  
{\overset{\mathfrak{su}_2}{2}} \, \,
{\overset{\mathfrak{g}_2}{3}} \, \,
1 \, \,
{\overset{\mathfrak{f}_4}{5}} \, \,
1 \, \,
{\overset{\mathfrak{g}_{2}}{3}} \, \,
{\overset{\mathfrak{su}_2}{2}} \, \,
2 \, \,
1 \, \,
{\overset{\mathfrak{e}_{8}}{12}} \, \,}_{\times (M-1)}  \, \,
1 \, \,
2 \, \,
{\overset{\mathfrak{su}_2}{2}} \, \,
{\overset{\mathfrak{g}_{2}}{3}} \, \,
1 \, \,
{\overset{\mathfrak{so}_{9}}{4}} \, \,
{\overset{\mathfrak{sp}_{1}}{1}} \, \,
{\overset{\mathfrak{so}_{11}}{4}} \, \,
{\overset{\mathfrak{sp}_{2}}{1}} \, \,
\lbrack \mathfrak{so}_{13} \rbrack   \, ,   \qquad \qquad
\lbrack  \mathfrak{sp}_{2}\rbrack     \, \,   
{\overset{\mathfrak{so}_{4M+8}}{4}} 
{\overset{\mathfrak{sp}_{4M-2}}{1}} \, \,  
\myoverset{\overset{\mathfrak{sp}_{3M-4}}{1}}
{\overset{\mathfrak{so}_{12M}}{4}} \, \, 
{\overset{\mathfrak{sp}_{5M-2}}{1}}  \, \,
{\overset{\mathfrak{so}_{8M+8}}{4}} \, \, 
{\overset{\mathfrak{sp}_{3M+12}}{1}} \, \, 
{\overset{\mathfrak{so}_{4M+16}}{4}} \, \, 
{\overset{\mathfrak{sp}_{M+6}}{1}} \, \,
\lbrack \mathfrak{so}_{24} \rbrack 
\end{align*}
  \end{minipage}
}  \\ \hline

      \resizebox{\columnwidth}{!}{
  \begin{minipage}{0.05\textwidth}\Huge
     \begin{align*}
       &\mathfrak{so}_{16}^{M} \\ \\
&14 \\ \\
       & 24M \\ \\
       & 14M+4
         \end{align*}
  \end{minipage}
}&
         \resizebox{16\textwidth}{!}{
  \begin{minipage}{0.1\textwidth}\Huge
   \begin{align*}
   \lbrack  \mathfrak{so}_{12}\rbrack     \, \,   
{\overset{\mathfrak{sp}_{3}}{1}}   \, \,  
\myoverset{1}
{\underset{{\left[\mathfrak{sp}_1\right]}}
{\overset{\mathfrak{so}_{16}}{4}}}  \, \,
{\overset{\mathfrak{sp}_4}{1}} \, \, 
\underbrace{{\overset{\mathfrak{so}_{16}}{4}} \, \,
{\overset{\mathfrak{sp}_4}{1}}}_{\times (M-2)}  \, \,
\myoverset{1}
{\underset{{\left[\mathfrak{sp}_{1}\right]}}
{\overset{\mathfrak{so}_{16}}{4}}} \, \,
{\overset{\mathfrak{sp}_3}{1}}  \, \,
\lbrack \mathfrak{so}_{12} \rbrack   \, ,   \qquad \qquad 
\textcolor{blue}{\lbrack \mathfrak{u}_{1,D_{1}} \times  \mathfrak{u}_{1,D_{2}} \rbrack } \times \bigg(
{\overset{\mathfrak{su}_{2M+2}}{2}}  \, \,   
\myoverset{\myoverset{\overset{\mathfrak{sp}_{2M-6}}{1}}{\overset{\mathfrak{su}_{4M-4}}{2}}}
{\underset{{\left[\mathfrak{su}_{2}\right]}}{\overset{\mathfrak{su}_{4M+4}}{2}}} \, \,     
{\overset{\mathfrak{su}_{2M+8}}{2}} \, \,  
\lbrack \mathfrak{su}_{12} \rbrack \bigg),  
\qquad \qquad 
{\lbrack \mathfrak{so}_{8}\rbrack }    \, \,   
{\overset{\mathfrak{sp}_{M-1}}{1}}   \, \,  
\myoverset{\overset{\mathfrak{sp}_{M-3}}{1}}
{\underset{{\left[\mathfrak{sp}_{2}\right]}}{\overset{\mathfrak{so}_{4M+4}}{4}}} \, \, 
{\overset{\mathfrak{sp}_{2M-2}}{1}}  \, \,
{\overset{\mathfrak{so}_{4M+4}}{4}} \, \, 
{\overset{\mathfrak{sp}_{2M-2}}{1}} \, \, 
\myoverset{\overset{\mathfrak{sp}_{M-3}}{1}}
{\overset{\mathfrak{so}_{4M+4}}{4}} \, \, 
{\overset{\mathfrak{sp}_{M+1}}{1}} \, \,
\lbrack \mathfrak{so}_{16} \rbrack 
 \end{align*}
  \end{minipage}
} \\ \hline

      \resizebox{1.7\columnwidth}{!}{
  \begin{minipage}{0.1\textwidth}\Huge
     \begin{align*}
       &\mathfrak{f}_4 -\mathfrak{e}_6^{M-2}-\mathfrak{f}_4 \\ \\
&14 \\ \\
       & 22M+4 \\ \\
       & 11M+6
              \end{align*}
  \end{minipage}
}&
         \resizebox{12\textwidth}{!}{
  \begin{minipage}{0.1\textwidth}\Huge
   \begin{align*}
   {\lbrack  \mathfrak{so}_{13}\rbrack}     \, \,   
{\overset{\mathfrak{sp}_1}{1}} \, \,   
{\overset{\mathfrak{g}_2}{3}}\, \,  
1 \, \,
{\overset{\mathfrak{f}_4}{5}} \, \,
1 \, \,
{\overset{\mathfrak{su}_3}{3}} \, \
1 \, \,
{\overset{\mathfrak{e}_6}{6}} \, \,
\underbrace{1 \, \,
{\overset{\mathfrak{su}_3}{3}} \, \,
1 \, \,
{\overset{\mathfrak{e}_6}{6}}}_{\times (M-3)} \, \, 
1 \, \,
{\overset{\mathfrak{su}_3}{3}} \, \,
1 \, \,
{\overset{\mathfrak{f}_{4}}{5}}  \, \,
1 \, \, 
{\overset{\mathfrak{g}_2}{3}} \, \,
{\overset{\mathfrak{su}_{2}}{2}} \, \,
2 \, \, 
1 \, \,
\lbrack \mathfrak{e}_{8} \rbrack   \, ,   \qquad
\qquad {\overset{\mathfrak{su}_{2M-2}}{2}}   \, \,  
{\overset{\mathfrak{su}_{4M-4}}{2}} \, \, 
{\overset{\mathfrak{sp}_{3M-3}}{1}}  \, \,
{\underset{{\left[\mathfrak{sp}_{2}\right]}}
{\overset{\mathfrak{so}_{4M+12}}{4}}} \, \,
{\overset{\mathfrak{sp}_{M+5}}{1}} \, \, 
\lbrack \mathfrak{so}_{24} \rbrack
 \end{align*}
  \end{minipage}
}
    \\ \hline

      \resizebox{1.4\columnwidth}{!}{
  \begin{minipage}{0.01\textwidth}\Huge
     \begin{align*}
       &\mathfrak{e}_7 -\mathfrak{e}_8^{M-1} \\ \\
&14 \\ \\
       & 120M-48 \\ \\
       & 30M+7
             \end{align*}
  \end{minipage}
} 
        &  \resizebox{18\textwidth}{!}{
  \begin{minipage}{0.1\textwidth}\Huge
   \begin{align*}
    {\lbrack  \mathfrak{so}_{12}\rbrack}    \, \,   
{\overset{\mathfrak{sp}_1}{1}} \, \,   
{\overset{\mathfrak{so}_7}{3}}\, \, 
{\overset{\mathfrak{su}_2}{2}}\, \, 
1 \, \,
{\overset{\mathfrak{e}_7}{8}} \, \,
1 \, \,
{\overset{\mathfrak{su}_2}{2}} \, \
{\overset{\mathfrak{g}_2}{3}}\, \, 
1 \, \,
{\overset{\mathfrak{f}_4}{5}}\, \,
1 \, \,
{\overset{\mathfrak{g}_2}{3}}\, \, 
{\overset{\mathfrak{su}_2}{2}}\, \,
2 \, \,
1 \, \,
\underbrace{{\overset{\mathfrak{e}_{8}}{12}} \, \,
1  \, \,   
2 \, \,  
{\overset{\mathfrak{su}_2}{2}} \, \,
{\overset{\mathfrak{g}_2}{3}} \, \,
1 \, \,
{\overset{\mathfrak{f}_4}{5}} \, \,
1 \, \,
{\overset{\mathfrak{g}_{2}}{3}} \, \,
{\overset{\mathfrak{su}_2}{2}} \, \,
2 \, \,
1 \, \,
}_{\times (M-2)}  \, \, 
\myoverset{1}
{\overset{\mathfrak{e}_8}{12}} \, \,
1  \, \,   
2 \, \,  
{\overset{\mathfrak{su}_2}{2}} \, \,
{\overset{\mathfrak{g}_2}{3}} \, \,
1 \, \,
{\overset{\mathfrak{f}_4}{5}} \, \,
1 \, \,
{\overset{\mathfrak{g}_{2}}{3}} \, \,
{\overset{\mathfrak{su}_2}{2}} \, \,
2 \, \,
1 \, \,
\lbrack \mathfrak{e}_{8} \rbrack   \, ,   \qquad \qquad
    \underset{{\left[\mathfrak{sp}_{1}\right]}}
{\overset{\mathfrak{so}_{4M+4}}{4}}   \, \, 
{\overset{\mathfrak{sp}_{4M-5}}{1}}   \, \, 
\myoverset{\overset{\mathfrak{sp}_{3M-6}}{1}}
{\overset{\mathfrak{so}_{12M-8}}{4}}   \, \,  
{\overset{\mathfrak{sp}_{5M-5}}{1}} \, \, 
{\overset{\mathfrak{so}_{8M+4}}{4}}  \, \,
{\overset{\mathfrak{sp}_{3M+1}}{1}} \, \, 
\underset{{\left[\mathfrak{sp}_{1}\right]}}
{\overset{\mathfrak{so}_{4M+6}}{4}}  \, \,
{\overset{\mathfrak{sp}_{M+6}}{1}} \, \, 
\lbrack \mathfrak{so}_{24} \rbrack
 \end{align*}
  \end{minipage}
} \\ \hline

      \resizebox{2.3\columnwidth}{!}{
  \begin{minipage}{0.05\textwidth}\Huge
     \begin{align*}
       &\mathfrak{so}_{12} -\mathfrak{e}_7^{M-2}-\mathfrak{so}_{12} \\ \\
&14 \\ \\
       & 48M-48 \\ \\
       & 18M-9
       \end{align*}
  \end{minipage}
} 
         &  \resizebox{18\textwidth}{!}{
  \begin{minipage}{0.1\textwidth}\Huge
   \begin{align*}
\lbrack  \mathfrak{so}_{12}\rbrack     \, \,   
{\overset{\mathfrak{sp}_{2}}{1}}   \, \,  
\underset{{\left[\mathfrak{sp}_{1}\right]}}
{\overset{\mathfrak{so}_{12}}{4}}  \, \,
{\overset{\mathfrak{sp}_{1}}{1}} \, \,
{\overset{\mathfrak{so}_7}{3}} \, \, 
{\overset{\mathfrak{su}_2}{2}} \, \,  
1 \, \,
{\overset{\mathfrak{e}_7}{8}} \, \, 
\underbrace{1 \, \,
{\overset{\mathfrak{su}_2}{2}} \, \,
{\overset{\mathfrak{so}_7}{3}} \, \,
{\overset{\mathfrak{su}_2}{2}} \, \,
1 \, \,
{\overset{\mathfrak{e}_7}{8}}}_{\times (M-3)} \, \, 
1 \, \,
{\overset{\mathfrak{su}_{2}}{2}} \, \,
{\overset{\mathfrak{so}_{7}}{3}} \, \,
{\overset{\mathfrak{sp}_{1}}{1}} \, \,
\underset{{\left[\mathfrak{sp}_{1}\right]}}
{\overset{\mathfrak{so}_{12}}{4}} \, \,
{\overset{\mathfrak{sp}_2}{1}}  \, \,
\lbrack \mathfrak{so}_{12} \rbrack   \, ,   \qquad \qquad
\textcolor{blue}{\lbrack  \mathfrak{u}_{1,D_{1}} \times \mathfrak{u}_{1,D_{2}} \rbrack } \times \left(
{\overset{\mathfrak{so}_{4M}}{4}}  \, \,   
{\overset{\mathfrak{sp}_{4M-8}}{1}}  \, \,     
{\overset{\mathfrak{su}_{6M-8}}{2}} \, \,  
\underset{\left[{\mathfrak{su}_{2}}\right]} 
{\overset{\mathfrak{su}_{4M}}{2}}  \, \,
{\overset{\mathfrak{su}_{2M+6}}{2}} \, \,
\lbrack \mathfrak{su}_{12} \rbrack \right), 
\qquad \qquad
{\lbrack  \mathfrak{so}_{8}\rbrack}     \, \,   
{\overset{\mathfrak{sp}_{M-1}}{1}}   \, \,  
\underset{{\left[\mathfrak{sp}_{2}\right]}}{\overset{\mathfrak{so}_{4M+4}}{4}} \, \, 
{\overset{\mathfrak{sp}_{3M-5}}{1}}  \, \,
\myoverset{\overset{\mathfrak{sp}_{2M-6}}{1}}
{\overset{\mathfrak{so}_{8M-8}}{4}} \, \, 
{\overset{\mathfrak{sp}_{3M-5}}{1}} \, \, 
{\overset{\mathfrak{so}_{4M+4}}{4}} \, \, 
{\overset{\mathfrak{sp}_{M+1}}{1}} \, \,
\lbrack \mathfrak{so}_{16} \rbrack     
 \end{align*}
  \end{minipage}
}\\ \hline

      \resizebox{1.8\columnwidth}{!}{
  \begin{minipage}{0.05\textwidth}\Huge
     \begin{align*}
       &\mathfrak{f}_{4}-\mathfrak{e}_{7}^{M-2}-\mathfrak{f}_{4} \\ \\
&12 \\ \\
       & 48M-54 \\ \\
       & 18M-13
             \end{align*}
  \end{minipage}
} 
          &  \resizebox{16\textwidth}{!}{
  \begin{minipage}{0.1\textwidth}\Huge
   \begin{align*}
    \lbrack  \mathfrak{so}_{10}\rbrack     \, \,   
{\overset{{{\left[\mathfrak{u}_{1}\right]}}}{1}} \, \,   
{\overset{\mathfrak{su}_{3}}{3}}  \, \,
1 \, \,
{\overset{\mathfrak{f}_{4}}{5}}  \, \,
1 \, \,
{\overset{\mathfrak{g}_{2}}{3}}  \, \,
{\overset{\mathfrak{su}_{2}}{2}}  \, \,
1 \, \,
{\overset{\mathfrak{e}_7}{8}} \, \,
\underbrace{1 \, \,
{\overset{\mathfrak{su}_2}{2}} \, \,
{\overset{\mathfrak{so}_7}{3}} \, \,
{\overset{\mathfrak{su}_2}{2}} \, \,
1 \, \,
{\overset{\mathfrak{e}_7}{8}} \, \,}_{\times (M-3)} \, \,
1 \, \,
{\overset{\mathfrak{su}_2}{2}} \, \,
{\overset{\mathfrak{g}_2}{3}} \, \,
1 \, \,
{\overset{\mathfrak{f}_4}{5}} \, \,
1 \, \,
{\overset{\mathfrak{su}_{3}}{3}}  \, \,
{\overset{{{\left[U(1)\right]}}}{1}}\, \,   
\lbrack \mathfrak{so}_{10} \rbrack   \, ,   \qquad \qquad
\myoverset{\overset{\mathfrak{sp}_{M-5}}{1}}
{\overset{\mathfrak{so}_{4M-4}}{4}}  \, \,   
{\overset{\mathfrak{sp}_{3M-7}}{1}}  \, \,     
\myoverset{\overset{\mathfrak{sp}_{2M-6}}{1}}
{\overset{\mathfrak{so}_{8M-8}}{4}} \, \,  
\underset{{\left[\mathfrak{so}_{4}\right]}}
{\overset{\mathfrak{sp}_{3M-3}}{1}}  \, \,
{\overset{\mathfrak{so}_{4M+8}}{4}} \, \,
\underset{{\left[\mathfrak{so}_{4}\right]}} 
{\overset{\mathfrak{sp}_{M+3}}{1}}  \, \,     
\lbrack \mathfrak{so}_{16} \rbrack    
 \end{align*}
  \end{minipage}
}
    \\ \hline

      \resizebox{2.4\columnwidth}{!}{
  \begin{minipage}{0.05\textwidth}\Huge
     \begin{align*}
       &\mathfrak{so}_{7}-\mathfrak{so}_{8}^{M-2}-\mathfrak{so}_{7} \\ \\
&16 \\ \\
       & 8M+2 \\ \\
       & 6M+2
              \end{align*}
  \end{minipage}
}  &  \resizebox{14\textwidth}{!}{
  \begin{minipage}{0.1\textwidth}\Huge
   \begin{align*}
\lbrack  \mathfrak{e}_{7}\rbrack     \, \,    
1 \, \,
{\overset{\mathfrak{su}_2}{2}} \, \,
\underset{{\left[\mathfrak{sp}_{1}\right]}}
{\overset{\mathfrak{so}_7}{3}} \, \,   
1 \, \, 
{\overset{\mathfrak{so}_8}{4}} \, \,
\underbrace{1 \, \,
{\overset{\mathfrak{so}_8}{4}} \, \,}_{\times (M-3)} \, \,
1 \, \,
\underset{{\left[\mathfrak{sp}_{1}\right]}}
{\overset{\mathfrak{so}_7}{3}} \, \, 
{\overset{\mathfrak{su}_{2}}{2}} \, \,
1 \, \,
\lbrack \mathfrak{e}_{7} \rbrack   \, ,   \qquad \qquad
\textcolor{blue}{\lbrack \mathfrak{u}_{1,D} \rbrack } \times \left({\overset{\mathfrak{su}_{2M-2}}{2}}  \, \,   
{\overset{\mathfrak{sp}_{2M-2}}{1}}  \, \,     
{\overset{\mathfrak{su}_{2M+6}}{2}} \, \,  
\lbrack \mathfrak{su}_{16} \rbrack \right),  \qquad \qquad  {\lbrack  \mathfrak{so}_{8}\rbrack }    \, \,   
{\overset{\mathfrak{sp}_{M-1}}{1}}   \, \,  
\myoverset{\overset{\mathfrak{sp}_{M-3}}{1}}
{\myunderset{\underset{\mathfrak{sp}_{M-3}}{1}}
{\overset{\mathfrak{so}_{4M+4}}{4}}}  \, \, 
{\overset{\mathfrak{sp}_{M+3}}{1}} \, \,
\lbrack \mathfrak{so}_{24} \rbrack   
 \end{align*}
  \end{minipage}
}
    \\ \hline

    \resizebox{2.5\columnwidth}{!}{
  \begin{minipage}{0.05\textwidth}\Huge
     \begin{align*}
       &\mathfrak{so}_{11}-\mathfrak{so}_{12}^{M-2}-\mathfrak{so}_{11} \\ \\
&15 \\ \\
       & 16M+25 \\ \\
       & 10M+18
                    \end{align*}
  \end{minipage}} & 
         \resizebox{14\textwidth}{!}{
  \begin{minipage}{0.1\textwidth}\Huge
   \begin{align*}
    \lbrack  \mathfrak{e}_{7}\rbrack     \, \,    
1 \, \,
{\overset{\mathfrak{su}_2}{2}} \, \,
{\overset{\mathfrak{so}_7}{3}} \, \,  
\underset{\left[N_{f}=1/2\right]}
{\overset{\mathfrak{sp}_1}{1}} \, \,
{\overset{\mathfrak{so}_{11}}{4}} \, \,
\underset{\left[N_{f}=1/2\right]}
{\overset{\mathfrak{sp}_2}{1}} \, \,
{\overset{\mathfrak{so}_{12}}{4}} \, \,
\underbrace{{{\overset{\mathfrak{sp}_2}{1}} \, \,
\overset{\mathfrak{so}_{12}}{4}} \, \,}_{\times (M-3)} 
\underset{\left[N_{f}=1/2\right]}
{\overset{\mathfrak{sp}_2}{1}} \, \,
{\overset{\mathfrak{so}_{11}}{4}} \, \, 
{\overset{\mathfrak{sp}_{1}}{1}} \, \,
{\overset{\mathfrak{so}_{9}}{4}} \, \, 
1 \, \,
{\overset{\mathfrak{g}_{2}}{3}} \, \, 
{\overset{\mathfrak{su}_{2}}{2}} \, \, 
2 \, \,
1 \, \,
\lbrack \mathfrak{e}_{8} \rbrack   \, ,   \qquad \qquad
    \lbrack  \mathfrak{su}_{2}\rbrack     \, \, 
\myoverset{\overset{\mathfrak{sp}_{M-3}}{1}}
{\myunderset{\underset{\mathfrak{sp}_{M-3}}{1}}
{\overset{\mathfrak{so}_{4M+4}}{4}}}  \, \,   
{\overset{\mathfrak{sp}_{2M+1}}{1}}  \, \,     
\myoverset{\overset{\mathfrak{sp}_{M}}{1}}
{\overset{\mathfrak{so}_{4M+16}}{4}} \, \, 
{\overset{\mathfrak{sp}_{M+7}}{1}} \, \,
\lbrack \mathfrak{so}_{28} \rbrack \\  
 \end{align*}
  \end{minipage}
}
     
     \\ \hline

           \end{tabular}}

\clearpage

    \resizebox{1\textwidth}{!}{
    \begin{tabular}{|p{1300pt}|c|}  \hline
       \resizebox{2.7\textwidth}{!}{
           ($\mathfrak{g}$,
            $\text{rk}(G_{F})$, 
            $\kappa_{R}$,
            $\text{Dim(CB)})^{T}$  
              }
      & \resizebox{2.7\textwidth}{!}{\Huge T-dual theory description} \\ \hline

      \resizebox{1.2\columnwidth}{!}{
  \begin{minipage}{0.05\textwidth}\Huge
     \begin{align*}
       &\mathfrak{so}_{16}^{M} \\ \\
&15 \\ \\
       & 48M+2 \\ \\
       & 18M+11
                 \end{align*}
  \end{minipage}} &  
         \resizebox{18\textwidth}{!}{
  \begin{minipage}{0.1\textwidth}\Huge
   \begin{align*}
   \textcolor{blue}{\lbrack  \mathfrak{u}_{1,D} \rbrack } \times \bigg(
\lbrack  \mathfrak{su}_{8}\rbrack     \, \,   
{\overset{\mathfrak{su}_{8}}{2}}   \, \,  
{\overset{\mathfrak{sp}_{4}}{1}}   \, \,  
\underbrace{{\overset{\mathfrak{so}_{16}}{4}} \, \,
{\overset{\mathfrak{sp}_{4}}{1}}}_{\times (M-1)}  \, \,
\myoverset{1}
{\underset{\left[\mathfrak{sp}_1\right]}
{\overset{\mathfrak{so}_{16}}{4}}}  \, \,
{\overset{\mathfrak{sp}_{3}}{1}} \, \,
\lbrack \mathfrak{so}_{12} \rbrack \bigg)  \, ,   \qquad \qquad
\textcolor{blue} {\lbrack  \mathfrak{u}_{1,D_{1}} \times  \mathfrak{u}_{1,D_{2}} \times \mathfrak{u}_{1,D_{3}} \rbrack } \times \bigg(
\lbrack  \mathfrak{su}_{2}\rbrack     \, \,
{\overset{\mathfrak{su}_{2M+4}}{2}}  \, \,   
\myoverset{\myoverset{\overset{\mathfrak{sp}_{2M-5}}{1}}{\overset{\mathfrak{su}_{4M-2}}{2}}}
{\overset{\mathfrak{su}_{4M+6}}{2}}  \, \,     
{\overset{\mathfrak{su}_{2M+9}}{2}} \, \,  
\lbrack \mathfrak{su}_{12} \rbrack \bigg), \qquad \qquad  \lbrack  \mathfrak{so}_{10}\rbrack     \, \,   
{\overset{\mathfrak{sp}_{M}}{1}}   \, \,  
\myoverset{\overset{\overset{{\left[\mathfrak{u}_{1} \right]}}{\mathfrak{sp}_{M-2}}}{1}}
{\underset{\left[\mathfrak{sp}_1\right]}
{\overset{\mathfrak{so}_{4M+6}}{4^{*}}}} \, \, 
{\overset{\mathfrak{sp}_{2M-1}}{1}}  \, \,
{\overset{\mathfrak{so}_{4M+6}}{4}} \, \, 
{\overset{\mathfrak{sp}_{2M-1}}{1}} \, \, 
\myoverset{\overset{\overset{{\left[ \mathfrak{u}_{1} \right]}}{\mathfrak{sp}_{M-2}}}{1}}
{\overset{\mathfrak{so}_{4M+6}}{4}} \, \, 
{\overset{\mathfrak{sp}_{M+1}}{1}} \, \,
\lbrack \mathfrak{so}_{14} \rbrack   
 \end{align*}
  \end{minipage}
}
  \\ \hline

      \resizebox{\columnwidth}{!}{
  \begin{minipage}{0.05\textwidth}\Huge
     \begin{align*}
       &\mathfrak{so}_{8}^{M-3}  \\ \\
    & 16 \\ \\
       & 8M+2 \\ \\
       & 6M+2
       \end{align*}
  \end{minipage}
}&  \resizebox{16\textwidth}{!}{
  \begin{minipage}{0.1\textwidth}\Huge
   \begin{align*}
\lbrack  \mathfrak{e}_{6}\rbrack     \, \,    
{\overset{{\textcolor{blue}{{\left[\mathfrak{u}_{1}\right]}}}}{1}} \, \,
{\overset{\mathfrak{su}_2}{2}} \, \,
\underset{\textcolor{blue}{\left[\mathfrak{su}_{2}\right]}}
{\overset{\mathfrak{so}_7}{3}} \, \, 
1 \, \, 
{\overset{\mathfrak{so}_8}{4}} \, \,
\underbrace{1 \, \,
{\overset{\mathfrak{so}_8}{4}} \, \,}_{\times (M-3)} \, \,
1 \, \, 
\underset{\textcolor{blue}{\left[\mathfrak{su}_{2}\right]}}
{\overset{\mathfrak{so}_7}{3}} \, \, 
{\overset{\mathfrak{su}_{2}}{2}} \, \,
{\overset{{\textcolor{blue}{{\left[\mathfrak{u}_{1}\right]}}}}{1}} \, \,\lbrack \mathfrak{e}_{6} \rbrack   \,,   \qquad \qquad 
\textcolor{blue}{\lbrack  \mathfrak{u}_{1,D}\rbrack } \times \left( 
{\overset{\mathfrak{su}_{2M-2}}{2}}  \, \,   
{\overset{\mathfrak{sp}_{2M-2}}{1}}  \, \,     
{\overset{\mathfrak{su}_{2M+6}}{2}} \, \,  
\textcolor{blue}{\lbrack \mathfrak{su}_{16} \rbrack }\right)  \, ,   \qquad \qquad
\textcolor{blue}{\lbrack  \mathfrak{so}_{24}\rbrack}     \, \,   
{\overset{\mathfrak{sp}_{M+3}}{1}}   \, \,  
\myoverset{\overset{\mathfrak{sp}_{M-3}}{1}}
{\myunderset{\overset{\mathfrak{sp}_{M-3}}{1}}
{\overset{\mathfrak{so}_{4M+4}}{4}}}  \, \, 
{\overset{\mathfrak{sp}_{M-1}}{1}} \, \,
\textcolor{blue}{\lbrack \mathfrak{so}_{8} \rbrack}   \, ,  
      \end{align*}
  \end{minipage}
} \\ \hline

    \resizebox{\columnwidth}{!}{
  \begin{minipage}{0.05\textwidth}\Huge
     \begin{align*}
       &\mathfrak{so}_{16}^{M} \\ \\
    &   14 \\ \\
       & 24M \\ \\
       & 14M+4
        \end{align*}
  \end{minipage}
}&  \resizebox{18\textwidth}{!}{
  \begin{minipage}{0.1\textwidth}\Huge
   \begin{align*}
    \lbrack  \mathfrak{so}_{12}\rbrack     \, \,   
{\overset{\mathfrak{sp}_{3}}{1}}   \, \,  
\myoverset{1}
{\underset{{\left[ \mathfrak{sp}_{1}\right]}}
{\overset{\mathfrak{so}_{16}}{4}}}  \, \,
{\overset{\mathfrak{sp}_4}{1}} \, \, 
\underbrace{{\overset{\mathfrak{so}_{16}}{4}} \, \,
{\overset{\mathfrak{sp}_4}{1}}}_{\times (M-2)}  \, \,
\myoverset{1}
{\underset{{\left[ \mathfrak{sp}_{1}\right]}}
{\overset{\mathfrak{so}_{16}}{4}}} \, \,
{\overset{\mathfrak{sp}_3}{1}} \, \,
\lbrack \mathfrak{so}_{12} \rbrack   \, \, ,   \qquad \qquad
\textcolor{blue}{\lbrack  \mathfrak{u}_{1,D_{1}} \times  \mathfrak{u}_{1,D_{2}} \rbrack } \times \bigg(
{\overset{\mathfrak{su}_{2M+2}}{2}}  \, \,   
\myoverset{\myoverset{\overset{\mathfrak{sp}_{2M-6}}{1^{*}}}{\overset{\mathfrak{su}_{4M-4}}{2^{*}}}}
{\underset{\textcolor{blue}{\left[\mathfrak{su}_2 \right]}}{\overset{\mathfrak{su}_{4M+4}}{2}}} \, \,     
{\overset{\mathfrak{su}_{2M+8}}{2}} \, \,  
\lbrack \mathfrak{su}_{12} \rbrack\bigg)   \, ,   \qquad  \qquad
\textcolor{blue}{\lbrack  \mathfrak{so}_{8}\rbrack }    \, \,   
{\overset{\mathfrak{sp}_{M-1}}{1}}   \, \,  
\myoverset{\overset{\mathfrak{sp}_{M-3}}{1}}
{\underset{\textcolor{blue}{\left[\mathfrak{sp}_2\right]}}{\overset{\mathfrak{so}_{4M+4}}{4}}} \, \, 
{\overset{\mathfrak{sp}_{2M-2}}{1}}  \, \,
{\overset{\mathfrak{so}_{4M+4}}{4}} \, \, 
{\overset{\mathfrak{sp}_{2M-2}}{1}} \, \, 
\myoverset{\overset{\mathfrak{sp}_{M-3}}{1^{*}}}
{\overset{\mathfrak{so}_{4M+4}}{4^{*}}} \, \, 
{\overset{\mathfrak{sp}_{M+1}}{1}} \, \,
\lbrack \mathfrak{so}_{16} \rbrack   \,
   \end{align*}
  \end{minipage}
}
    \\ \hline

   \resizebox{\columnwidth}{!}{
  \begin{minipage}{0.05\textwidth}\Huge
     \begin{align*}
       &\mathfrak{so}_{16}^{M} \\ \\
&16 \\ \\
       &24M+50 \\ \\
       & 14M+34
         \end{align*}
  \end{minipage}
}&     
         \resizebox{18\textwidth}{!}{
  \begin{minipage}{0.1\textwidth}\Huge
   \begin{align*}
   \lbrack  \mathfrak{e}_{7}\rbrack     \, \,   
1 \, \,
{\overset{\mathfrak{su}_{2}}{2}}  \, \,
{\overset{\mathfrak{so}_{7}}{3}}  \, \,
{\overset{\mathfrak{sp}_{1}}{1}}  \, \, 
{\overset{\mathfrak{so}_{12}}{4}}  \, \,
{\overset{\mathfrak{sp}_{3}}{1}}  \, \,
\underset{\textcolor{blue}{\left[\mathfrak{sp}_{1}\right]}}
{\overset{\mathfrak{so}_{16}}{4}}  \, \,
\underbrace{{{\overset{\mathfrak{sp}_{4}}{1}}  \, \,
\overset{\mathfrak{so}_{16}}{4}}}_{\times (M-2)}  \, \,
\overset{\mathfrak{sp}_{4}}{1} \, \,
\underset{\textcolor{blue}{\left[\mathfrak{sp}_{1}\right]}}
{\overset{\mathfrak{so}_{16}}{4}} \, \,
{\overset{\mathfrak{sp}_{3}}{1}}  \, \,
{\overset{\mathfrak{so}_{12}}{4}}  \, \,
{\overset{\mathfrak{sp}_{1}}{1}}  \, \,
{\overset{\mathfrak{so}_{7}}{3}}  \, \,
{\overset{\mathfrak{su}_{2}}{2}}  \, \,
{\overset{{\textcolor{blue}{\left[\mathfrak{u}_{1}\right]}}}{1}} \, \, 
\lbrack \mathfrak{e}_{6} \rbrack   \, ,   \qquad \qquad
    \textcolor{blue}{\lbrack  \mathfrak{u}_{1,D} \rbrack } \times \bigg(
{\overset{\mathfrak{su}_{2M+6}}{2}}  \, \,   
\myoverset{\myoverset{\overset{\mathfrak{sp}_{2M-2}}{1}}{\overset{\mathfrak{su}_{4M}}{2}}}
{\overset{\mathfrak{su}_{4M+12}}{2}}  \, \,     
{\overset{\mathfrak{su}_{2M+14}}{2}} \, \,  
{\lbrack \mathfrak{su}_{16}}  \rbrack, \qquad \qquad {\lbrack  \mathfrak{so}_{24}\rbrack }    \, \,   
{\overset{\mathfrak{sp}_{M+7}}{1}}   \, \,  
\myoverset{\overset{\mathfrak{sp}_{M+1}}{1}}
{\overset{\mathfrak{so}_{4M+20}}{4}} \, \, 
{\overset{\mathfrak{sp}_{2M+4}}{1}}  \, \,
{\overset{\mathfrak{so}_{4M+12}}{4}} \, \, 
{\overset{\mathfrak{sp}_{2M}}{1}} \, \, 
\myoverset{\overset{\mathfrak{sp}_{M-3}}{1}}
{\overset{\mathfrak{so}_{4M+4}}{4}} \, \, 
{\overset{\mathfrak{sp}_{M-1}}{1}} \, \,
{\lbrack \mathfrak{so}_{8} }\rbrack   
 \end{align*}
  \end{minipage}
}
     \\ \hline

   \resizebox{\columnwidth}{!}{
  \begin{minipage}{0.05\textwidth}\Huge
     \begin{align*}
       &\mathfrak{su}_{6}^{M} \\ \\
&17 \\ \\
       & 6M+14 \\ \\
       & 6M+13
          \end{align*}
  \end{minipage}
}&      
         \resizebox{18\textwidth}{!}{
  \begin{minipage}{0.1\textwidth}\Huge
   \begin{align*}
    \textcolor{blue}{\lbrack  \mathfrak{u}_{1,D} \rbrack } \times \bigg(
\lbrack  \mathfrak{e}_{6}\rbrack     \, \,   
1 \, \,
{\overset{\mathfrak{su}_{2}}{2}}  \, \,
{\overset{\mathfrak{su}_{4}}{2}}  \, \,
\underset{\textcolor{blue}{\left[\mathfrak{su}_{2}\right]}}
{\overset{\mathfrak{su}_{6}}{2}}  \, \,
\underbrace{{\overset{\mathfrak{su}_{6}}{2}}}_{\times (M-2)}  \, \,
\underset{\textcolor{blue}{\left[\mathfrak{su}_{2}\right]}}
{\overset{\mathfrak{su}_{6}}{2}}  \, \,
{\overset{\mathfrak{su}_{4}}{2}}  \, \,
{\overset{\mathfrak{su}_{2}}{2}}  \, \,
1 \, \,
\lbrack \mathfrak{e}_{6} \rbrack \bigg)   \, ,   \qquad \qquad
     \textcolor{blue}{\lbrack  \mathfrak{u}_{1,D_{1}} \times  \mathfrak{u}_{1,D_{2}} \rbrack } \times \bigg(
{\overset{\mathfrak{su}_{2M+2}}{1}}  \, \,
\myoverset{\overset{\mathfrak{su}_{2M+2}}{1}}
{\overset{\mathfrak{su}_{2M+10}}{2}} \, \, 
\lbrack \mathfrak{su}_{16} \rbrack \bigg),  \qquad \qquad  \textcolor{blue}{\lbrack  \mathfrak{u}_{1,D} \rbrack } \times \left(
{\lbrack  \mathfrak{so}_{24}\rbrack}     \, \,   
{\overset{\mathfrak{sp}_{M+5}}{1}}   \, \,   
{\overset{\mathfrak{su}_{2M+6}}{2}} \, \, 
{\overset{\mathfrak{su}_{2M+2}}{2}} \, \,
{\overset{\mathfrak{sp}_{M-1}}{1}} \, \,
\lbrack \mathfrak{so}_{8} \rbrack  \right)  
 \end{align*}
  \end{minipage}
}  \\ \hline

   \resizebox{\columnwidth}{!}{
  \begin{minipage}{0.05\textwidth}\Huge
     \begin{align*}
       &\mathfrak{so}_{16}^{M} \\ \\
&12 \\ \\
       &24M+22 \\ \\
       & 14M+16
              \end{align*}
  \end{minipage}
}& 
           \resizebox{14\textwidth}{!}{
  \begin{minipage}{0.1\textwidth}\Huge
   \begin{align*}
\lbrack  \mathfrak{e}_{7}\rbrack     \, \,   
1  \, \,   
{\overset{\mathfrak{su}_2}{2}} \, \,
{\overset{\mathfrak{so}_7}{3}} \, \,
{\overset{\mathfrak{sp}_1}{1}} \, \,
{\overset{\mathfrak{so}_{12}}{4}} \, \,
{\overset{\mathfrak{sp}_{3}}{1}} \, \,
\underbrace{\underset{\left[\mathfrak{sp}_{1} \right]}
{\overset{\mathfrak{so}_{16}}{4}} \, \,
{\overset{\mathfrak{sp}_4}{1}}}_{\times (M-1)}  \, \,
\myoverset{1}
{\myunderset{1}
{\overset{\mathfrak{so}_{16}}{4}}} \, \,
\lbrack \mathfrak{sp}_{4}\rbrack   \, ,   \qquad 
\lbrack  \mathfrak{so}_{17}\rbrack     \, \,   
{\overset{\mathfrak{sp}_{M+3}}{1}}   \, \,  
\myoverset{\overset{\overset{\left[N_{f}=1/2 \right]}{\mathfrak{sp}_{M-1}}}{1}}
{\overset{\mathfrak{so}_{4M+11}}{4}} \, \, 
\underset{\left[N_{f}=1/2 \right]}
{\overset{\mathfrak{sp}_{2M+2}}{1}}   \, \,  
{\overset{\mathfrak{so}_{4M+8}}{4}} \, \, 
\underset{\left[N_{f}=1/2 \right]}
{\overset{\mathfrak{sp}_{2M-1}}{1}}   \, \,  
\myoverset{\overset{\mathfrak{sp}_{M-3}}{1^{*}}}
{\overset{\mathfrak{so}_{4M+3}}{4^{*}}} \, \, 
{\overset{\mathfrak{sp}_{M-1}}{1}}   \, \,  
\lbrack \mathfrak{so}_{9} \rbrack  
  \end{align*}
  \end{minipage}
}
    \\ \hline

                  \resizebox{\columnwidth}{!}{
  \begin{minipage}{0.05\textwidth}\Huge
     \begin{align*}
       &\mathfrak{e}_{7}^{M} \\ \\
&12 \\ \\
       & 48M+46 \\ \\
       & 18M+25
              \end{align*}
  \end{minipage}
}& 
          \resizebox{16\textwidth}{!}{
  \begin{minipage}{0.1\textwidth}\Huge
   \begin{align*}
\lbrack  \mathfrak{so}_{13}\rbrack \, \,     
{\overset{\mathfrak{sp}_2}{1}}  \, \,   
{\overset{\mathfrak{so }_{11}}{4}} \, \,   
\underset{\left[N_f=1 / 2\right]} 
{\overset{\mathfrak{sp}_1}{1}} \, \, 
{\overset{\mathfrak{so}_7}{3}} \, \,
{\overset{\mathfrak{su}_2}{2}} \, \,
1 \, \,
{\overset{\mathfrak{e}_7}{8}} \, \,
\underbrace{1 \, \,
{\overset{\mathfrak{su}_2}{2}} \, \,
{\overset{\mathfrak{so}_7}{3}} \, \,
.{\overset{\mathfrak{su}_2}{2}} \, \,
1\, \,
{\overset{\mathfrak{e}_7}{8}}}_{\times (M-1)}\, \,
1 \, \,
{\overset{\mathfrak{su}_2}{2}} \, \,
{\overset{\mathfrak{so}_7}{3}} \, \,
\underset{\left[N_f=1 / 2\right]}
{\overset{\mathfrak{sp}_1}{1}} \, \,
{\overset{\mathfrak{so}_{11}}{4}} \, \,
{\overset{\mathfrak{sp}_2}{1}} \, \,
\lbrack \mathfrak{so}_{13} \rbrack   \, ,   \qquad \qquad
   \lbrack  \mathfrak{sp}_{4}\rbrack     \, \, 
\myoverset{{\overset{\mathfrak{sp}_{M-1}}{1}}}
{\overset{\mathfrak{so}_{4M+12}}{4}}
{\overset{\mathfrak{sp}_{3M+1}}{1}}  \, \,     
\myoverset{{\overset{\mathfrak{sp}_{2M-2}}{1}}}
{\overset{\mathfrak{so}_{8M+8}}{4}} \, \,   
{\overset{\mathfrak{sp}_{3M+1}}{1}}  \, \,
{\overset{\mathfrak{so}_{4M+12}}{4}} \, \,
{\overset{\mathfrak{sp}_{M+3}}{1}} \, \,
\lbrack \mathfrak{so}_{16} \rbrack   \, ,  
  \end{align*}
  \end{minipage}
}
     
     \\ \hline

    \resizebox{\columnwidth}{!}{
  \begin{minipage}{0.05\textwidth}\Huge
     \begin{align*}
       &\mathfrak{e}_{7}^{M} \\ \\
&16 \\ \\
       & 48M+50 \\ \\
       & 18M+29
         \end{align*}
  \end{minipage}
}  
          &   
         \resizebox{18\textwidth}{!}{
  \begin{minipage}{0.1\textwidth}\Huge
   \begin{align*}
   \lbrack  \mathfrak{so}_{16}\rbrack     \, \,     {\overset{\mathfrak{sp}_3}{1}}  \, \,   {\overset{\mathfrak{so }_{12}}{4}}  \, \,      {\overset{\mathfrak{sp}_1}{1}} \, \,   {\overset{\mathfrak{so}_7}{3}}  \, \,
{\overset{\mathfrak{su}_2}{2}} \, \,
1 \, \, {\overset{\mathfrak{e}_7}{8}} \, \,
\underbrace{1 \, \,
{\overset{\mathfrak{su}_2}{2}} \, \,
{\overset{\mathfrak{so}_7}{3}} \, \,
{\overset{\mathfrak{su}_2}{2}} \, \,
1 \, \,
{\overset{\mathfrak{e}_7}{8}} }_{\times (M-1)}\, \,
1
{\overset{\mathfrak{su}_2}{2}} \, \,
{\overset{\mathfrak{so}_7}{3}}  \, \,
{\overset{\mathfrak{sp}_1}{1}} \, \, 
{\overset{\mathfrak{so }_{12}}{4}}  \, \,
{\overset{\mathfrak{sp}_3}{1}}  \, \, 
\lbrack \mathfrak{so}_{16} \rbrack,   \   \qquad \qquad
\textcolor{blue}{\lbrack  \mathfrak{u}_{1,D} \rbrack } \times \left(
     {\overset{\mathfrak{so}_{4M+8}}{4}}  \, \,   {\overset{\mathfrak{sp}_{4M}}{1}}  \, \,      {\overset{\mathfrak{su}_{6M+4}}{2}} \, \,   {\overset{\mathfrak{su}_{4M+8}}{2}}  \, \,
{\overset{\mathfrak{su}_{2M+12}}{2}} \, \,
\lbrack \mathfrak{su}_{16} \rbrack \right)  \, ,   \qquad \qquad  
\lbrack  \mathfrak{so}_{16}\rbrack     \, \,
{\overset{\mathfrak{sp}_{M+3}}{1}}  \, \,   {\overset{\mathfrak{so}_{4M+12}}{4}}  \, \,      {\overset{\mathfrak{sp}_{3M+1}}{1}} \, \, 
\myoverset{{\overset{\mathfrak{sp}_{2M-2}}{1}}}
{\overset{\mathfrak{so}_{8M+8}}{4}}  \, \,
{\overset{\mathfrak{sp}_{3M+1}}{1}} \, \,
{\overset{\mathfrak{so}_{4M+12}}{4}} \, \,
{\overset{\mathfrak{sp}_{M+3}}{1}} \, \,
\lbrack \mathfrak{so}_{16} \rbrack   \, ,  
 \end{align*}
  \end{minipage}
}
  \\ \hline

   \resizebox{\columnwidth}{!}{
  \begin{minipage}{0.18\textwidth}\Huge
     \begin{align*}
       &\mathfrak{e}_{8} \\ \\
        &16 \\ \\
       & 194\\ \\
       & 64
      \end{align*}
  \end{minipage}
}&   
         \resizebox{16\textwidth}{!}{
  \begin{minipage}{0.1\textwidth}\Huge
   \begin{align*}
   \lbrack  \mathfrak{e}_{8}\rbrack     \, \,   
1 \, \,   
2 \, \,  
{\overset{\mathfrak{su}_2}{2}} \, \,
{\overset{\mathfrak{g}_2}{3}} \, \,
1 \, \,
{\overset{\mathfrak{f}_4}{5}} \, \
1 \, \,
{\overset{\mathfrak{g}_2}{3}} \, \,
{\overset{\mathfrak{su}_2}{2}} \, \,
2 \, \,
1 \, \,
{\overset{\mathfrak{e}_8}{12}} \, \,
1 \, \,
2 \, \,
{\overset{\mathfrak{su}_{2}}{2}}  \, \,
{\overset{\mathfrak{g}_2}{3}} \, \, 
1 \, \,
{\overset{\mathfrak{so}_{9}}{4}} \, \,
{\overset{\mathfrak{sp}_1}{1}} \, \,
{\overset{\mathfrak{so}_{11}}{4}} \, \,
{\overset{\mathfrak{sp}_2}{1}} \, \,
{\overset{\mathfrak{so}_{13}}{4}} \, \,
{\overset{\mathfrak{sp}_3}{1}} \, \,
\lbrack \mathfrak{so}_{15} \rbrack   \, ,   \qquad \qquad
    \lbrack  \mathfrak{so}_{7}\rbrack     \, \,   
{\overset{\mathfrak{su}_{2}}{2}}   \, \,  
1 \, \,
{\overset{\mathfrak{e}_7}{8}} \, \,
1 \, \,
{\overset{\mathfrak{su}_{2}}{2}} \, \, 
{\overset{\mathfrak{so}_{7}}{3}}  \, \,
{\overset{\mathfrak{sp}_{1}}{1}} \, \, 
{\overset{\mathfrak{so}_{12}}{4}} \, \, 
{\overset{\mathfrak{sp}_{3}}{1}} \, \, 
{\overset{\mathfrak{so}_{16}}{4}} \, \,
{\overset{\mathfrak{sp}_5}{1}} \, \,
{\overset{\mathfrak{so}_{20}}{4}} \, \,
{\overset{\mathfrak{sp}_7}{1}} \, \,
\lbrack \mathfrak{so}_{24} \rbrack   
\end{align*}
  \end{minipage}
} 
     \\ \hline

   \resizebox{\columnwidth}{!}{
  \begin{minipage}{0.18\textwidth}\Huge
     \begin{align*}
       & \mathfrak{e}_{8} \\ \\
&11 \\ \\
       & 166 \\ \\
       & 52
          \end{align*}
  \end{minipage}
}&  
           
         \resizebox{16\textwidth}{!}{
  \begin{minipage}{0.1\textwidth}\Huge
   \begin{align*}
    \lbrack  \mathfrak{f}_{4}\rbrack     \, \,   
1 \, \,  
{\overset{\mathfrak{g}_2}{3}} \, \,
{\overset{\mathfrak{su}_2}{2}} \, \,
2 \, \,
1 \, \,
\myoverset{1}
{\overset{\mathfrak{e}_8}{12}} \, \,
1 \, \,
2 \, \,
{\overset{\mathfrak{su}_{2}}{2}}  \, \,
{\overset{\mathfrak{g}_2}{3}} \, \, 
1 \, \,
{\overset{\mathfrak{so}_{9}}{4}} \, \,
{\overset{\mathfrak{sp}_1}{1}} \, \,
{\overset{\mathfrak{so}_{11}}{4}} \, \,
{\overset{\mathfrak{sp}_2}{1}} \, \,
{\overset{\mathfrak{so}_{13}}{4}} \, \,
{\overset{\mathfrak{sp}_3}{1}} \, \,
\lbrack \mathfrak{so}_{15} \rbrack   \, ,   \qquad \qquad
\lbrack  \mathfrak{su}_{4}\rbrack     \, \,   
{\overset{\mathfrak{su}_{2}}{2}}   \, \,  
1 \, \,
{\overset{\mathfrak{e}_7}{8}} \, \,
1 \, \,
{\overset{\mathfrak{su}_{2}}{2}} \, \, 
{\overset{\mathfrak{so}_{7}}{3}}  \, \,
\underset{\left[N_{f}=1/2\right]}
{\overset{\mathfrak{sp}_{1}}{1}} \, \, 
{\overset{\mathfrak{so}_{11}}{4}} \, \, 
{\overset{\mathfrak{sp}_{2}}{1}} \, \, 
{\overset{\mathfrak{so}_{13}}{4}} \, \,
{\overset{\mathfrak{sp}_3}{1}} \, \,
{\overset{\mathfrak{so}_{15}}{4}} \, \,
{\overset{\mathfrak{sp}_4}{1}} \, \,
\lbrack \mathfrak{so}_{17} \rbrack   \, ,  
 \end{align*}
  \end{minipage}
}  \\ \hline
\end{tabular}}

%%%%%%%%%%%%%%%%%%%%%%%%% Bibliography
\clearpage
\bibliographystyle{ytphys}
\bibliography{refs}
\end{document}